\documentclass[onecolumn,longbibliography,nofootinbib,superscriptaddress,10pt,aps,prx]{revtex4-2}
\usepackage[T1]{fontenc}
\usepackage{amsmath,amssymb,amsthm,mathtools,bm}
\usepackage{graphicx,booktabs,enumitem,microtype,needspace}
\usepackage{xcolor}
\usepackage{algorithm}
\usepackage[normalem]{ulem}
\makeatletter
\renewcommand{\ftype@algorithm}{8}
\makeatother
\usepackage{algpseudocode}
\usepackage[colorlinks, linkcolor=blue, anchorcolor=blue, citecolor=blue]{hyperref}

\makeatletter
\newif\ifatoc@sublist
\newcommand{\atoc@endsublist}{%
	\ifatoc@sublist
	\par\endgroup
	\atoc@sublistfalse
	\fi
}
\newcommand{\l@atocsection}[2]{%
	\expandafter\atoc@section\expandafter{\Hy@tocdestname}{#1}{#2}%
}
\newcommand{\atoc@section}[3]{%
	\atoc@endsublist
	\addvspace{0.4\baselineskip}%
	\begingroup
	\def\Hy@tocdestname{#1}%
	\@dottedtocline{1}{0em}{2em}{\bfseries#2}{\bfseries#3}%
	\endgroup
	\nopagebreak[4]%
}
\newcommand{\l@atocsubsection}[2]{%
	\ifatoc@sublist
	\unskip\nobreakspace\textbullet\penalty0\hspace{0.4em}%
	\else
	\atoc@sublisttrue
	\begingroup
	\leftskip=2em\relax
	\small
	\rightskip=0pt plus 2em\relax
	\parfillskip=0pt plus 1fil\relax
	\parindent=0pt
	\let\numberline\@gobble
	\leavevmode
	\fi
	#1\ignorespaces
}
\newcommand{\l@atocsubsubsection}[2]{}
\newcommand{\l@atocparagraph}[2]{}
\newcommand{\l@atocsubparagraph}[2]{}
\newcommand{\printappendixtoc}{%
	\begingroup
	\parskip=0pt
	\parindent=0pt
	\setcounter{tocdepth}{2}%
	\hypersetup{linktoc=all}%
	\let\atoc@saved@post\toc@post
	\def\toc@post{\atoc@endsublist\atoc@saved@post}%
	\@starttoc{atoc}%
	\endgroup
}
\makeatother

\newtheorem{theorem}{Theorem}
\newtheorem{lemma}{Lemma}
\newtheorem{proposition}{Proposition}

\theoremstyle{definition}
\newtheorem{definition}{Definition}
\theoremstyle{remark}

\newcommand{\tr}{\operatorname{Tr}}
\newcommand{\TV}{\operatorname{TV}}
\newcommand{\E}{\mathbb E}

\newcommand{\HS}{\mathrm{HS}}

\newcommand{\col}{\operatorname{col}}

\newcommand{\ket}[1]{\lvert#1\rangle}
\newcommand{\bra}[1]{\langle#1\rvert}
\newcommand{\ip}[2]{\langle#1\mid#2\rangle}
\newcommand{\norm}[1]{\lVert#1\rVert}
\newcommand{\cO}{\mathcal{O}}

\setlist{topsep=3pt,itemsep=3pt,leftmargin=1.8em}
\allowdisplaybreaks[2]

\begin{document}
	\title{
		Exponential advantage in quantum observation with one extra qubit
	}

\author{Zhenyu Du}
\affiliation{Center for Quantum Information, Institute for Interdisciplinary Information Sciences, Tsinghua University, Beijing 100084, China}

\author{Xiongfeng Ma}
\email{xma@tsinghua.edu.cn}
\affiliation{Center for Quantum Information, Institute for Interdisciplinary Information Sciences, Tsinghua University, Beijing 100084, China}

\author{Vlatko Vedral}
\affiliation{Clarendon Laboratory, University of Oxford, Parks Road, Oxford, OX1 3PU, United Kingdom}

\author{Jinzhao Sun}
\email{jinzhao.sun.phys@gmail.com}
\affiliation{School of Physical and Chemical Sciences, Queen Mary University of London, London, E1 4NS, United Kingdom}

\begin{abstract}
    While quantum technologies promise to revolutionise how we observe the physical world, achieving substantial advantages with minimal overhead remains a central pursuit. 
    Here, we show that augmenting a quantum machine with just one extra qubit yields an exponential advantage in probing quantum many-body dynamics.
    We propose an algorithm that uses this qubit to coherently interfere dynamical trajectories, simultaneously estimating exponentially many multi-time correlators with exponentially fewer evolutions than any ancilla-free scheme. 
    Remarkably, this advantage holds even when the ancilla-free machine has unrestricted coherence time and arbitrary adaptive control, and persists even for a single correlator.
    We further prove that extending coherence time exponentially enhances observational power, establishing our protocol's optimality in both space and time.
    Finally, we numerically demonstrate our approach by predicting higher-order probe signals in a many-body system with orders-of-magnitude fewer experimental runs than previous methods.
    These results highlight the broad utility of our framework across quantum learning, sensing, and simulation. 
\end{abstract}
\maketitle

\section{Introduction}
Scientific discovery relies on our ability to observe and extract information from the physical world. Within quantum many-body systems, this information is encoded in the complex correlations across their extensive degrees of freedom. 
Accessing these correlations has been a primary goal since the inception of quantum technologies~\cite{feynman1982simulating, lloyd1996universal}, underpinning efforts to simulating quantum dynamics~\cite{kim2023evidence, king2025beyond, haghshenas2026digital, Google2025ConstructiveInterference}, characterising quantum devices~\cite{Crooker2004SpectroscopySpin, Laraoui2013CorrelationSpectroscopyNV, Norris2016SpectroNonGaussian, Sung2019NongaussianSuperconducting, Pfender2019SpectroscopyNuclearWeak, Meinel2022NonlinearSpectroscopyNuclear, Shen2023FreeClassicalNoise}, and probing material properties~\cite{Dorfman2016NonlinearOptical, Lu2017CollectiveSpinWaves, Wan2019SpinEcho, Choi2020TheoryNonlinearKitaev, Zhang2025MolecularGeometry}.
To this end, modern experiments have established a transformative paradigm: deploying controlled ancillary probes that couple coherently to target systems and can themselves be measured with high fidelity. Now ubiquitous across atomic~\cite{cetina_ultrafast_2016}, solid-state~\cite{boss_quantum_2017, Meinel2022NonlinearSpectroscopyNuclear, Rezai2025ProbingDynamics}, and photonic platforms~\cite{Dorfman2016NonlinearOptical, Xia2023OptomechanicalSensing}, this approach has proven highly effective for quantum computing~\cite{Knill1998PowerOneBit}, learning~\cite{Huang2022QuantumAdvantage}, and sensing~\cite{Crooker2004SpectroscopySpin, Laraoui2013CorrelationSpectroscopyNV, Norris2016SpectroNonGaussian, Sung2019NongaussianSuperconducting, Pfender2019SpectroscopyNuclearWeak, Shen2023FreeClassicalNoise, kannan2026SingleQubitAdvantage}, as well as for diagnosing quantum chaos and thermalization~\cite{mi_information_2021, landsman_verified_2019, Kuffer2025SensingOutofEquilibrium}.

Despite their central role, the fundamental advantage of ancillary probes in quantum observation remains elusive. On one hand, compared to conventional approaches, probe-based protocols can achieve an exponential reduction in the number of experiments~\cite{Chen2022Memory, aharonov2022quantum, Huang2022QuantumAdvantage, Liu2025ScalablePhotonic}. Yet realising these reductions requires a large number of ancillary qubits---often comparable to the size of the target system---along with highly entangled measurements, imposing formidable experimental demands. On the other hand, coupling a small, controllable probe to the target system is an experimentally natural and important paradigm across quantum simulation, sensing, and many-body physics~\cite{Pedernales2014EfficientCorrelationFunction, Yao2016InterferometricApproach, Dressel2018WeakMeasurementOTOC, DelRe2024RobustnPointCorrelation, kannan2026SingleQubitAdvantage, Dong2026ProbeAdvantage}. However, whether such minimal ancillas can genuinely outperform direct measurements on the system itself remains an open question. 
Indeed, measuring the system directly already provides access to its dynamics and can resolve complex correlations across multiple spacetime locations~\cite{Dressel2018WeakMeasurementOTOC,DelRe2024RobustnPointCorrelation,Wang2025ComputingnTimeCorrelation}, making it profoundly challenging to establish a rigorous advantage for small probes.
This leads to a central question: 
\begin{center}
  \emph{Can a small quantum probe provide an exponential advantage in observing quantum many-body dynamics?}  
\end{center}

\begin{figure*}[tbp]
\centering
\includegraphics[width=0.98\linewidth]{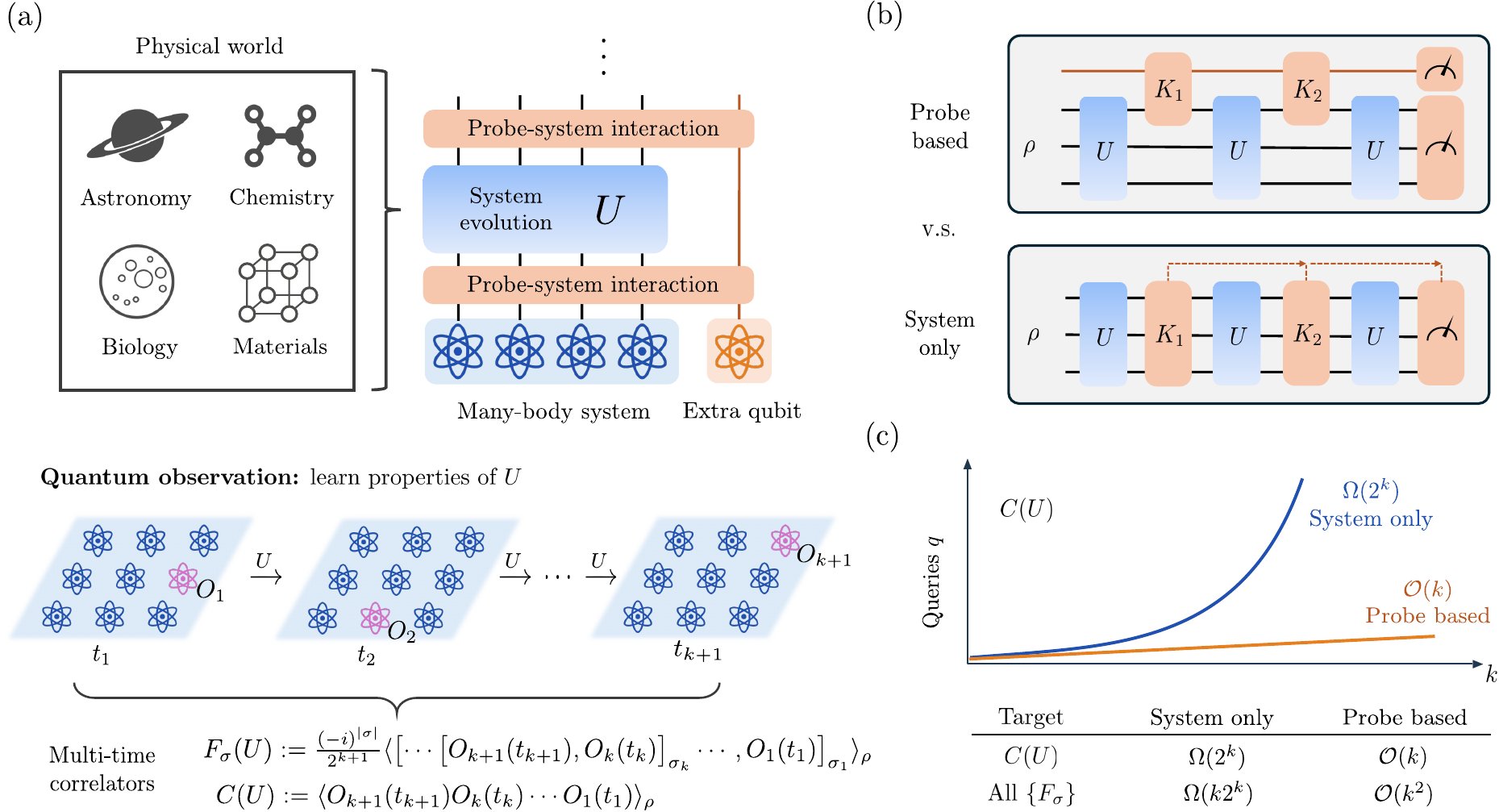}
\caption{
One extra qubit enables exponential advantages in observing quantum dynamics.
(a) Quantum observation framework. Complex dynamics across the physical world or within quantum simulators are governed by an unknown unitary evolution $U$ acting on a many-body system. 
The goal is to extract correlations among observables $O_1,\ldots,O_{k+1}$ across multiple spacetime locations.
We consider two tasks: estimating the full family of $2^k$ nested commutator and anticommutator correlators $F_\sigma(U)$, and estimating a single ordered multi-time correlator $C(U)$.
(b) Comparison of the two operational models.
In both circuits, the unknown unitary $U$ acts solely on the native $n$-qubit system.
In our probe-based protocol (top), a single probe qubit remains coherent between queries; the probe-system interactions $K_j$ are chosen before each run, and the probe and system are measured separately at the end.
In contrast, the system-only model (bottom) allows unrestricted coherence time and arbitrary adaptive mid-circuit operations $K_j$, with dashed arrows indicating classical feedforward, while confining all quantum memory and operations to the native system.
(c) Exponential separations in query complexity.
The schematic plot illustrates the separation for estimating $C(U)$, while the table summarises both tasks.
For total evolution time $T=k\le n/5$, probe-based protocols require $O(k)$ queries to estimate $C(U)$ and $O(k^2)$ queries to estimate all $2^k$ nested correlators, whereas any system-only protocol requires $\Omega(2^k)$ and $\Omega(k2^k)$ queries, respectively.
}
\label{fig:model}
\end{figure*}

Here, we answer this question in the affirmative by showing that augmenting a quantum machine with a single ancillary qubit can exponentially reduce the experimental cost of observing quantum many-body dynamics.
We establish these advantages by formalizing the task of characterising an unknown $n$-qubit unitary evolution $U$, a scenario ubiquitous in native physical environments~\cite{Abobeih2019AtomicImaging, Huang2022QuantumAdvantage, van_de_stolpe_2024_mapping, depradenne2026longtimes} and quantum simulation~\cite{Pedernales2014EfficientCorrelationFunction, Daley2022PracticalAdvantage, Browaeys2020Rydberg} (Fig.~\ref{fig:model}a). 
By comparing two operational models distinguished by a single-qubit probe (Fig.~\ref{fig:model}b), we reveal a stark separation in query complexity, defined as the total number of interactions between the quantum machine and $U$ required across all experimental runs (Fig.~\ref{fig:model}c). 
First, we propose a \emph{probe-based} quantum algorithm that uses this single ancilla to extract a full family of $2^k$ dynamical correlators across $k+1$ spacetime locations, with query complexity that scales linearly in $k$ and the evolution time.
Second, we prove that any \emph{system-only} protocol restricted to the native $n$-qubit register inevitably requires exponentially many queries in $k$ to learn these same properties. 
Proving this lower bound represents a major technical challenge, as it must hold even when the system-only machine is granted unrestricted coherence time and arbitrary adaptive mid-circuit operations (Fig.~\ref{fig:model}b).
Compared with previously established quantum memory advantages~\cite{Chen2022Memory, Huang2022QuantumAdvantage, aharonov2022quantum, Cotler2026SensingClassical, kannan2026SingleQubitAdvantage, Liu2026TimeReversalSymmetry}, our separation reduces the ancillary memory requirement to its single-qubit minimum while holding against a substantially stronger system-only baseline.

We further show that, beyond enabling the simultaneous estimation of exponentially many properties, a single-qubit probe yields an exponential advantage even when extracting a single multi-time correlator~\cite{Pedernales2014EfficientCorrelationFunction, Wang2025ComputingnTimeCorrelation}.
Conceptually, these advantages stem from using the probe to coherently interfere with dynamical trajectories, thereby identifying quantum interference as a key resource for observing many-body dynamics~\cite{Google2025ConstructiveInterference}.
Moreover, our probe-based protocol is optimal in both space and time (ancillary qubits and coherence time): we prove that any protocol that cannot maintain coherence throughout the required interactions incurs exponential query complexity, even with unbounded quantum memory size.
Extending the coherence time of a quantum machine thus provides an exponential advantage in its own right.
Finally, we numerically demonstrate the practical utility of our framework in a many-body quantum sensing task, in which our method predicts higher-order probe signals using two orders of magnitude fewer experimental runs than the system-only baseline~\cite{Dressel2018WeakMeasurementOTOC, DelRe2024RobustnPointCorrelation, Wang2025ComputingnTimeCorrelation}.

\section{Results}

Throughout this work, we consider characterising the physical evolution $U$ of an $n$-qubit many-body system (Fig.~\ref{fig:model}a). 
In system-only protocols, we interface with $U$ exclusively through this native $n$-qubit register. 
To establish a rigorous advantage, we benchmark against the strongest possible system-only baseline: the experimenter is granted unrestricted coherence time and can perform arbitrary adaptive mid-circuit operations---where subsequent operations depend on prior classical outcomes---provided that all retained quantum memory remains strictly confined to the native system itself (Fig.~\ref{fig:model}b). 
In contrast, our probe-based protocols couple the system to a single ancillary qubit that remains coherent across multiple queries to $U$, requiring only simple, non-adaptive probe-system interactions and separate final measurements on the probe and system.
In both scenarios, the experiment concludes with final measurements, and the experimenter must accurately estimate the properties of $U$ solely from the obtained classical data (see Methods for formal mathematical definitions).

\subsection{Exponential advantage in estimating many correlators}
\label{nest:sec:statement}

While characterising quantum dynamics typically begins with standard single-time observables~\cite{lloyd1996universal, kim2023evidence, king2025beyond}, gaining the full picture of many-body dynamics requires extending this paradigm to multi-time correlators. 
By evaluating a sequence of observables $O_1,\ldots, O_{k+1}$ across successive times $0 \le t_1 \le \cdots \le t_{k+1} = T$ (Fig.~\ref{fig:model}a), one can uncover essential dynamical fingerprints that remain hidden from single-time measurements~\cite{Dorfman2016NonlinearOptical, LewisSwan2019DynamicsReview}.
These multi-time correlators frequently manifest as nested commutators and anticommutators, underpinning non-linear spectroscopy~\cite{Lu2017CollectiveSpinWaves, Wan2019SpinEcho, Choi2020TheoryNonlinearKitaev, Meinel2022NonlinearSpectroscopyNuclear, Wu2024SelectiveDetection, Xia2025GaussianToNonGaussian} and non-equilibrium physics~\cite{Kuffer2025SensingOutofEquilibrium, Geier2022NonHermitianLinearResponse}.
Formally, consider a system initialised in a state $\rho$, with observables expressed in the Heisenberg picture as $O_j(t_j)=(U^{t_j})^\dagger O_j U^{t_j}$. 
Denoting the anticommutator and commutator by $[A,B]_0=AB+BA$ and $[A,B]_1=AB-BA$, respectively, each binary string $\sigma=(\sigma_1,\ldots,\sigma_k)\in\{0,1\}^k$ specifies a $(k+1)$-time nested correlator:
\begin{equation}
\label{nest:eq:targets}
    F_\sigma(U)=\frac{(-i)^{|\sigma|}}{2^{k+1}}
    \tr\!\left(
    \rho\,\bigl[\cdots
    \bigl[O_{k+1}(t_{k+1}),O_k(t_k)\bigr]_{\sigma_k}
    \cdots,O_1(t_1)\bigr]_{\sigma_1}
    \right),
\end{equation}
where $|\sigma|=\sum_{j=1}^k\sigma_j$ counts the number of commutators. By construction, $F_{\sigma}(U)$ is a real number satisfying $|F_\sigma(U)|\le 1/2$. For instance, $\sigma = (0,\dots,0)$ and $\sigma = (1,\dots,1)$ reduce Eq.~\eqref{nest:eq:targets} to purely nested anticommutators and commutators, respectively.

\begin{figure*}[!htbp]
\centering
\includegraphics[width=0.7\linewidth]{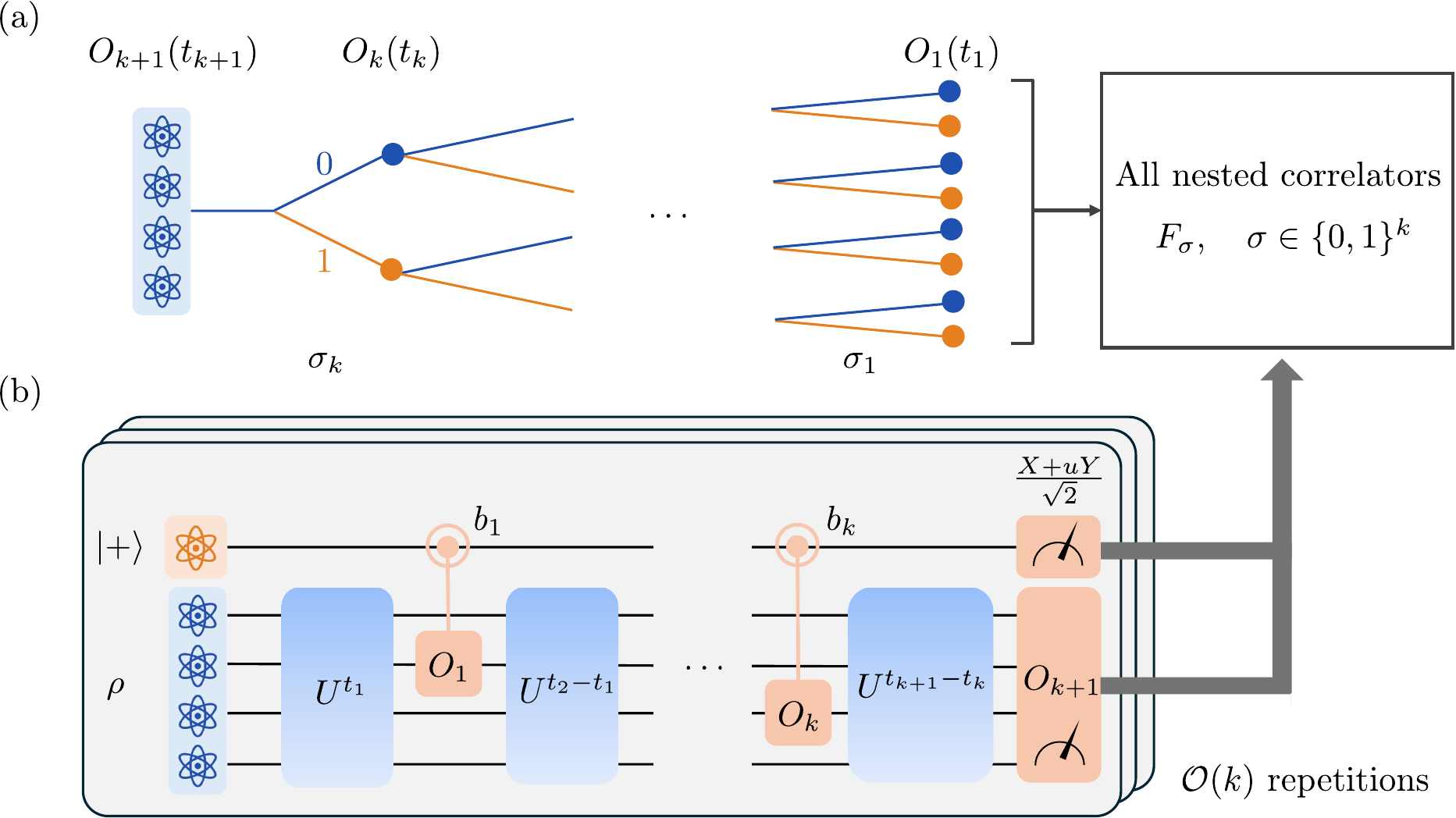}
\caption{Protocol for simultaneously estimating many nested correlators. (a) Structure of nested correlators. The system's dynamics generate a full family of $2^k$ nested commutators and anticommutators. Each correlator is specified by a binary string $\sigma$, so the family size grows exponentially with the number of time segments $k$. (b) The probe-based quantum circuit. In each experimental run, a random trajectory configuration $b$ is sampled, and the system and ancillary probe qubit are initialised in $\rho$ and $\ket{+}$, respectively. Evolution under $U$ is interleaved with conditional probe-system interactions governed by $b$ and the observables $O_j$, followed by final measurements on both registers. Repeating this single randomised procedure $\mathcal{O}(k)$ times suffices to reconstruct all $2^k$ nested correlators simultaneously.}
\label{fig:algo}
\end{figure*}

Our first task is to characterise the dynamics by accurately estimating the entire family of $M=2^k$ nested correlators (Fig.~\ref{fig:algo}a). 
We focus on the widely considered scenario in which each observable $O_j$ is unitary, such as a local Pauli operator. 
While extensive efforts have been devoted to measuring these quantities, most existing protocols are tailored to a specific configuration $\sigma$~\cite{Dressel2018WeakMeasurementOTOC, DelRe2024RobustnPointCorrelation,Wu2024SelectiveDetection,Wang2025ComputingnTimeCorrelation} and therefore require a separate measurement procedure for each target. 
Even with snapshot-based protocols that estimate multiple correlators simultaneously~\cite{Wang2024Snapshotting}, reconstructing all $M=2^k$ correlators incurs exponential overhead in $k$.

Here, we overcome this experimental bottleneck by proposing a quantum algorithm that couples the system to a single ancillary probe qubit and reconstructs all $2^k$ correlators simultaneously using only $\cO(k)$ experimental runs (Fig.~\ref{fig:algo}b). Our results establish that this probe-based approach (1) improves upon existing schemes for estimating correlators and (2) achieves an exponential advantage over \emph{all} system-only protocols.

\begin{theorem}[Exponential single-qubit advantage in estimating many correlators, informal]
    \label{nest:thm:advantage_main}
    Consider estimating all $2^k$ nested correlators $\{F_\sigma(U)\}_{\sigma\in\{0,1\}^k}$ of an unknown $n$-qubit unitary evolution $U$ over $k+1$ successive times up to total evolution time $T$, each to within additive error $\varepsilon$ with high probability.
    \begin{enumerate}
        \item Probe-based: With one ancillary qubit, Algorithm~\ref{alg:nested} achieves this using $q = \mathcal{O}\!\left(T\varepsilon^{-2}k\right)$ queries.
        \item System-only: For $T=k\le n/5$ and constant error $\varepsilon$, any ancilla-free protocol requires $q = \Omega(k 2^k)$ queries.
    \end{enumerate}
\end{theorem}

The formal statements and proofs are provided in Appendices~\ref{nest:sec:proof} and~\ref{nest:sec:upper}, with the detailed procedure presented in Algorithm~\ref{alg:nested}. 
The core insight of our algorithm is to use the probe qubit to render measurements of distinct nested correlators compatible via their Walsh--Hadamard spectra.
Specifically, for each binary string $b = (b_1, \ldots, b_k) \in \{0,1\}^k$ and its bitwise complement $\bar b = 1 - b$, we define the dynamical trajectory $W_b$ and the complementary-path overlap $Q_b$ as
\begin{equation}
    \label{nest:eq:main-paths}
\begin{split}
    W_b &= U^{t_{k+1}-t_k} O_k^{b_k} \cdots U^{t_2-t_1} O_1^{b_1} U^{t_1}, \\
    Q_b &= \tr\left(O_{k+1} W_b \rho W_{\bar b}^\dagger\right).
\end{split}
\end{equation}
For a pure initial state $\rho = \ket\psi\bra\psi$, the overlap $Q_b = \bra{\psi_{\bar b}} O_{k+1} \ket{\psi_b}$ represents the interference amplitude between two complementary histories $\ket{\psi_b} = W_b\ket\psi$ and $\ket{\psi_{\bar b}} = W_{\bar b}\ket\psi$. 
Our key observation is that the target nested correlators are given by the Walsh--Hadamard transform of these trajectory overlaps (see Lemma~\ref{nest:lem:expansion}):
\begin{equation}
\label{main:eq:fourier}
    F_\sigma(U)=\frac{i^{|\sigma|}}{2}\,
    \E_{b\sim\{0,1\}^k}\!\left[(-1)^{\sigma\cdot b}Q_b\right].
\end{equation}

Crucially, the interference amplitudes $Q_b$ depend solely on the path string $b$ and are independent of the target configuration $\sigma$.
Consequently, the experimental procedure need not be tailored to any specific nested correlator, thereby achieving measurement compatibility.
Coupling the native system to a single probe qubit makes these amplitudes directly accessible: in each experimental run, we uniformly sample $b$ and interleave forward queries to $U$ with coherent, probe-controlled applications of the known observables $O_j$ according to $b$ (Fig.~\ref{fig:algo}b).
Separate final measurements on the probe and system then yield an unbiased \emph{dynamical snapshot} of $Q_b$, which encodes rich information about the underlying correlations.
From $\cO(k)$ such independent snapshots, the algorithm simultaneously reconstructs the entire family of $2^k$ nested correlators via the empirical Walsh--Hadamard transform.
One may draw an analogy to the classical shadows of a quantum state, which also aim to predict many observables~\cite{Huang2020Predicting}.

Establishing the matching exponential lower bound is our main technical contribution, as it requires ruling out all possible system-only protocols, including those equipped with arbitrary adaptive mid-circuit operations. 
To achieve this, we construct a hard ensemble of random unitary evolutions that hides the target signal in a single, unknown correlator configuration $\theta\in\{0,1\}^k$. 
Physically, the signal is encoded in the relative phase between extended dynamical trajectories whose intermediate states are obscured by Haar-random bases. 
While a single-qubit probe can coherently interfere these hidden paths, any system-only protocol attempting to navigate the random bases blindly inevitably suffers a contraction of quantum coherence. 
We formalize this intuition by reducing any system-only protocol to a drastically simplified \emph{ideal process} (see Methods for the physical picture and Appendices~\ref{layer:section}--\ref{ideal:sec:comparison} for the full reduction). 
Within this ideal process, we prove that the information any ancilla-free measurement can extract about $\theta$ decays exponentially with $k$, solidifying the indispensable role of the extra qubit (Appendix~\ref{nest:sec:proof}).

For $T=k$, these results establish an exponential single-qubit advantage---$\cO(k^2)$ versus $\Omega(k 2^k)$ queries---in probing multi-time dynamical correlations, highlighting that a single extra qubit enables the reconstruction of exponentially more dynamical properties than any system-only protocol.

\subsection{Exponential advantage in estimating a single correlator}
\label{main:overview}

While learning exponentially many properties simultaneously yields a profound information-theoretic advantage, investigations of large-scale many-body systems often aim to isolate specific target quantities. 
Yet, if we restrict our focus to estimating a single, predetermined nested correlator $F_\sigma(U)$, this exponential separation appears to vanish, as this specific quantity can be efficiently extracted in $\cO(T\varepsilon^{-2})$ queries by adapting ancilla-free sequential weak-measurement protocols~\cite{Dressel2018WeakMeasurementOTOC,DelRe2024RobustnPointCorrelation,Wang2025ComputingnTimeCorrelation} (Appendix~\ref{nest:sec:single-free}).
To demonstrate the advantage on actual devices, an immediate question arises: when the task is to estimate just a single dynamical property, does a single-qubit probe still provide an exponential advantage?

We answer this question in the affirmative, showing that the exponential single-qubit advantage persists even when estimating a single multi-time correlation function. 
In contrast to the nested correlators analyzed above, this $(k+1)$-time correlator is defined as
\begin{equation}
\label{main:target}
    C(U)
    =
    \tr\!\left[
        O_{k+1}(t_{k+1})O_k(t_k)\cdots O_1(t_1)\rho
    \right]
    =
    Q_{\mathbf 1},
\end{equation}
where $\mathbf 1=(1,\ldots,1)$.
Such correlation functions are also central to diagnosing quantum dynamics~\cite{Sung2019NongaussianSuperconducting, Laraoui2013CorrelationSpectroscopyNV, Dorfman2016NonlinearOptical, Crooker2004SpectroscopySpin, Schweigler2017ExperimentalHigherOrder, Pedernales2014EfficientCorrelationFunction, Wang2025ComputingnTimeCorrelation, Norris2016SpectroNonGaussian} and demonstrating practical quantum advantages~\cite{Neill2021Accurately, Kokcu2024LinearResponse, Sun2025ProbingSpectral}.

Physically, $C(U) = Q_{\mathbf 1}$ represents the interference amplitude between the fully perturbed trajectory $W_{\mathbf 1}$, which contains all intermediate operator insertions, and the unperturbed trajectory $W_{\mathbf 0}=U^T$, probed by the final observable $O_{k+1}$ at time $T=t_{k+1}$ (Eq.~\eqref{nest:eq:main-paths}). 
Its probe-based estimation thus follows from the interference primitive of Algorithm~\ref{alg:nested}. 
Specifically, one can estimate $C(U)$ by fixing the path configuration to $b=\mathbf 1$ and averaging the snapshots $\widehat Q$. 
This procedure recovers the single-qubit interferometric scheme of Ref.~\cite{Pedernales2014EfficientCorrelationFunction} (Appendix~\ref{corr:sec:upper}).

While this interferometric approach is known to be highly efficient and outperforms existing ancilla-free algorithms, whether it achieves a strict exponential advantage over \emph{all} possible system-only strategies has remained an open question. 
This requires a general lower bound on system-only strategies.
The theorem below resolves this question by establishing an exponential separation.

\begin{theorem}[Exponential single-qubit advantage in estimating a single correlator, informal]
    \label{main:thm:advantage_single}
    Consider estimating the $(k+1)$-time correlator $C(U)$ of an unknown $n$-qubit unitary evolution $U$ across $k+1$ successive times up to total evolution time $T$, within additive error $\varepsilon$ with high probability. 
    \begin{enumerate}
        \item Probe-based: With one ancillary qubit, $C(U)$ can be estimated using $q = \mathcal{O}\!\left(T\varepsilon^{-2}\right)$ total queries.
        \item System-only: For $T=k\le n/5$ and constant error $\varepsilon$, any ancilla-free protocol requires $q = \Omega(2^k)$ queries.
    \end{enumerate}
\end{theorem}

Proving this exponential lower bound is highly challenging because it requires ruling out all system-only strategies, even when targeting a single correlator. 
To achieve this, we apply the reduction framework developed above to map any ancilla-free protocol to a simplified ideal process.
Within this framework, we construct two ensembles of unitary dynamics with opposite correlator values, $C(U)=\pm 1$, encoded solely in the relative phase between two length-$k$ dynamical trajectories whose intermediate states remain hidden. 
We prove that any system-only protocol produces nearly indistinguishable measurement records for these two ensembles unless it expends exponentially many queries (see Appendix~\ref{hard:section} for the formal derivation).

For $T=k$, this $\cO(k)$ versus $\Omega(2^k)$ query separation arises from extracting just a single interference amplitude, identifying quantum interference itself as the fundamental resource provided by the ancillary probe.

\subsection{Temporal optimality: Exponential advantage from quantum memory time}

The exponential advantages established above could be understood by preserving coherence between two dynamical trajectories and interfering them at readout. We now complete the last jigsaw by addressing the temporal optimality of the probe-based approach. 
In other words, does extending the memory time of a quantum machine also exponentially enhance its observational power compared to machines with shorter coherence times?


We answer this question in the affirmative: extending the coherence time to encompass just one additional interaction with the evolution $U$ yields an exponential advantage. 
To formalize this, we define the \emph{coherent query number} as the maximum number of queries to $U$ permitted within a single experimental run before all quantum registers must be measured or reset. 
Using this metric, we prove a sharp complexity threshold that demonstrates the temporal optimality of our protocol: if an experimenter is limited to a coherent query number of $L < k$ per run, estimating either the nested correlators $F_\sigma(U)$ or the single multi-time correlator $C(U)$ inevitably requires an exponentially larger number of experiments, regardless of the size of the available quantum memory. 
Notably, when the ancillary register consists of only a few qubits, as in our protocol, this coherent query number directly dictates the required memory time.

\begin{theorem}[Optimality: Exponential lower bound with restricted coherence, informal]
    \label{thm:query_depth_informal}
    Consider estimating either the single multi-time correlator $C(U)$ or a prescribed nested correlator $F_\sigma(U)$ across $k+1$ successive times of an unknown $n$-qubit unitary evolution $U$, within constant additive error with high probability. 

    For $k\le n / 5$, any protocol restricted to at most $L < k$ coherent queries per run requires a total query complexity of $q = 2^{\Omega(n)}$. This exponential overhead persists even with unbounded ancillary quantum memory and arbitrary adaptive mid-circuit operations.
\end{theorem}

The formal statement is given in Theorem~\ref{depth:thm}. 
For consecutive observation times $t_j=j-1$, our probe-based algorithm (Algorithm~\ref{alg:nested}) requires only $k$ forward queries per run to probe correlations across $k+1$ time steps. It thus achieves simultaneous optimality in both quantum memory size and coherence time, yielding an exponential advantage over any protocol operating below this coherence threshold. 

Practically speaking, this exponential separation holds even when available coherence time is modest (e.g., $k=2$), offering a pathway for demonstrating quantum observational advantages on near-term devices~\cite{Google2025ConstructiveInterference,Neill2021Accurately, Kokcu2024LinearResponse, Sun2025ProbingSpectral}, where coherence time remains the dominant bottleneck~\cite{Brock2025Breakeven, Google2026BelowThreshold}.

\subsection{Numerical experiments: predicting probe signals}

A key task in quantum sensing is understanding how external perturbations affect a coupled probe's signal, providing insight into environmental noise and its effects on quantum coherence~\cite{Wang2021QuantumQuenches} and guiding quantum control~\cite{Wu2024SelectiveDetection}. In particular, resolving higher-order components of these signals enables applications such as distinguishing quantum nuclear spins from classical fields~\cite{Meinel2022NonlinearSpectroscopyNuclear} and revealing non-Gaussian noise in qubits~\cite{Sung2019NongaussianSuperconducting}.

\begin{figure*}[!t]
\centering
\includegraphics[width=0.98\linewidth]{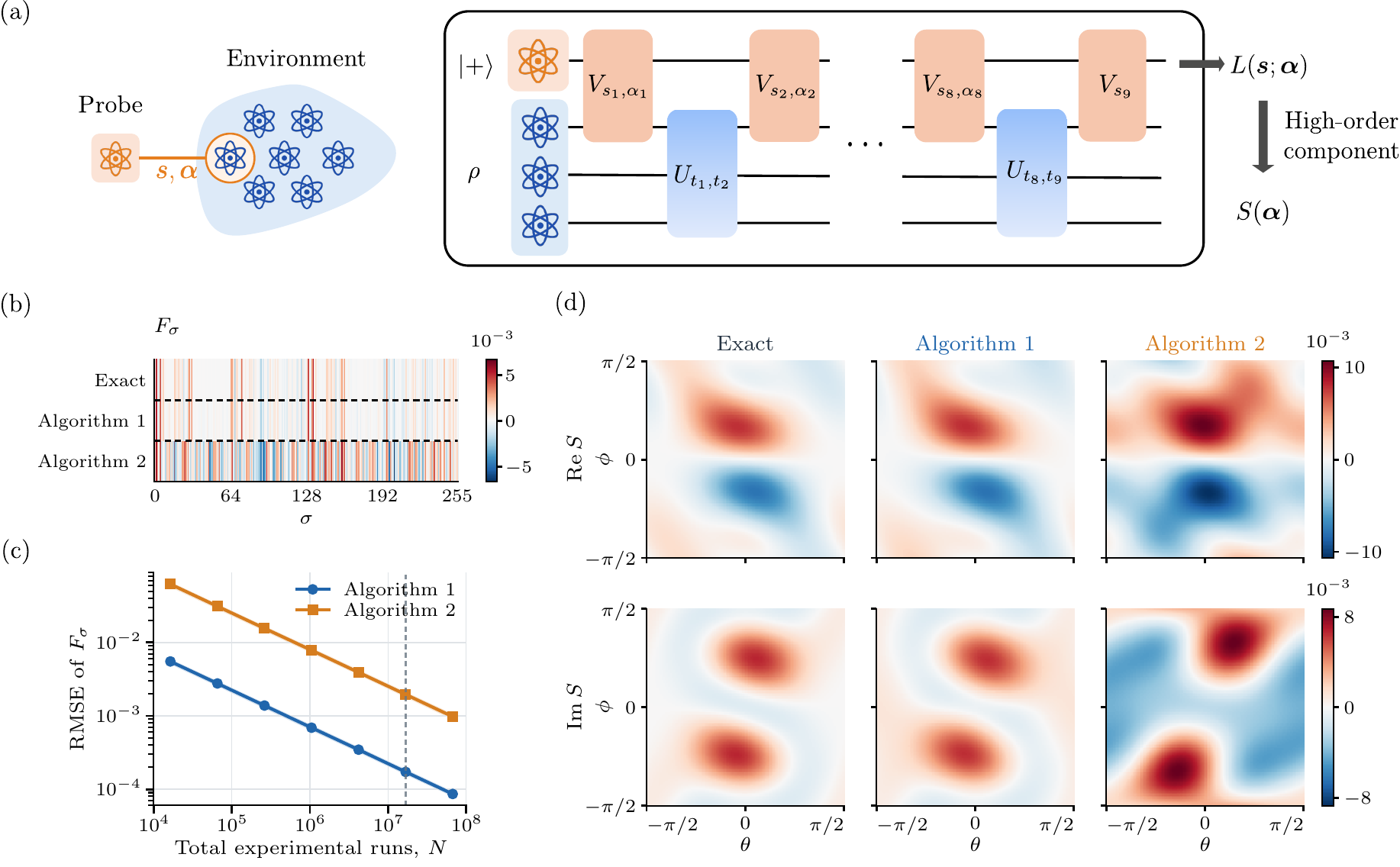}
\caption{
Learning nested correlators to predict higher-order probe signals.
(a) Target sensing experiment.
A single-qubit probe interacts with a boundary spin of a ten-spin
mixed-field Ising environment.
Control operations $V_{s_j,\alpha_j}$, parameterized by pulse signs $s_j$
and angles $\alpha_j$, are interleaved with environment evolution
$U_{t,t'}=e^{-iH\Delta(t'-t)}$.
The prediction target is the sign-weighted higher-order component
$S(\boldsymbol{\alpha})$ of the probe coherence
$L(\boldsymbol{s};\boldsymbol{\alpha})$.
(b) Exact values and estimates of all $256$ nested correlators
$F_\sigma$, using $N=2^{24}$ experimental runs per algorithm.
Algorithm~\ref{alg:nested} uses shared measurement records
to estimate the entire family, whereas
Algorithm~\ref{alg:single-nested} allocates $N/256$ runs
to each correlator.
(c) RMSE over all correlators and $1024$ independent
measurement datasets at each budget $N$.
Shaded bands indicate the central $95\%$ interval of the
per-dataset RMSE.
The vertical dashed line marks $N=2^{24}$, the budget used
in (b) and (d).
(d) Exact and predicted real (top) and imaginary (bottom)
parts of $S(\theta,\phi)$, with
$\alpha_{1,3,5,7}=\theta$ and $\alpha_{2,4,6,8}=\phi$.
The predictions are reconstructed from the correlator
estimates in (b), without additional experiments. 
}
\label{fig:num_exp}
\end{figure*}

Motivated by these capabilities, we numerically demonstrate that our learning algorithm predicts higher-order probe signals across different perturbations, outperforming the system-only baseline~\cite{Dressel2018WeakMeasurementOTOC,DelRe2024RobustnPointCorrelation,Wang2025ComputingnTimeCorrelation} (Appendix~\ref{nest:sec:single-free}).
We consider detecting the many-body environment as a $n$-spin system (with $n = 10$) and take its unknown evolution $U=e^{-iH\Delta}$ to be generated by the widely studied, nonintegrable mixed-field Ising Hamiltonian,
\begin{equation}
H=\sum_{i=1}^{n-1}Z_iZ_{i+1}
    +1.05\sum_{i=1}^{n}X_i
    +0.5\sum_{i=1}^{n}Z_i, \nonumber
\end{equation}
with time step $\Delta = 0.4$. 
The environment is initialised in the nonequilibrium state $\rho=(I+X_1)/2^{10}$ and probed using an ancillary qubit $Q$ prepared in $\ket{+}$. 
The probe interacts with the boundary spin at times $(t_1,\ldots,t_9)=(0,1,3,4,5,7,9,10,12)$. 
These interactions are parameterized by control angles $\boldsymbol{\alpha}=(\alpha_1,\ldots,\alpha_8)$ and independent, uniformly sampled pulse signs $\boldsymbol{s}=(s_1,\ldots,s_9) \in \{\pm 1\}^9$. 
At each of the first eight steps ($j=1,\ldots,8$), we apply a boundary rotation $e^{-is_j\alpha_j Z_1/2}$ and a probe--system coupling $e^{-is_j(\pi/2-\alpha_j)Z_Q\otimes Z_1/2}$. 
At the final step, we apply the coupling $e^{-is_9\pi Z_Q\otimes Z_1/4}$ (Fig.~\ref{fig:num_exp}a).

Our task is to predict how the resulting probe signal varies with the control parameters. Specifically, we consider the probe's coherence term, $L(\mathbf{s};\boldsymbol{\alpha})=\langle X_Q\rangle-i\langle Y_Q\rangle$, and focus on its sign-weighted component, which can be expressed as (Appendix~\ref{app:numerical})
\begin{equation}
\label{eq:numerical-signal}
S(\boldsymbol{\alpha})
:=\mathbb E_{\boldsymbol{s}}
\left[
\left(\prod_{j=1}^{9}s_j\right)
L(\boldsymbol{s};\boldsymbol{\alpha})
\right] = 2\sum_{\sigma\in\{0,1\}^{8}}
w_\sigma(\boldsymbol{\alpha})F_\sigma(U).
\end{equation}
Physically, this high-order coherence component $S(\boldsymbol{\alpha})$ of a single-qubit probe coupled to a many-body environment can reveal noise that cannot be resolved from low-order environmental spectra.
Importantly, the second equality expresses the signal as a linear combination of nested correlators, where $F_\sigma(U)$ are the correlators defined in Eq.~\eqref{nest:eq:targets} with $O_j=Z_1$.
The control dependence of the signal is captured by the coefficients $w_\sigma(\boldsymbol{\alpha})$, and therefore estimating these nested correlators allows us to predict the probe signal $S(\boldsymbol{\alpha})$ for arbitrary unseen $\boldsymbol{\alpha}$ without requiring new experiments.

We apply Algorithm~\ref{alg:nested} to estimate all $2^8=256$ nested correlators from $N$ experimental runs. As a benchmark, we use the system-only protocol (Algorithm~\ref{alg:single-nested}, Appendix~\ref{nest:sec:single-free}), which estimates each correlator separately using $N/256$ experimental runs. Both methods use $t_9=12$ queries to $U$ per run, ensuring an equal total query budget of $q=12N$. At $N=2^{24}$, our algorithm closely reproduces the full correlator spectrum, while the system-only benchmark exhibits substantially larger errors (Fig.~\ref{fig:num_exp}b). Fig.~\ref{fig:num_exp}c quantifies this improvement via the root-mean-square error (RMSE), averaged over the correlator family and $D=1024$ independent datasets. The approximately ten-fold reduction in RMSE implies that our algorithm requires roughly two orders of magnitude fewer experimental runs to achieve the same accuracy.

We then use these learned correlators to predict the probe signal without additional experiments. For visualization, we set $\alpha_j=\theta$ for odd $j$ and $\alpha_j=\phi$ for even $j$. Using the same correlator estimates as in Fig.~\ref{fig:num_exp}b, our algorithm successfully captures the structure and amplitude of both the real and imaginary parts of $S(\theta,\phi)$, whereas the system-only reconstruction shows substantial distortions (Fig.~\ref{fig:num_exp}d). This demonstrates that our probe-based approach outperforms traditional system-only baselines in predicting such nonlinear signals. 
The above analysis can be naturally applied to nonlinear spectroscopy of many-body systems, whose response functions depend on the higher-order structure of nested multi-time correlators, which reveals features obscured in conventional linear-response spectra~\cite{Lu2017CollectiveSpinWaves, Wan2019SpinEcho, Choi2020TheoryNonlinearKitaev}.
Details of the numerical experiments are provided in Appendix~\ref{app:numerical}.

\section{Discussion}

We have shown that augmenting a quantum machine with just one qubit yields an exponential advantage in observing quantum many-body dynamics. 
This advantage is threefold. 
First, the single-qubit probe renders the measurements of distinct dynamical properties compatible, enabling the reconstruction of exponentially many nested correlators. 
Second, by coherently interfering extended dynamical trajectories, the probe achieves an exponential reduction in query complexity even when estimating a single multi-time correlator. 
Finally, sustaining this coherence across the full trajectory is optimal, establishing an exponential separation over any protocol restricted to shorter memory times.

Our framework is highly relevant to state-of-the-art experimental architectures, such as probe-based quantum technologies~\cite{cetina_ultrafast_2016, Rezai2025ProbingDynamics, Xia2025GaussianToNonGaussian} and hybrid digital-analogue quantum simulators~\cite{Bluvstein2022CoherentTransport, Andersen2025Thermalization}. 
While recent experimental breakthroughs have significantly advanced quantum memory times~\cite{Brock2025Breakeven, Google2026BelowThreshold, He2025LeakageSuppresion}, performing fault-tolerant quantum computation remains challenging. 
Our probe-based protocol bridges this gap by leveraging modest memory sizes and coherence times to realise exponential advantages without prohibitive resource overhead.
Going forward, it is intriguing to investigate whether introducing additional probes can unlock further capabilities~\cite{Dong2026ProbeAdvantage}, such as estimating out-of-time-order correlators, and thus establish a finer-grained hierarchy of advantages based on probe size~\cite{Hsieh2026DepthHierarchy}.
Our framework also provides a pathway to characterise the fundamental spacetime trade-offs between ancillary memory and coherence time in quantum learning and sensing.

The maturation of digital quantum simulation will grant experimenters significantly richer control over the underlying evolution. 
Specifically, access to an explicit gate decomposition of $U$ enables the realisation of advanced query types, such as time-reversed evolutions $U^\dagger$ or coherently controlled unitaries, which may offer further advantages in quantum observation~\cite{Tang2026ConjugateQueries, Tang2026ControlledUnitaries, King2024ConjugateStates}. 
Furthermore, because practical many-body systems are typically governed by known locality structures or symmetries, incorporating such prior physical knowledge can further reduce the overhead of learning or certification~\cite{Haah2024LearningHighTemperature, Hu2025AnsatzFreeLearning, Gao2026HamiltonianCertification, Liu2026TimeReversalSymmetry}.
Ultimately, fault-tolerant architectures will enable the integration of these observational tasks with advanced quantum algorithms, raising the compelling question of whether this synthesis can yield unprecedented speedups~\cite{Allen2025EnhancedSensing, liu2026optimalquantumsensingalgorithms,  kannan2026algorithmicquantumsensing}.

\begin{acknowledgments}
Z.D. and X.M. are supported by the National Natural Science Foundation of China (Grant No. 12575023), the Quantum Science and Technology-National Science and Technology Major Project (Grants No.~2021ZD0300804 and No.~2021ZD0300702), the CCF-QuantumCtek Superconducting Quantum Computing Special Cooperation Program (Grant No.~CCF-QC2025005), and the Turing AI Institute of Nanjing. V.V. thanks the Gordon and Betty Moore Foundation and the Conjecture Institute for financial support.

\end{acknowledgments}
	
\bibliographystyle{ref_style}
\bibliography{ref}

\section*{Methods}
\subsection{Operational models}
To capture the most general adaptive strategy in the system-only model, we consider a native $n$-qubit register $\mathcal H=\mathbb C^d$ ($d=2^n$) and a classical measurement history $h$. The operations between successive queries are represented by a history-dependent quantum instrument $\{K_{z\mid h}\}$ satisfying $\sum_z K_{z\mid h}^\dagger K_{z\mid h}=I$, which absorbs any initialisations and intermediate operations. Given a history $h$ and a new outcome $z$, the unnormalised branch state $\sigma$ of the quantum register evolves across the $j$-th query as
\begin{equation}
    \sigma_{hz}^{(j+1)}(U)
    =U K_{z\mid h}^{(j)}\sigma_h^{(j)}(U)
    (K_{z\mid h}^{(j)})^\dagger U^\dagger.
\end{equation}
After $q$ queries, the final operations are absorbed into a history-dependent POVM $\{M_{y|h}\}$, yielding the output distribution
\begin{equation}
    p(y\mid U)=\sum_h\tr\bigl(M_{y\mid h}\sigma_h^{(q)}(U)\bigr).
\end{equation}

By contrast, the probe-based model introduces one additional qubit, on which the unknown evolution $U$ never acts, that can remain coherent between successive queries (Fig.~\ref{fig:model}b).
Notably, while our system-only lower bounds hold for arbitrary adaptive instruments $\{K_{z|h}\}$, our probe-based protocols require only non-adaptive probe--system interactions chosen before each run and separate final measurements on the probe and system.

In the time-constrained setting (Theorem~\ref{thm:query_depth_informal}), we focus on the \emph{coherent query number} $L$. Here, the experimenter is granted arbitrary and unbounded ancillary quantum registers. However, each experimental run is restricted to at most $L$ forward queries to $U$. Between runs, all quantum memory must be measured or completely discarded, retaining only the classical records to adaptively configure subsequent experiments.

\subsection{Algorithm for estimating many nested correlators}
Our probe-based protocol utilizes a single ancillary qubit to coherently interfere complementary dynamical trajectories and access the Walsh--Hadamard spectrum of the correlator family, as detailed in Algorithm~\ref{alg:nested}.

\begin{algorithm}[!htbp]
    \caption{Simultaneous estimation of nested correlators with one extra qubit}
    \label{alg:nested}
    \begin{algorithmic}[1]
        \Require Initial state $\rho$, known observables
        $O_1,\ldots,O_{k+1}$, observation times
        $0\le t_1\le\cdots\le t_{k+1}$,
        and number of repetitions $N$.

        \For{$r=1,\ldots,N$}
            \State Prepare the system in $\rho$ and the probe in $\ket+$.
            \Comment{Initialise and randomise}
            \State Draw $b\in\{0,1\}^k$ uniformly
            and an independent fair sign $u\in\{\pm1\}$.
            \State Evolve the system for $t_1$ steps under $U$.

            \For{$j=1,\ldots,k$}
                \Comment{Interfere complementary trajectories}
                \State Apply
                $\ket0\bra0\otimes O_j^{b_j}
                +\ket1\bra1\otimes O_j^{1-b_j}$.
                \State Evolve the system for
                $t_{j+1}-t_j$ steps under $U$.
            \EndFor

            \State Measure $(X+uY)/\sqrt2$ on the probe,
            obtaining $a\in\{\pm1\}$.
            \State Measure $O_{k+1}$ on the system,
            obtaining $x\in\{\pm1\}$.
            \State Store $b_r=b$ and
            $\widehat Q_r=\sqrt2\,ax(1-iu)$.
        \EndFor

        \For{$\sigma\in\{0,1\}^k$}
            \Comment{Walsh--Hadamard transform}
            \State Compute
            $\displaystyle
            \widehat F_\sigma
            =\frac{1}{2N}\sum_{r=1}^{N}
            (-1)^{\sigma\cdot b_r}
            \operatorname{Re}\!\left[
                i^{|\sigma|}\widehat Q_r
            \right]$.
        \EndFor
        \State \Return $\{\widehat F_\sigma\}_{\sigma\in\{0,1\}^k}$.
    \end{algorithmic}
\end{algorithm}

In this algorithm, each experimental run measures a single-shot unbiased estimator of the complementary-path overlap. Specifically, by preparing the probe in a superposition state and performing conditional evolutions, the final joint measurement of the probe and the system yields a snapshot $\widehat Q$ satisfying $\mathbb{E}[\widehat Q \mid b] = Q_b$. As detailed in Appendix~\ref{nest:sec:upper}, this interferometric primitive ensures that the variance of each snapshot remains bounded. By applying standard concentration inequalities (such as Hoeffding's inequality) and a union bound over all $M=2^k$ bracket patterns, we establish that $N = \mathcal{O}(\varepsilon^{-2} \log(M/\delta))$ total experimental repetitions are sufficient to guarantee an additive error of $\varepsilon$ across all nested correlators simultaneously with high probability.
    
    \subsection{Physical intuition underlying the lower bounds}

    Here, we outline the essential mechanism underlying the exponential lower bounds for system-only protocols (Theorems~\ref{nest:thm:advantage_main} and~\ref{main:thm:advantage_single}). Both constructions employ similar hard instances, encoding the target signal within the interference between hidden trajectories. To build intuition, we focus here on the simpler single-correlator case, deferring the simultaneous estimation of multiple correlators to Appendix~\ref{nest:sec:proof}.

    \subsubsection{Encoding the signal in complementary trajectories}
    
    To establish the exponential lower bound, we construct an ensemble of unitary dynamics where the target correlator $C(U) = s \in \{\pm 1\}$ is encoded exclusively in the relative phase between two complementary dynamical trajectories. We initialise the system in a known root state $\ket{a_0} = \ket{\psi}$ and define an orthogonal counterpart $\ket{b_0} = O\ket{\psi}$ using a known dichotomic observable $O=Z_1$. At each internal layer $1\le j < k$, the unitary $U$ propagates these states deeper into the system:
    \begin{equation}
        U\ket{a_{j-1}} = \ket{a_j}, \qquad
        U\ket{b_{j-1}} = O\ket{b_j}.
    \end{equation}
    Notice that the first trajectory evolves purely under $U$, while the second requires an interleaved insertion of $O$ at every step. At the final depth $k$, the two trajectories reach $\ket{a_k}$ and $sO\ket{a_k}$, respectively. 
    A final measurement of $O$ generates an interference amplitude exactly equal to $C(U) = s$. Importantly, altering $s$ only flips the relative phase, leaving the density matrix of either individual path completely unchanged. A probe-based protocol naturally reads out this relative sign by coherently routing the system along these two paths.
    
    To make this task intractable for a system-only machine, we make these intermediate trajectories opaque. Rather than using fixed and known states, we draw $\ket{a_j}$ and $\ket{b_j}$ from independent Haar-random orthonormal frames at each internal layer. As a result, the states $\ket{a_j}, O\ket{a_j}, \ket{b_j}, O\ket{b_j}$ become mutually orthogonal and entirely unknown to the experimenter. We complete these prescribed transitions into a random unitary oracle, ensuring that the target sign is deeply buried (see Appendices~\ref{app:planted_family} and~\ref{app:hard_oracle_multi_time} for detailed constructions).
    
    \subsubsection{Irreversible contraction of quantum coherence}
    
    To bound the information a system-only protocol can extract, we map its physical evolution to a highly simplified \emph{ideal process} (Theorem~\ref{layer:thm:simulation}). This framework accounts for any arbitrary adaptive strategy, including intermediate measurements and feedback controls, by revealing that attempting to blindly navigate the random frames induces a severe effective decoherence mechanism.
    
    In the ideal process, a contribution can only reach the final target by successfully surviving all $k-1$ internal layers. At each step, a measurement outcome $z$ acts on the paths via an effective control $\alpha_z I + \beta_z O$. Due to the orthogonality of the random bases, only the $I$ component advances the first trajectory ($\ket{a_j}$), while only the $O$ component advances the second ($O\ket{b_j}$). Consequently, any off-diagonal quantum coherence $c$ between the two paths is attenuated to $c_z = \alpha_z \overline{\beta_z} \, c$ in the subsequent layer. Because the physical instrument must satisfy the completeness relation $\sum_z(|\alpha_z|^2+|\beta_z|^2)\le 1$, the surviving coherence strictly contracts by at least a factor of two (Lemma~\ref{sign:lem:contraction}):
    \[
        \sum_z |c_z| = |c|\sum_z|\alpha_z\beta_z| \le \frac{1}{2} |c|.
    \]
    
    Since a trajectory originates at the root with a coherence magnitude bounded by half its probability weight, $k-1$ consecutive contractions exponentially suppress the available coherent signal to at most $2^{-k}$ upon reaching the endpoint. Flipping the target sign $s$ solely reverses this exponentially small off-diagonal element, perturbing the final ideal output distribution by at most $2^{1-k}$ in total variation distance. Summing this error across a protocol with $q$ queries yields a collective bound of $q\,2^{1-k}$. Transferring this bound back to the true physical ensembles gives
    \[
        \TV(\mathsf{Out}_+,\mathsf{Out}_-)
        \le q\,2^{1-k}
        +\mathcal O\!\left(\frac{q^2+q\sqrt{k}}{\sqrt d}\right),
    \]
    where the second term accounts for the error introduced by replacing the physical dynamics with the ideal process. 
    An accurate estimator of $C(U)$ with additive error $1/3$ and success probability $2/3$ distinguishes the two signs, strictly requiring a total variation distance of at least $1/3$. For $n\ge4k$, this forces the query complexity to be $q=\Omega(2^k)$ (Theorem~\ref{hard:thm:formal}).

    \clearpage
	\appendix
	\onecolumngrid
	
	\begin{center}
		{\large \textbf{Supplementary Material}}
	\end{center}
	\vspace{0.5\baselineskip}
	
	\printappendixtoc
	
	\makeatletter
	\let\oldaddcontentsline\addcontentsline
	\renewcommand{\addcontentsline}[3]{%
		\def\target{#1}%
		\def\toclabel{toc}%
		\ifx\target\toclabel
		\oldaddcontentsline{atoc}{atoc#2}{#3}%
		\else
		\oldaddcontentsline{#1}{#2}{#3}%
		\fi
	}
	\makeatother
	\vspace{0.5\baselineskip}
	
	\renewcommand{\thetheorem}{S\arabic{theorem}}
	\renewcommand{\thelemma}{S\arabic{lemma}}
	\renewcommand{\theproposition}{S\arabic{proposition}}
	\renewcommand{\thecorollary}{S\arabic{corollary}}
	\renewcommand{\thedefinition}{S\arabic{definition}}
	\renewcommand{\theremark}{S\arabic{remark}}
	\setcounter{theorem}{0}
	\setcounter{lemma}{0}
	\setcounter{proposition}{0}
	\setcounter{corollary}{0}
	\setcounter{definition}{0}
	\setcounter{remark}{0}
	\renewcommand{\theHtheorem}{supp.\arabic{theorem}}
	\renewcommand{\theHlemma}{supp.\arabic{lemma}}
	\renewcommand{\theHproposition}{supp.\arabic{proposition}}
	\renewcommand{\theHcorollary}{supp.\arabic{corollary}}
	\renewcommand{\theHdefinition}{supp.\arabic{definition}}
	\renewcommand{\theHremark}{supp.\arabic{remark}}
	\numberwithin{equation}{section}
	
	\section{Layered hidden oracles and their ideal process}\label{layer:section}
	
	This appendix presents the construction shared by the two lower bounds.
	We first construct a family of random unitaries using the eigenspaces of the known observable $O$.
	We then introduce a simpler ideal process that replaces the oracle queries and the instruments applied between queries.
	Finally, the simulation theorem  (Theorem~\ref{layer:thm:simulation}) shows that the classical output distribution of the original protocol, averaged over the random unitary, is close to that of the ideal process.

    Throughout, $\norm{\cdot}$, $\norm{\cdot}_1$, and
	$\norm{\cdot}_{\HS}$ denote the operator, trace, and Hilbert--Schmidt
	norms, respectively. We write
	$\TV(p,p')=\frac12\norm{p-p'}_1$ for total variation distance and
	$\col_z A_z:x\mapsto\bigoplus_z A_zx$ for a column operator.
	
	\subsection{The planted family}
	\label{app:planted_family}
	Let $O$ be a known dichotomic observable on $\mathcal H=\mathbb C^d$
	satisfying $O^2=I$ and $\tr O=0$.
	Choose one known unit vector in each eigenspace, and call their span
	the root subspace.
	Let $J:\mathbb C^2\to\mathcal H$ be an isometry onto this subspace,
	with projector $R_0=JJ^\dagger$.
	For $\sigma\in\{+,-\}$, let $\mathcal H_\sigma$ be the orthogonal
	complement of the chosen vector within the corresponding eigenspace,
	and let $\Pi_\sigma$ be its projector.
	Then
	\[
	\mathcal H=\operatorname{Ran}J\oplus\mathcal H_+\oplus\mathcal H_-,
	\qquad \dim\mathcal H_\sigma=N=d/2-1.
	\]
	For $1\le r\le N$, draw independent Haar-distributed $r$-frames
	$F_\sigma:\mathbb C^r\to\mathcal H_\sigma$.
	Define
	\[
	\mathcal L=\mathcal L_+\oplus\mathcal L_-,
	\qquad \mathcal L_\sigma=\mathbb C^r,
	\qquad F=F_+\oplus F_-.
	\]
	The space $\mathcal L$ provides fixed labels for the hidden states,
	while the isometry $F$ embeds them into the physical space $\mathcal H$.
	Let $Z_{\mathcal L}$ act as $+I$ on $\mathcal L_+$ and $-I$ on
	$\mathcal L_-$, so that $OF=FZ_{\mathcal L}$.
	
	Choose an orthogonal decomposition of the label space into layers:
	\begin{equation}\label{layer:eq:layers}
		\mathcal L=\bigoplus_{j=1}^{\ell}\mathcal L_j,
		\qquad Z_{\mathcal L}\mathcal L_j\subseteq\mathcal L_j,
		\qquad \ell\ge1.
	\end{equation}
	Each layer is invariant under $Z_{\mathcal L}$ and therefore splits
	into its $+$ and $-$ sectors.
	Its image $F\mathcal L_j$ is the corresponding hidden physical layer.
	Set $\mathcal L_0=\mathbb C^2$ for the root.
	For each intermediate layer $1\le j<\ell$, choose a finite-dimensional
	input coordinate space $\mathcal I_j$ and an isometric embedding
	$V_j:\mathcal I_j\to\mathcal L_j$.
	Write
	\[
	\Gamma_j=V_jV_j^\dagger
	\]
	for the projector onto $\operatorname{Ran}V_j$.
	This is the subspace on which we will prescribe a transition to
	layer $j+1$, and it need not fill $\mathcal L_j$.
	At the root, take $\mathcal I_0=\mathcal L_0$, $V_0=I$, and
	$\Gamma_0=I$.
	At the endpoint, set $\Gamma_\ell=0$, since no prescribed transition
	leaves the last layer.
	
	Let $\vartheta$ index the oracle family.
	The label spaces, their layer decomposition, and the input embeddings
	$V_j$ are fixed independently of both $\vartheta$ and the random frames.
	The dependence on $\vartheta$ enters through prescribed transitions:
	for each $\vartheta$, choose partial isometries
	$S_{\vartheta,j}:\mathcal L_j\to\mathcal L_{j+1}$, independently
	of the frames, satisfying
	\begin{equation}\label{layer:eq:partial-isometries}
		S_{\vartheta,j}^\dagger S_{\vartheta,j}=\Gamma_j,
		\qquad 0\le j<\ell.
	\end{equation}
	Thus $S_{\vartheta,j}$ acts isometrically on
	$\operatorname{Ran}V_j$ and vanishes on its orthogonal complement.
	Each prescribed transition advances by one layer, and none returns
	to the root.
	
	Collect the input coordinate spaces into
	\[
	\mathcal I=\bigoplus_{j=0}^{\ell-1}\mathcal I_j,
	\qquad p=\dim\mathcal I
	=2+\sum_{j=1}^{\ell-1}\dim\mathcal I_j\le2+2r.
	\]
	The combined transition map $S_\vartheta:\mathcal I\to\mathcal L$
	and the physical input and output embeddings
	$V_F,W_{F,\vartheta}:\mathcal I\to\mathcal H$ are defined by
	\begin{equation}\label{layer:eq:column-maps}
		\begin{aligned}
			S_\vartheta|_{\mathcal I_j}
			&=S_{\vartheta,j}V_j
			&&(0\le j<\ell),\\
			V_F|_{\mathcal I_0}
			&=J,\\
			V_F|_{\mathcal I_j}
			&=FV_j
			&&(1\le j<\ell),\\
			W_{F,\vartheta}
			&=FS_\vartheta.
		\end{aligned}
	\end{equation}
	These maps are isometries: their blocks preserve inner products and
	have mutually orthogonal ranges.
	For $V_F$, this also uses the orthogonality of the root and the hidden
	physical layers.
	
	We now complete the prescribed transitions to a unitary.
	Set $D=d-p$.
	Choose isometries $C_F,B_{F,\vartheta}:\mathbb C^D\to\mathcal H$
	whose ranges are the orthogonal complements of
	$\operatorname{Ran}V_F$ and $\operatorname{Ran}W_{F,\vartheta}$,
	respectively.
	Draw $H$ independently of the frames and define
	\begin{equation}\label{layer:eq:oracle}
		U_{F,H,\vartheta}
		=W_{F,\vartheta}V_F^\dagger+B_{F,\vartheta}HC_F^\dagger,
		\qquad H\sim\operatorname{Haar}(\mathrm U(D)).
	\end{equation}
	The first term implements the prescribed transitions, while the second
	maps the remaining input space unitarily onto the remaining output space.
	The two terms have orthogonal input and output spaces, so
	$U_{F,H,\vartheta}$ is unitary and satisfies
	\[
	U_{F,H,\vartheta}V_F=W_{F,\vartheta}.
	\]
	The pair $(F,H)$ is sampled once, and the resulting unitary is used
	at every query.
	By Haar invariance, its distribution does not depend on the choice of bases for the complementary spaces.
	
	\subsection{The ideal process}
	
	The ideal process simplifies the physical dynamics by removing explicit
	dependence on the random frames $F$ and completion $H$.
	It tracks transitions in the label spaces, retaining quantum coherence within each layer but not between different layers.
	A residual block $\ket{\perp}\bra{\perp}$ records only probability weight.
	The intuition is that the random frames and completion spread untracked
	components over a large physical space, making their detailed structure
	difficult to exploit with few queries.
	Theorem~\ref{layer:thm:simulation} below quantifies this simplification by comparing the final
	classical output distributions, with the physical distribution averaged
	over the single initial draw of $(F,H)$.
	
	Specifically, an ideal state is block diagonal on
	\begin{equation}\label{layer:eq:state}
		\mathcal L_0\oplus\mathcal L_1\oplus\cdots\oplus\mathcal L_\ell
		\oplus\mathbb C\ket{\perp},
		\qquad
		\omega=\rho_0\oplus\cdots\oplus\rho_\ell
		\oplus p_\perp\ket{\perp}\bra{\perp}.
	\end{equation}
	Here $\rho_j\succeq0$, $p_\perp\ge0$, and
	$\sum_{j=0}^{\ell}\tr\rho_j+p_\perp=1$.
	For a known physical initial state $\rho$, initialise the ideal process as
	\begin{equation}\label{layer:eq:initial}
		\omega_{\mathrm{in}}
		=J^\dagger\rho J\ \oplus\ 0_{\mathcal L}
		\oplus\bigl(1-\tr(R_0\rho)\bigr)\ket{\perp}\bra{\perp}.
	\end{equation}
	Thus, the root block retains the compression of $\rho$ to the known root
	subspace, while the remaining probability is assigned to the residual block.
	
	On a layer-$j$ input, the ideal query acts as
	\begin{equation}\label{layer:eq:query}
		\mathcal Q_\vartheta(\rho_j)
		=S_{\vartheta,j}\rho_jS_{\vartheta,j}^\dagger
		+\tr[(I-\Gamma_j)\rho_j]\ket{\perp}\bra{\perp},
		\qquad 0\le j<\ell.
	\end{equation}
	The component supported on $\operatorname{Ran}\Gamma_j$ advances to
	layer $j+1$, while the remaining weight becomes residual.
	All endpoint weight is sent to $\perp$, and the existing residual weight remains there.
	Summing these contributions defines the query on a general ideal state.
	By~\eqref{layer:eq:partial-isometries}, this map is completely positive and trace-preserving.
	No query outputs into the root, so the root block is zero immediately after every query.
	
	Between queries, the ideal process uses transition rules determined by the physical protocol's instruments.
	It follows the same adaptive classical policy: at each history, the
	chosen physical instrument determines the corresponding ideal instrument.
	These instruments may return weight to the root before the next query.
	
	Consider a physical instrument $\{K_z\}_z$ satisfying
	$\sum_zK_z^\dagger K_z=I$, with history dependence suppressed.
	Define
	\begin{equation}\label{layer:eq:instrument-coefficients}
		\begin{aligned}
			\kappa_{z\sigma}
			&=\frac{\tr(\Pi_\sigma K_z)}N,
			&D_z
			&=\bigoplus_{\sigma=\pm}
			\kappa_{z\sigma}I_{\mathcal L_\sigma},\\
			p_{z\sigma}
			&=\frac{\tr(\Pi_\sigma K_z^\dagger K_z)}N,
			&Q_{z\sigma}
			&=\frac{J^\dagger K_z\Pi_\sigma K_z^\dagger J}{N},\\
			w_{z\sigma}
			&=p_{z\sigma}-\tr Q_{z\sigma}-|\kappa_{z\sigma}|^2.
		\end{aligned}
	\end{equation}
	These coefficients are determined by $K_z$ and the known sector
	projectors, and no sampled frame appears in them.
	The operator $D_z$ describes the component that stays in its current
	layer, while $Q_{z\sigma}$ and $w_{z\sigma}$ describe transfer to the
	root and residual blocks.
	Specifically, let $P_{j\sigma}$ be the projector onto the $\sigma$ sector of
	$\mathcal L_j$.
	For a hidden input block $\rho_j$, $1\le j\le\ell$, the unnormalised
	contributions associated with outcome $z$ are
	\begin{equation}\label{layer:eq:hidden-instrument}
		\begin{aligned}
			\text{same layer: }&
			D_z\rho_jD_z^\dagger,\\
			\text{root: }&
			\sum_{\sigma=\pm}\tr(P_{j\sigma}\rho_j)Q_{z\sigma},\\
			\text{residual weight: }&
			\sum_{\sigma=\pm}\tr(P_{j\sigma}\rho_j)w_{z\sigma}.
		\end{aligned}
	\end{equation}
	Sector invariance ensures that $D_z$ preserves each layer.
	For unit residual input $\ket{\perp}\bra{\perp}$, the corresponding contributions are
	\begin{equation}\label{layer:eq:residual-instrument}
		\begin{aligned}
			\text{root: }&
			\frac{J^\dagger K_zK_z^\dagger J}{d},\\
			\text{residual weight: }&
			\frac{\tr(K_z^\dagger K_z)
				-\tr(J^\dagger K_zK_z^\dagger J)}d.
		\end{aligned}
	\end{equation}
	For residual weight $p_\perp$, multiply these expressions by $p_\perp$.
	There is no residual-to-hidden transition.
	
	We write $\Phi_K=\{\Phi_{K,z}\}_z$ for this ideal instrument.
	Summing the contributions from all input blocks gives the unnormalised ideal state for outcome $z$.
	Its trace is the outcome probability.
	Append $z$ to the classical history and select the next instrument according to the physical protocol's policy.
	
	Each outcome map is completely positive, and their sum preserves trace on states with no root block, as occurs immediately after a query.
	To verify this, first observe that $Q_{z\sigma}\succeq0$ and
	\begin{equation}\label{layer:eq:positivity}
		\begin{aligned}
			p_{z\sigma}-\tr Q_{z\sigma}
			&=\frac{\norm{(I-R_0)K_z\Pi_\sigma}_{\HS}^2}{N}\\
			&\ge\frac{\norm{\Pi_\sigma K_z\Pi_\sigma}_{\HS}^2}{N}
			\ge\frac{|\tr(\Pi_\sigma K_z)|^2}{N^2}
			=|\kappa_{z\sigma}|^2.
		\end{aligned}
	\end{equation}
	Hence $w_{z\sigma}\ge0$.
	The same-layer contribution is a Kraus map.
	The other hidden-input contributions measure a sector and prepare
	a positive root matrix or residual weight.
	For input supported in sector $\sigma$, their combined output trace,
	including the same-layer contribution, is $p_{z\sigma}$ times the
	input trace.
	Kraus completeness gives $\sum_zp_{z\sigma}=1$.
	The residual-input contributions are also positive, and their total
	trace for outcome $z$ is $\tr(K_z^\dagger K_z)/d$, which sums to one.
	
	It remains to specify the final measurement.
	For an operator $K$ on $\mathcal H$, define its sector-averaged action
	on the hidden label space by
	\[
	T(K)=\bigoplus_{\sigma=\pm}
	\frac{\tr(\Pi_\sigma K)}N I_{\mathcal L_\sigma}.
	\]
	Since the root block is zero after the last query, the terminal effects need only act on the hidden and residual blocks.
	For a physical POVM $\{M_y\}_y$, the ideal effects are
	\begin{equation}\label{layer:eq:terminal}
		M_y^{\mathrm{id}}
		=T(M_y)\oplus
		\frac{\tr M_y}{d}\ket{\perp}\bra{\perp}.
	\end{equation}
	They are positive and sum to the identity on
	$\mathcal L\oplus\mathbb C\ket{\perp}$.
	For a history-dependent final POVM, apply this rule separately at each
	history.
	Together, these rules define a normalised ideal output distribution.
	
	The following theorem justifies using this simpler process for lower
	bounds: bounds on ideal output distinguishability transfer to the
	physical oracle, up to a small simulation error.
	
	\begin{theorem}[Simulation for layered hidden oracles]
		\label{layer:thm:simulation}
		There are universal constants $c,C>0$ for which the following holds.
		Consider any oracle family~\eqref{layer:eq:oracle}, with the fixed root
		and input embeddings, sector-invariant layers, and prescribed transitions
		specified in~\eqref{layer:eq:layers}--\eqref{layer:eq:column-maps}.
		Assume $r\le N=d/2-1$ and $p\le2+2r$.
		For every integer $q\ge1$ satisfying
		\begin{equation}\label{layer:eq:delta}
			\Delta(q,r,d)=\frac{q^2+q\sqrt{r+1}}{\sqrt d}\le c,
		\end{equation}
		every adaptive protocol retaining only the original $d$-dimensional
		quantum system and making at most $q$ forward queries obeys
		\begin{equation}\label{layer:eq:comparison}
			\TV(\mathsf{Out}_\vartheta,
			\mathsf{Out}^{\mathrm{id}}_\vartheta)
			\le C\Delta(q,r,d)
		\end{equation}
		for each fixed $\vartheta$.
		Here $\mathsf{Out}_\vartheta$ is the physical output distribution
		averaged over one initial draw of $(F,H)$, and
		$\mathsf{Out}^{\mathrm{id}}_\vartheta$ is the output distribution
		of the corresponding ideal process.
	\end{theorem}
	
	The proof uses a bosonic record model to connect the physical oracle to the ideal process.
	We first replace the Haar completion and the hidden Haar frames with Gaussian matrices and bound the resulting change in the protocol's average output.
	Gaussian polynomial moments admit an exact representation as inner products of bosonic record states, so tracing out the records reproduces the Gaussian-averaged output.
	Within this record representation, we approximate the Gaussian frame operators by an isometric encoder.
	Finally, we propagate measurement effects backward through the record
	evolution and show that they obey the update rules of the ideal process up to a small accumulated error.
	
	Appendix~\ref{enc:section} constructs the record oracle and establishes the operator estimates needed for the proof.
	Appendix~\ref{sim:section} compares the physical and record outputs, while Appendix~\ref{ideal:sec:comparison} compares the record and ideal outputs.
	Combining these two bounds proves Theorem~\ref{layer:thm:simulation}.
	
	\section{Bosonic records and operator estimates}
	\label{enc:section}
	
	This appendix prepares the record model used to connect the physical oracle to the ideal process. 
	Bosonic records encode polynomial dependence on Gaussian variables, allowing the average over those variables to be expressed as a partial trace over the records. 
	We first explain this correspondence and specify the record spaces. 
	We then construct an isometric encoder for hidden labels and a recorder for the random completion, and combine them into a record oracle.
	
	The estimates have two roles. 
	They control the approximation from Gaussian frame fields to the isometric encoder in Appendix~\ref{sim:section}, and they identify the sector averages and residual dynamics that appear in the ideal process in Appendix~\ref{ideal:sec:comparison}. 
	The record spaces are auxiliary spaces used only in the proof.
	
	\subsection{Fock space as a representation of polynomial moments}
	Bosonic records allow us to express Gaussian averages in terms of partial traces.
	We first recall the relevant Fock-space operators and then establish this correspondence for polynomial amplitudes.
	
	For a one-particle Hilbert space $\mathfrak h$, define the bosonic
	Fock space
	\[
	\mathcal F(\mathfrak h)
	=\bigoplus_{n=0}^\infty\operatorname{Sym}^n\mathfrak h.
	\]
	Here $\operatorname{Sym}^n\mathfrak h$ is the permutation-invariant
	subspace of $\mathfrak h^{\otimes n}$, and
	$\operatorname{Sym}^0\mathfrak h=\mathbb C\ket\Omega$ is the vacuum sector.
	Choose an orthonormal mode basis $\{\ket j\}_j$ of $\mathfrak h$.
	The normalised occupation state $\ket{\boldsymbol n}$ has $n_j$
	particles in mode $j$, and the creation and annihilation operators act as
	\[
	a_j^\dagger\ket{\boldsymbol n}
	=\sqrt{n_j+1}\ket{\boldsymbol n+\boldsymbol e_j},
	\qquad
	a_j\ket{\boldsymbol n}
	=\sqrt{n_j}\ket{\boldsymbol n-\boldsymbol e_j}.
	\]
	The annihilation operator gives zero when $n_j=0$.
	
	For later norm estimates, introduce the  annihilation and creation maps
	\begin{equation}\label{enc:boson-maps}
		\begin{aligned}
			\mathsf A_n:\operatorname{Sym}^n\mathfrak h
			&\longrightarrow
			\mathfrak h\otimes\operatorname{Sym}^{n-1}\mathfrak h,
			&\mathsf A_n\xi
			&=\sum_j\ket j\otimes a_j\xi,
			&\norm{\mathsf A_n}
			&=\sqrt n,\\
			\mathsf T_n:\mathfrak h\otimes\operatorname{Sym}^n\mathfrak h
			&\longrightarrow
			\operatorname{Sym}^{n+1}\mathfrak h,
			&\mathsf T_n(\ket j\otimes\xi)
			&=a_j^\dagger\xi,
			&\norm{\mathsf T_n}
			&=\sqrt{n+1}.
		\end{aligned}
	\end{equation}
	Here $\mathsf A_n$ is defined for $n\ge1$ and $\mathsf T_n$ for $n\ge0$.
	The first norm identity follows from
	$\sum_j a_j^\dagger a_j=nI$ on the $n$-particle sector.
	The second follows from $\mathsf T_n=\mathsf A_{n+1}^\dagger$.
	Equivalently, $\mathsf T_n$ is $\sqrt{n+1}$ times symmetrization.
	
	Now let $g_j$ be independent standard complex Gaussian variables.
	The correspondence
	\[
	\prod_j g_j^{n_j}
	\longmapsto
	\prod_j(a_j^\dagger)^{n_j}\ket\Omega
	=\sqrt{\prod_j n_j!}\ket{\boldsymbol n}
	\]
	preserves inner products: distinct monomials are orthogonal under the
	Gaussian measure, and each monomial has squared norm $\prod_j n_j!$,
	matching its record vector.
	Extend this map linearly to holomorphic polynomials, meaning polynomials
	in the variables $g_j$ without their conjugates.
	Writing $\ket p_{\mathrm{rec}}$ for the record vector of $p$, we have
	\[
	\ip{p_j}{p_i}_{\mathrm{rec}}=\E[p_i(g)\overline{p_j(g)}]
	\]
	
	For a system vector with polynomial amplitudes, use the same record
	space for every amplitude:
	\[
	\ket{v(g)}=\sum_i p_i(g)\ket i
	\quad\longmapsto\quad
	\ket{\widehat v}
	=\sum_i\ket i\otimes\ket{p_i}_{\mathrm{rec}}.
	\]
	The inner-product identity gives
	\[
	\tr_{\mathrm{rec}}\ket{\widehat v}\bra{\widehat v}
	=\sum_{i,j}\E\!\left[p_i(g)\overline{p_j(g)}\right]\ket i\bra j
	=\E_g\!\left[\ket{v(g)}\bra{v(g)}\right].
	\]
	Thus, tracing out the records reproduces the full averaged output
	operator, including its off-diagonal entries.
	
	To handle frame amplitudes that involve both Gaussian variables and their conjugates, we introduce a second family of modes.
	The corresponding representation is established in Appendix~\ref{sim:section}.
	
	\subsection{Record spaces}
	
	Recall the sector label spaces from Appendix~\ref{layer:section}:
	\[
	\mathcal L=\mathcal L_+\oplus\mathcal L_-,
	\qquad
	Z_{\mathcal L}\ket{\alpha,\sigma}
	=\sigma\ket{\alpha,\sigma}.
	\]
	Choose orthonormal bases $\{\ket{i,\sigma}\}_i$ of $\mathcal H_\sigma$
	and $\{\ket{\alpha,\sigma}\}_\alpha$ of $\mathcal L_\sigma$.
	Each frame entry is indexed by a physical row $i$, a label column
	$\alpha$, and a common sector $\sigma$.
	The corresponding one-particle record space is
	\begin{equation}\label{enc:record-space}
		\mathfrak h_f
		=(\mathcal H_+\otimes\mathcal L_+)
		\oplus(\mathcal H_-\otimes\mathcal L_-).
	\end{equation}
	The direct sum reflects the sector structure of the frames:
	entries connecting opposite sectors vanish and require no record modes.
	
	The full record space consists of three bosonic mode families:
	\begin{equation}\label{enc:full-fock}
		\mathcal F
		=\mathcal F(\mathfrak h_f)_a
		\otimes\mathcal F(\mathfrak h_f)_b
		\otimes\mathcal F(\mathbb C^d_X\otimes\mathbb C^d_Y)_c.
	\end{equation}
	The $a$ and $b$ families jointly represent the Gaussian frame entries
	and their conjugates.
	The $c$ family represents a full $d\times d$ Gaussian matrix $\Gamma$,
	whose entries $\Gamma_{yx}$ have variance $1/D$.
	Here $x$ indexes the input coordinate and $y$ the output coordinate.
	Although the completion acts between $D$-dimensional subspaces, using a full $d\times d$ matrix lets us define its records in fixed physical coordinates, independently of the frames.
	For fixed frames, the compression
	$B_{F,\vartheta}^\dagger\Gamma C_F$ has independent entries with law
	$\mathcal N_{\mathbb C}(0,1/D)$.
	It therefore gives the Gaussian replacement of the original $D$-dimensional Haar completion.
	This representation is established in Appendix~\ref{sim:section}.
	
	Let $n_a,n_b,n_c$ denote the total occupation numbers of the three mode families.
	Write $\Pi_{\le R}^{\mathrm{rec}}$ for the projector onto $n_a+n_b+n_c\le R$.
	When restricting only one family, write $\Pi_{n_a\le R}$ or
	$\Pi_{n_c\le q}$.
	In an expression such as $A\Pi_{\le R}^{\mathrm{rec}}$, the projector restricts only the input of $A$, while the output may occupy higher particle-number sectors.
	All identities involving unbounded bosonic operators are understood on finite-particle vectors.
	The norm estimates below use the indicated restricted input domains, on which the relevant operators are bounded.
	
	\subsection{The hidden encoder and compression of a system operation}
	
	We now construct an isometric encoding of hidden labels into the physical system and frame records. 
	Its key property is that compressing a known system operation through this encoding gives approximately the sector-averaged action used in the ideal process.
	
	Start with the creation-operator matrix, defined on finite-particle inputs,
	\begin{equation}\label{enc:raw}
		E_0=\frac1{\sqrt N}\sum_{\sigma,i,\alpha}
		\ket{i,\sigma}\bra{\alpha,\sigma}\otimes a_{i\alpha\sigma}^\dagger.
	\end{equation}
	It maps $\mathcal L\otimes\mathcal F$ into
	$\mathcal H\otimes\mathcal F$ and creates one $a$-particle.
	In particular, on a label with vacuum records,
	\begin{equation}\label{enc:vacuum}
		E_0(\ket{\alpha,\sigma}\otimes\ket\Omega)
		=\frac1{\sqrt N}\sum_i
		\ket{i,\sigma}\otimes\ket{i,\alpha,\sigma}_a,
	\end{equation}
	where $\ket{i,\alpha,\sigma}_a$ denotes the corresponding one-particle
	record state.
	
	For a system operator $K$, recall its sector-averaged action on labels:
	\begin{equation}\label{enc:twirl}
		T(K)=\bigoplus_{\sigma=\pm}\kappa_\sigma I_{\mathcal L_\sigma},
		\qquad
		\kappa_\sigma=\frac{\tr(\Pi_\sigma K)}N.
	\end{equation}
	Tensor products with record identities are implicit.
	The next lemma normalises $E_0$ to an exact isometry and bounds the
	error in replacing the compressed operator $E^\dagger K E$ by $T(K)$.

	\begin{lemma}[An isometric encoder with sector-averaged compression]
		\label{enc:compression}
		Assume $r\le N$.
		There is a globally defined isometry
		$E:\mathcal L\otimes\mathcal F\to\mathcal H\otimes\mathcal F$
		that raises $n_a$ by one, leaves the $b,c$ records unchanged, and satisfies
		\begin{equation}\label{enc:intertwine}
			E^\dagger E=I,
			\qquad
			(O\otimes I)E=E(Z_{\mathcal L}\otimes I).
		\end{equation}
		For every system operator $K$ and $0\le R\le N/2$,
		\begin{align}
			\norm{(E_0^\dagger K E_0-T(K))\Pi_{n_a\le R}}
			&\le\frac RN\norm K,
			\label{enc:raw-compression}\\
			\norm{(E-E_0)\Pi_{n_a\le R}}
			&\le\frac{2R}N,
			\label{enc:polar-error}\\
			\norm{(E^\dagger K E-T(K))\Pi_{n_a\le R}}
			&\le\frac{8R}N\norm K.
			\label{enc:normalised-compression}
		\end{align}
		These estimates allow arbitrary $b,c$ records, including entanglement
		with them.
	\end{lemma}
	
	\begin{proof}
		\emph{Compression of the creation map.}
		Normal ordering gives
		\[
		a_{i\alpha\sigma}a_{j\beta\tau}^\dagger
		=\delta_{ij}\delta_{\alpha\beta}\delta_{\sigma\tau}
		+a_{j\beta\tau}^\dagger a_{i\alpha\sigma}.
		\]
		Hence the block of $E_0^\dagger K E_0$ from label $(\beta,\tau)$
		to label $(\alpha,\sigma)$ is
		\[
		(E_0^\dagger K E_0)_{\alpha\sigma,\beta\tau}
		=\delta_{\sigma\tau}\delta_{\alpha\beta}
		\frac{\tr(\Pi_\sigma K)}N I
		+\frac1N\sum_{i,j}K_{i\sigma,j\tau}
		a_{j\beta\tau}^\dagger a_{i\alpha\sigma}.
		\]
		The first term is $T(K)$.
		On the $n_a$-particle sector, the remaining operator can be written as
		\[
		\frac1N\sum_{\ell=1}^{n_a}R_K^{(\ell)},
		\qquad
		\norm{R_K^{(\ell)}}\le\norm K.
		\]
		
		To obtain this bound without a factor depending on the number of labels,
		embed $\mathfrak h_f$ into
		$(\mathcal H_+\oplus\mathcal H_-)\otimes\mathcal L$.
		Each record particle then has a physical coordinate and a label
		coordinate.
		The operator $R_K^{(\ell)}$ swaps the external label with the label
		of particle $\ell$, applies the transpose of the hidden-space
		compression of $K$ to that particle's physical coordinate, and projects
		back onto the allowed sector pairs.
		These operations reproduce the matrix entries above and have norms
		at most $1$, $\norm K$, and $1$, respectively.
		
		Each $R_K^{(\ell)}$ acts on the ambient tensor product with numbered
		particles. Although an individual summand need not preserve the
		symmetric subspace, their sum does.
		Its restriction to that subspace therefore has norm at most
		$n_a\norm K$.
		This proves~\eqref{enc:raw-compression}.
		Taking $K=I$ also gives
		\begin{equation}\label{enc:gram}
			(1-R/N)I\preceq E_0^\dagger E_0\preceq(1+R/N)I
			\qquad\text{on }n_a\le R.
		\end{equation}
		
		\emph{Construction of the isometric encoder.}
		We define the polar normalisation separately on every particle-number sector. 
		To this end, we first show that $E_0$ is injective on each sector.
		
		Identify occupation vectors with polynomials through
		$\ket{\boldsymbol n}\mapsto
		z^{\boldsymbol n}/\sqrt{\boldsymbol n!}$.
		Under this identification, creation is multiplication by the
		corresponding variable.
		For each label sector $\sigma$, the equation $E_0f=0$ becomes
		$Z_\sigma f_\sigma=0$, where $Z_\sigma$ is the $N\times r$ matrix
		of coordinate indeterminates.
		Since $N\ge r$, its first $r$ rows form a square matrix whose determinant
		is a nonzero polynomial.
		Multiplying these first $r$ equations by the adjugate shows that this
		determinant times every coordinate of $f_\sigma$ is zero.
		The polynomial ring has no zero divisors, so $f_\sigma=0$.
		Applying this argument in both label sectors proves injectivity.
		
		Now fix $n_a$ and omit the unchanged $b,c$ factors.
		On the finite-dimensional space
		$\mathcal L\otimes\operatorname{Sym}^{n_a}(\mathfrak h_f)_a$,
		the Gram matrix $E_0^\dagger E_0$ is positive and invertible.
		Define
		\begin{equation}\label{enc:polar}
			E=E_0(E_0^\dagger E_0)^{-1/2}
		\end{equation}
		on this sector.
		Taking the direct sum over $n_a$ and tensoring with the identity on the $b,c$ records gives a bounded global isometry.
		The map $E$ raises $n_a$ by one.
		Moreover, $OE_0=E_0Z_{\mathcal L}$ implies that
		$E_0^\dagger E_0$ commutes with $Z_{\mathcal L}$.
		Its inverse square root therefore also commutes with $Z_{\mathcal L}$,
		which proves~\eqref{enc:intertwine}.
		
		\emph{Low-occupation estimates.}
		Restrict to $n_a\le R$ and put $s=R/N\le1/2$.
		By~\eqref{enc:gram}, the spectrum of $G=E_0^\dagger E_0$ lies in
		$[1-s,1+s]\subseteq[1/2,3/2]$.
		Therefore, 
		\[
		\norm{G^{-1/2}}\le\sqrt2,
		\qquad
		\norm{G^{-1/2}-I}\le2s,
		\qquad
		\norm{G^{1/2}-I}\le2s.
		\]
		Since $E_0=EG^{1/2}$ and $E$ is an isometry, the last bound proves
		\eqref{enc:polar-error}.
		
		For the compressed action, write
		\[
		E^\dagger K E-T(K)
		=G^{-1/2}(E_0^\dagger K E_0-T(K))G^{-1/2}
		+[G^{-1/2}T(K)G^{-1/2}-T(K)].
		\]
		The first term has norm at most $2s\norm K$ by
		\eqref{enc:raw-compression}.
		Using $\norm{T(K)}\le\norm K$, the second term has norm at most
		\[
		\norm{G^{-1/2}-I}\,\norm{T(K)}
		\bigl(\norm{G^{-1/2}}+1\bigr)
		\le2(1+\sqrt2)s\norm K.
		\]
		Their sum is at most $8s\norm K$, proving
		\eqref{enc:normalised-compression}.
	\end{proof}
	
	For an adaptive protocol, the estimate must jointly control all instrument outcomes. 
	The following column bound does so with an error independent of the number of outcomes.
	
	\begin{lemma}[Compression of an entire instrument]
		\label{enc:instrument}
		Let $\{K_z\}_z$ satisfy $\sum_zK_z^\dagger K_z\preceq I$, and set
		$D_z=T(K_z)$.
		For $0\le R\le N/2$,
		\begin{equation}\label{enc:column-bound}
			\norm{\col_z(E^\dagger K_zE-D_z)\Pi_{n_a\le R}}
			\le\frac{8R\sqrt d}{N}
			\le C\frac R{\sqrt N},
		\end{equation}
		where $C$ is universal.
		Moreover, both $\col_zE^\dagger K_zE$ and $\col_zD_z$ are contractions.
		In particular,
		\begin{equation}\label{enc:kappa-completeness}
			\sum_z|\kappa_{z,\sigma}|^2\le1
			\qquad(\sigma=\pm).
		\end{equation}
	\end{lemma}
	
	\begin{proof}
		Apply~\eqref{enc:normalised-compression} to each $K_z$.
		The squared norm of an operator column is at most the sum of the
		squared norms of its entries, and
		\[
		\sum_z\norm{K_z}^2
		\le\sum_z\norm{K_z}_{\HS}^2
		=\tr\sum_zK_z^\dagger K_z
		\le d.
		\]
		This gives the first inequality in~\eqref{enc:column-bound}.
		The second follows from $d=2N+2\le4N$.
		
		The compressed Kraus column factors as
		\[
		\col_z(E^\dagger K_zE)
		=\left(\bigoplus_zE^\dagger\right)(\col_zK_z)E.
		\]
		Each factor is a contraction, so the column isas welln.
		Finally, Hilbert--Schmidt Cauchy--Schwarz gives
		\[
		|\kappa_{z,\sigma}|^2
		\le\frac1N\tr(\Pi_\sigma K_z^\dagger K_z).
		\]
		Summing over $z$ proves~\eqref{enc:kappa-completeness}.
		Since each $D_z$ acts as the scalar $\kappa_{z,\sigma}$ on sector
		$\sigma$, this also shows that $\col_zD_z$ is a contraction.
	\end{proof}
	
	\subsection{The completion recorder and return to hidden labels}
	
	The completion recorder represents the full Gaussian matrix used for the random completion. We need to control both its action on system operators and its coupling to the hidden encoding.
	
	On finite-particle inputs in $\mathcal H\otimes\mathcal F$, define
	\begin{equation}\label{comp:def}
		\mathsf C=\frac1{\sqrt D}\sum_{x,y=1}^d
		\ket y\bra x\otimes c_{xy}^\dagger.
	\end{equation}
	This operator creates one completion particle and leaves the frame
	records unchanged.
	The next lemma shows that $\mathsf C^\dagger K\mathsf C$ is a trace
	average plus a correction controlled by the number of completion
	particles. It also bounds the amplification caused by each use of
	$\mathsf C$.

	\begin{lemma}[Trace-average identity and norm bounds]
		\label{comp:norm}
		For a system operator $K$, on the $n_c$-particle sector,
		\begin{equation}\label{comp:identity}
			\mathsf C^\dagger K\mathsf C
			=\frac{\tr K}{D}I
			+\frac1D\sum_{\ell=1}^{n_c}
			\mathsf S_\ell(K^{\mathsf T})_{Y_\ell}.
		\end{equation}
		Here $\mathsf S_\ell$ swaps the system coordinate with the input
		coordinate $X_\ell$ of particle $\ell$, and $K^{\mathsf T}$ acts on
		its output coordinate $Y_\ell$.
		The transpose is taken in the physical basis used in~\eqref{comp:def}.
		The sum is understood as its restriction to the symmetric
		$n_c$-particle subspace.
		
		Consequently,
		\begin{align}
			\norm{\mathsf C\Pi_{n_c\le q}}
			&\le c_q:=\sqrt{\frac{d+q}{D}},
			\label{comp:norm-bound}\\
			\norm{\left(\mathsf C^\dagger K\mathsf C
				-\frac{\tr K}{D}I\right)\Pi_{n_c\le q}}
			&\le\frac qD\norm K.
			\label{comp:contraction}
		\end{align}
	\end{lemma}

	\begin{proof}
		Expanding~\eqref{comp:def} and normal ordering gives
		\[
		\mathsf C^\dagger K\mathsf C
		=\frac1D\sum_{x,u,y,v}\ket x\bra u\,K_{yv}
		\bigl(\delta_{xu}\delta_{yv}
		+c_{uv}^\dagger c_{xy}\bigr).
		\]
		The contraction term is $(\tr K/D)I$.
		The remaining term removes a particle with coordinates $(x,y)$,
		creates one with coordinates $(u,v)$, and changes the system
		coordinate from $u$ to $x$.
		In the numbered-particle representation, this is precisely
		$\mathsf S_\ell(K^{\mathsf T})_{Y_\ell}$, summed over all particles.
		This proves~\eqref{comp:identity}.
		
		Each summand has norm at most $\norm K$: the swap has norm one,
		and transposition preserves the operator norm.
		Although individual summands need not preserve the symmetric subspace,
		their sum does. Its restriction therefore has norm at most
		$n_c\norm K$, proving~\eqref{comp:contraction}.
		
		Taking $K=I$ gives
		\[
		\mathsf C^\dagger\mathsf C
		=\frac dD I+\frac1D\sum_{\ell=1}^{n_c}\mathsf S_\ell,
		\qquad
		\norm{\sum_{\ell=1}^{n_c}\mathsf S_\ell}\le n_c.
		\]
		Thus $\mathsf C^\dagger\mathsf C\preceq(d+q)I/D$ on $n_c\le q$,
		which proves~\eqref{comp:norm-bound}.
	\end{proof}
	
	We next bound the return from the completion term to the hidden encoding.
	The relevant operator is $E^\dagger K\mathsf C$: it applies the completion recorder, performs a system operation, and compresses the result back to hidden labels. 
	This estimate will control the cross terms between the planted and completion parts of the record oracle.

	\begin{lemma}[Small return from completion to hidden labels]
		\label{comp:return}
		For a system operator $K$, on inputs with $n_a\le R\le N/2$ and
		$n_c\le q$,
		\begin{equation}\label{comp:return-single}
			\norm{E^\dagger K\mathsf C}
			\le\sqrt{\frac{2R(q+1)}{ND}}\norm K.
		\end{equation}
		For a Kraus family satisfying $\sum_zK_z^\dagger K_z\preceq I$,
		on the same input domain,
		\begin{equation}\label{comp:return-column}
			\norm{\col_z E^\dagger K_z\mathsf C}
			\le\sqrt{\frac{2dR(q+1)}{ND}}.
		\end{equation}
		Both maps vanish on inputs with no $a$-particles.
	\end{lemma}

	\begin{proof}
		Before polar normalisation, the operator is
		\[
		E_0^\dagger K\mathsf C
		=\frac1{\sqrt{ND}}\sum_{\sigma,\alpha,i,x,y}
		K_{i\sigma,y}\ket{\alpha,\sigma}\bra x
		\otimes a_{i\alpha\sigma}c_{xy}^\dagger.
		\]
		This expression factors into three operations.
		First, annihilate an $a$-particle and extract its physical coordinate
		$(i,\sigma)$ and label $(\alpha,\sigma)$.
		Next, apply $K^{\mathsf T}$ to the extracted physical coordinate.
		Finally, create a $c$-particle whose input coordinate is the original
		system coordinate $x$ and whose output coordinate is the resulting
		physical coordinate $y$.
		The extracted label $(\alpha, \sigma)$ becomes the output label.
		
		By~\eqref{enc:boson-maps}, these three operations have norms at most
		$\sqrt R$, $\norm K$, and $\sqrt{q+1}$, respectively.
		Including the normalisation gives
		\[
		\norm{E_0^\dagger K\mathsf C}
		\le\sqrt{\frac{R(q+1)}{ND}}\norm K.
		\]
		The expression vanishes on $a$-vacuum inputs because it contains an
		$a$-annihilation operator.
		
		For nonzero contributions, the annihilation leaves at most $R-1$
		$a$-particles.
		The remaining polar factor in
		\[
		E^\dagger=(E_0^\dagger E_0)^{-1/2}E_0^\dagger
		\]
		therefore has norm at most $\sqrt2$ by~\eqref{enc:gram}.
		This proves~\eqref{comp:return-single}.
		Summing the squared single-operator bounds and using
		$\sum_z\norm{K_z}^2\le d$ proves~\eqref{comp:return-column}.
		
	\end{proof}

	\subsection{The record oracle with an isometric encoder}
	
	We now combine the hidden encoder $E$ and completion recorder $\mathsf C$ into a record oracle.
	Define the input and output embeddings $V_E,W_{E,\vartheta}:\mathcal I\otimes\mathcal F
	\to\mathcal H\otimes\mathcal F$ by replacing $F$ with $E$ in~\eqref{layer:eq:column-maps}:
	\begin{equation}\label{layer:eq:encoders}
		\begin{aligned}
			V_E|_{\mathcal I_0\otimes\mathcal F}
			&=J\otimes I_{\mathcal F},\\
			V_E|_{\mathcal I_j\otimes\mathcal F}
			&=E(V_j\otimes I_{\mathcal F})
			&&(1\le j<\ell),\\
			W_{E,\vartheta}
			&=E(S_\vartheta\otimes I_{\mathcal F}).
		\end{aligned}
	\end{equation}
	Both maps are global isometries.
	For $V_E$, this follows from $E^\dagger E=I$, the orthogonality of the input layer subspaces, and $J^\dagger E=0$.
	For $W_{E,\vartheta}$, it follows because both $S_\vartheta$ and $E$
	are isometries.
	
	Let $P_V=V_EV_E^\dagger$ and $P_{W,\vartheta}=W_{E,\vartheta}W_{E,\vartheta}^\dagger$.
	Define the record query by
	\begin{equation}\label{layer:eq:record-oracle}
		\mathcal U_{E,\vartheta}
		=W_{E,\vartheta}V_E^\dagger
		+(I-P_{W,\vartheta})\mathsf C(I-P_V).
	\end{equation}
	The first term implements the prescribed transitions through the
	hidden encoding. The second applies the completion recorder between
	the complementary input and output spaces.
	As in the physical construction, the output projector may depend
	on $\vartheta$.
	
	The next lemma controls the amplification of approximation errors and the record occupation reached after finitely many queries.
	
	\begin{lemma}[Norm and record degree]\label{rec:norm-degree}
		On inputs with $n_c\le q$,
		\begin{equation}\label{rec:oracle-norm}
			\norm{\mathcal U_{E,\vartheta}\Pi_{n_c\le q}}
			\le c_q:=\sqrt{\frac{d+q}{D}}.
		\end{equation}
		A record query maps the total-occupation sector $n$ into the sum of
		sectors $n$ and $n+1$.
		Its adjoint never increases total occupation.
		Known system operations preserve all record occupations.
	\end{lemma}
	
	\begin{proof}
		The projections $P_V$ and $P_{W,\vartheta}$ preserve each of
		$n_a,n_b,n_c$.
		The planted term is a partial isometry, while the completion term
		has norm at most $c_q$ on inputs with $n_c\le q$, by
		Lemma~\ref{comp:norm}.
		The two terms have orthogonal input supports and orthogonal output
		ranges, so their sum has norm at most $\max\{1,c_q\}=c_q$.
		This proves~\eqref{rec:oracle-norm}.
		
		For the occupation bound, a planted transition from the root creates one $a$-particle.
		A planted transition between hidden layers applies $E^\dagger$
		followed by $E$, preserving $n_a$.
		The completion term creates one $c$-particle, and its surrounding
		projections preserve all three occupation numbers.
		Thus, the only total-occupation shifts are $0$ and $1$.
		Taking adjoints reverses these shifts.
		Known system operations act as the identity on the records and
		therefore preserve their occupations.
	\end{proof}
	
	For a given physical protocol, define its record evolution by starting the records in vacuum, replacing each query by $\mathcal U_{E,\vartheta}$, and tensoring every known system operation with the record identity.
	Use the same record space throughout, together with the protocol's original history-dependent instruments and final readout.
	After the readout and the partial trace over records, denote the positive output operator by $\rho_{E,\vartheta}$.
	
	Although the encoder $E$ is isometric, the full record oracle need not
	preserve trace, so $\rho_{E,\vartheta}$ need not be normalised.
	Appendix~\ref{sim:section} bounds its trace-norm distance from the
	averaged physical output.
	Appendix~\ref{ideal:sec:comparison} then compares its event weights
	with the probabilities of the normalised ideal process.
	
	\section{From the Haar oracle to the record model}
	\label{sim:section}
	
	This appendix compares the averaged physical output with the output of the record oracle constructed in Appendix~\ref{enc:section}.
	Fix $\vartheta$ and set $\Delta=\Delta(q,r,d)$ as defined in~\eqref{layer:eq:delta}.
	Let $\rho_{\mathrm{true},\vartheta}$ denote the physical output
	averaged over a single initial draw of $(F,H)$, reused at every query.
	Let $\rho_{E,\vartheta}$ denote the positive, possibly unnormalised
	record output defined in Appendix~\ref{enc:section}.
	The following proposition bounds their difference.
	
	\begin{proposition}[Physical-to-record comparison]\label{sim:comparison}
		There exist universal constants $c_0,C>0$ such that every protocol
		considered in Theorem~\ref{layer:thm:simulation} with $\Delta\le c_0$
		satisfies
		\begin{equation}\label{sim:main-bound}
			\norm{\rho_{\mathrm{true},\vartheta}
				-\rho_{E,\vartheta}}_1
			\le C\Delta.
		\end{equation}
	\end{proposition}
	
	The proof proceeds in three steps.
	First, we replace the Haar completion by a Gaussian matrix and represent it using the $c$ records.
	Next, we replace the hidden Haar frames by Gaussian matrices and represent their entries and conjugates using the $a,b$ records.
	Finally, we approximate the resulting Gaussian frame operators by the isometric encoder $E$.
	Both Gaussian-to-record representations preserve polynomial inner
	products exactly, so the approximation error comes entirely from the two Haar-to-Gaussian replacements and the replacement of the frame operators by $E$.
	
	\subsection{Positive moment comparison for forward queries}
	\label{sim:haar-subsection}
	
	We first replace the Haar completion with a single Gaussian matrix, which is reused throughout the protocol.
	The following lemma controls the resulting change in the averaged output.
	
	\begin{lemma}[Haar completion versus Gaussian completion]
		\label{sim:haar-gaussian}
		Let $A$ be a partial isometry of rank $d-D$ on $\mathcal H$.
		Let $B_0,C_0:\mathbb C^D\to\mathcal H$ be isometries onto the
		orthogonal complements of its output and input spaces, respectively.
		Consider the oracle
		\[
		U(H)=A+B_0HC_0^\dagger.
		\]
		For a protocol making at most $q<D$ forward queries, let $\rho_H$
		denote its output averaged over a single draw
		$H\sim\operatorname{Haar}(\mathrm U(D))$.
		Replace $H$ by a single matrix $Z$ with independent entries
		$\mathcal N_{\mathbb C}(0,1/D)$, and denote the resulting averaged
		positive output by $\rho_Z$.
		In both cases, the sampled matrix is reused at every query.
		
		With
		\[
		a_q=\frac{q(q-1)}{D-q+1},
		\]
		the outputs satisfy
		\begin{equation}\label{sim:haar-order}
			e^{-a_q}\rho_Z\preceq\rho_H\preceq e^{a_q}\rho_Z,
			\qquad
			\norm{\rho_H-\rho_Z}_1\le e^{a_q}(e^{a_q}-1).
		\end{equation}
		This comparison also holds for protocols retaining additional quantum
		registers.
	\end{lemma}

	\begin{proof}
		We first compare homogeneous moments of $H$ and $Z$, then apply
		this comparison to the protocol's polynomial amplitudes.
		
		\emph{Homogeneous moments.}
		Use the unnormalised vectorization $\ket{X}\!\rangle=\sum_{i,j=1}^D X_{ij}\ket i\ket j$, and define
		\[
		M_{H,t}
		=\E_H\bigl[(\ket{H}\!\rangle\langle\!\bra{H})^{\otimes t}\bigr],
		\qquad
		M_{Z,t}
		=\E_Z\bigl[(\ket{Z}\!\rangle\langle\!\bra{Z})^{\otimes t}\bigr].
		\]
		Both operators are supported on
		$\operatorname{Sym}^t(\mathbb C^D\otimes\mathbb C^D)$.
		On this subspace, Wick's rule gives
		\begin{equation}\label{sim:gaussian-homogeneous-moment}
			M_{Z,t}=t!D^{-t}I.
		\end{equation}
		
		To evaluate the Haar moment, fix $1\le t\le q<D$ and regroup
		the tensor factors so that all row indices precede all column indices.
		On $\mathcal V_t=(\mathbb C^D)^{\otimes t}$, define the permutation
		operators by
		\[
		T_\pi\ket{i_1,\ldots,i_t}
		=\ket{i_{\pi^{-1}(1)},\ldots,i_{\pi^{-1}(t)}},
		\qquad \pi\in S_t.
		\]
		Write $c(\pi)$ for the number of cycles of $\pi$, including fixed
		points. The permutation Gram matrix is
		\[
		G_{\pi,\sigma}
		=\tr(T_\pi^\dagger T_\sigma)
		=D^{c(\pi^{-1}\sigma)}.
		\]
		For $t\le D$, the operators $T_\pi$ are linearly independent:
		applying them to $\ket{1,\ldots,t}$ gives distinct orthonormal
		vectors. Thus, $G$ is invertible, and its inverse defines the
		Weingarten coefficients:
		\[
		(G^{-1})_{\pi,\sigma}
		=\operatorname{Wg}_D(\pi^{-1}\sigma).
		\]
		The Weingarten integration formula
		\cite[Eqs.~(35)--(36)]{Mele2024HaarTools} gives
		\[
		M_{H,t}
		=\sum_{\pi,\sigma\in S_t}
		\operatorname{Wg}_D(\pi^{-1}\sigma)
		T_\pi\otimes T_\sigma.
		\]
		To see the tensor convention, $M_{H,t}$ is the unnormalised Choi
		operator of the twirling map
		\[
		X\longmapsto
		\E_H[H^{\otimes t}X(H^\dagger)^{\otimes t}].
		\]
		The Choi operator of
		$X\mapsto\tr(T_\sigma^\dagger X)T_\pi$ is
		$T_\pi\otimes(T_\sigma^\dagger)^{\mathsf T}
		=T_\pi\otimes T_\sigma$, because permutation matrices are real.
		
		Define
		\[
		A_t(D)=\sum_{\pi\in S_t}D^{c(\pi)}T_\pi,
		\qquad
		\Pi_{\rm sym}
		=\frac1{t!}\sum_{\sigma\in S_t}T_\sigma\otimes T_\sigma.
		\]
		Here $\Pi_{\rm sym}$ projects onto the original symmetric tensor
		space after regrouping.
		The identity $GG^{-1}=I$ yields
		\[
		\sum_{\pi\in S_t}D^{c(\pi)}
		\operatorname{Wg}_D(\pi^{-1}\tau)
		=\mathbf1_{\{\tau=e\}},
		\]
		where $e$ is the identity permutation. Multiplying the corresponding
		sums of permutation operators therefore gives
		\[
		A_t(D)^{-1}
		=\sum_{\pi\in S_t}\operatorname{Wg}_D(\pi)T_\pi.
		\]
		The Weingarten coefficients are invariant under conjugation and
		inversion, so regrouping the double sum for $M_{H,t}$ gives
		\[
		M_{H,t}
		=t!\bigl(A_t(D)^{-1}\otimes I\bigr)\Pi_{\rm sym}.
		\]
		Moreover, $A_t(D)$ commutes with every permutation operator because
		its coefficients depend only on cycle structure.
		Hence $A_t(D)^{-1}\otimes I$ commutes with $\Pi_{\rm sym}$.
		
		\emph{Bounding the Haar moment.}
		It remains to bound $A_t(D)$.
		For $t\ge2$, set
		\[
		B_t=\sum_{i=1}^{t-1}T_{(i\,t)}.
		\]
		Every permutation of $t$ elements either fixes $t$ or is uniquely
		of the form $\sigma(i\,t)$ with $\sigma(t)=t$ and $i<t$.
		In the first case, $t$ contributes one additional cycle.
		In the second, inserting $t$ into an existing cycle leaves the
		cycle count unchanged. Consequently,
		\[
		A_t(D)
		=\bigl(A_{t-1}(D)\otimes I\bigr)(DI+B_t),
		\qquad A_1(D)=DI.
		\]
		Each transposition operator is Hermitian with norm one, so
		\[
		\norm{B_t}\le t-1,
		\qquad
		(D-t+1)I\preceq DI+B_t\preceq(D+t-1)I.
		\]
		Conjugation by a permutation of the first $t-1$ positions merely
		permutes the summands of $B_t$.
		Thus the two factors in the recursion commute.
		Since $t\le D$, induction from $A_1(D)=DI$ gives
		\[
		\left[\prod_{j=0}^{t-1}(D-j)\right]I
		\preceq A_t(D)\preceq
		\left[\prod_{j=0}^{t-1}(D+j)\right]I.
		\]
		Taking inverses and restricting to $\operatorname{Ran}\Pi_{\rm sym}$
		yields
		\[
		\frac{t!}{\prod_{j=0}^{t-1}(D+j)}I
		\preceq M_{H,t}\preceq
		\frac{t!}{\prod_{j=0}^{t-1}(D-j)}I.
		\]
		Combining this with~\eqref{sim:gaussian-homogeneous-moment}, we obtain
		\[
		\left[\prod_{j=0}^{t-1}(1+j/D)^{-1}\right]M_{Z,t}
		\preceq M_{H,t}\preceq
		\left[\prod_{j=0}^{t-1}(1-j/D)^{-1}\right]M_{Z,t}.
		\]
		Using $\log(1+x)\le x$ and
		$-\log(1-x)\le x/(1-x)$, the absolute logarithm of either scalar
		factor is at most
		\[
		\sum_{j=0}^{t-1}\frac{j}{D-j}
		\le\frac{t(t-1)}{2(D-t+1)}
		\le a_q.
		\]
		Therefore
		\begin{equation}\label{sim:moment-order}
			e^{-a_q}M_{Z,t}\preceq M_{H,t}\preceq e^{a_q}M_{Z,t}.
		\end{equation}
		The degree-zero moments are identical, so this comparison holds
		for every $0\le t\le q$.
		
		\emph{From moments to protocol outputs.}
		For the analysis, purify the initial state and instruments, keeping
		orthogonal records of their outcomes.
		Every ket amplitude after at most $q$ forward queries is a
		holomorphic polynomial of degree at most $q$ in the entries of
		the sampled matrix.
		Indeed, each query applies $A+B_0XC_0^\dagger$, whose entries are
		affine functions of $X$, while all known operations have coefficients
		independent of $X$.
		Adaptivity changes these coefficients according to the recorded
		history but does not introduce conjugated entries of $X$.
		
		The resulting vector can therefore be written as
		\[
		\ket{\Phi(X)}
		=\sum_{t=0}^q L_t\ket{X}\!\rangle^{\otimes t}
		\]
		for fixed linear maps $L_t$.
		Different degrees occur because a query may contribute either the
		fixed term $A$ or a term linear in $X$.
		For both $X=H$ and $X=Z$, invariance under
		$X\mapsto e^{i\theta}X$ implies
		\[
		\E\bigl[
		\ket{X}\!\rangle^{\otimes t}
		\langle\!\bra{X}^{\otimes u}
		\bigr]=0
		\qquad(t\ne u).
		\]
		Hence
		\[
		\E\ket{\Phi(X)}\bra{\Phi(X)}
		=\sum_{t=0}^q L_tM_{X,t}L_t^\dagger.
		\]
		Applying~\eqref{sim:moment-order} to each term proves the desired
		positive order comparison for the averaged purified state.
		Partial traces, dephasing of the history records, and the final
		readout preserve positive order, so
		\[
		e^{-a_q}\rho_Z\preceq\rho_H\preceq e^{a_q}\rho_Z.
		\]
		
		Finally, $\tr\rho_H=1$ and the lower order bound imply
		$\tr\rho_Z\le e^{a_q}$.
		The two order bounds also give
		\[
		-(e^{a_q}-1)\rho_Z
		\preceq\rho_H-\rho_Z
		\preceq(e^{a_q}-1)\rho_Z.
		\]
		For Hermitian $Y$ and positive $X$, the inequalities
		$-X\preceq Y\preceq X$ imply $\norm{Y}_1\le\tr X$.
		Applying this fact proves the trace-norm bound
		in~\eqref{sim:haar-order}.
	\end{proof}
	
	We now apply the lemma to the completion in \eqref{layer:eq:oracle}.
	Since $q\ge1$ and $\Delta\le c_0$,
	\[
	p\le2(r+1)\le2c_0^2d,
	\qquad
	q^2\le c_0\sqrt d.
	\]
	For sufficiently small $c_0$, these inequalities give
	$D=d-p\ge d/2$ and $q\le D/2$, so $a_q=O(q^2/d)$.
	Condition on the frames $F$, apply
	Lemma~\ref{sim:haar-gaussian}, and then average over $F$.
	The complement bases may depend on $F$ and $\vartheta$, since the
	lemma applies uniformly to every fixed choice.
	The resulting trace-norm error is $O(q^2/d)$.
	
	\subsection{Gaussian completion records}
	\label{sim:completion-subsection}
	
	We now represent the Gaussian completion using the $c$ records.
	To keep the record coordinates independent of the frames, first
	express the completion through a Gaussian matrix on the full
	physical space.
	
	Let $\Gamma$ be a $d\times d$ matrix with independent entries
	$\mathcal N_{\mathbb C}(0,1/D)$, independent of $F$.
	Define the planted input and output projections by
	\[
	P_{\mathrm{in},F}=V_FV_F^\dagger,
	\qquad
	P_{\mathrm{out},F,\vartheta}
	=W_{F,\vartheta}W_{F,\vartheta}^\dagger.
	\]
	For fixed $F$, Gaussian unitary invariance implies that
	$B_{F,\vartheta}^\dagger\Gamma C_F$ has the same distribution as
	the $D\times D$ Gaussian matrix $Z$. Hence, conditional on $F$,
	\[
	B_{F,\vartheta}ZC_F^\dagger
	\ \stackrel{\mathrm{law}}{=}\
	(I-P_{\mathrm{out},F,\vartheta})
	\Gamma(I-P_{\mathrm{in},F}).
	\]
	Thus a single full-space matrix $\Gamma$, independent of the frames,
	represents the Gaussian completion for every fixed choice of $F$.
	Its entry variance is $1/D$.
	
	For a scalar $g\sim\mathcal N_{\mathbb C}(0,1/D)$,
	\[
	\E[\overline g^{\,m}g^n]
	=\delta_{mn}\frac{n!}{D^n}.
	\]
	Consequently, the linear map
	\begin{equation}\label{sim:holomorphic-map}
		g^n\longmapsto D^{-n/2}(c^\dagger)^n\ket{\Omega}
		=\sqrt{\frac{n!}{D^n}}\ket n
	\end{equation}
	preserves the $L^2$ inner product of holomorphic polynomials.
	Assigning one mode $c_{xy}$ to each entry $\Gamma_{yx}$ gives
	the corresponding multivariate map:
	\[
	\prod_{x,y}\Gamma_{yx}^{n_{xy}}
	\longmapsto
	D^{-\frac12\sum_{x,y}n_{xy}}
	\prod_{x,y}(c_{xy}^\dagger)^{n_{xy}}\ket{\Omega}.
	\]
	Under this map, multiplication by $\Gamma_{yx}$ becomes
	$c_{xy}^\dagger/\sqrt D$.
	Applying $\Gamma$ to a system vector is therefore represented by
	the completion recorder
	\begin{equation}\label{sim:completion-creator}
		\mathsf C
		=\frac1{\sqrt D}\sum_{x,y=1}^d
		\ket y\bra x\otimes c_{xy}^\dagger,
	\end{equation}
	as defined in~\eqref{comp:def}.
	
	Let $T_c$ denote the polynomial-to-record map above.
	For a vector with holomorphic amplitudes,
	\[
	\ket{v(\Gamma)}=\sum_i f_i(\Gamma)\ket i,
	\qquad
	\ket{\widehat v}=\sum_i\ket i\otimes T_cf_i,
	\]
	preservation of inner products gives
	\begin{equation}\label{sim:partial-trace-gaussian}
		\tr_c\ket{\widehat v}\bra{\widehat v}
		=\sum_{i,j}\E_\Gamma[f_i(\Gamma)\overline{f_j(\Gamma)}]
		\ket i\bra j
		=\E_\Gamma\ket{v(\Gamma)}\bra{v(\Gamma)}.
	\end{equation}
	Thus, tracing out the records reproduces the entire averaged matrix, including its off-diagonal entries.
	Using the same modes throughout the protocol preserves the correlations arising from reusing the same Gaussian matrix.
	
	For each fixed realisation of $F$, the Gaussian-completed oracle therefore has the exact record representation
	\begin{equation}\label{sim:haar-frame-record}
		\mathcal U_{F,\vartheta}
		=W_{F,\vartheta}V_F^\dagger
		+(I-W_{F,\vartheta}W_{F,\vartheta}^\dagger)
		\mathsf C(I-V_FV_F^\dagger).
	\end{equation}
	All factors other than $\mathsf C$ act only on the physical system, and their tensor products with the record identity are implicit.
	
	We will also need a norm bound for this query.
	By Lemma~\ref{comp:norm},
	\begin{equation}\label{sim:c-norm}
		\norm{\mathsf C\Pi_{n_c\le q}}
		\le c_q:=\sqrt{\frac{d+q}{D}}.
	\end{equation}
	The planted term is a partial isometry, and the two terms
	in~\eqref{sim:haar-frame-record} have orthogonal input supports
	and orthogonal output ranges.
	Since $c_q\ge1$, it follows that
	\begin{equation}\label{sim:haar-frame-norm}
		\norm{\mathcal U_{F,\vartheta}\Pi_{n_c\le q}}
		\le c_q.
	\end{equation}
	
	\subsection{Coupling Haar frames to Gaussian frames}
	\label{sim:frames-subsection}
	
	We next replace the hidden Haar frames by Gaussian matrices, while keeping the completion represented by the $c$ records.
	We couple each Haar frame to a Gaussian matrix through polar normalisation, so that their difference can be bounded directly.
	
	In each sector $\sigma\in\{+,-\}$, independently draw
	$G_\sigma:\mathcal L_\sigma\to\mathcal H_\sigma$ with entries
	$\mathcal N_{\mathbb C}(0,1/N)$, and set
	\begin{equation}\label{sim:polar-coupling}
		\begin{aligned}
			F_\sigma&=G_\sigma(G_\sigma^\dagger G_\sigma)^{-1/2},\\
			\delta_\sigma
			&=\norm{F_\sigma-G_\sigma}
			=\norm{(G_\sigma^\dagger G_\sigma)^{1/2}-I}.
		\end{aligned}
	\end{equation}
	Since $r\le N$, each $G_\sigma$ has full column rank almost surely,
	and $F_\sigma$ is an isometry.
	Polar normalisation commutes with left multiplication by a unitary:
	\[
	(QG_\sigma)
	\bigl((QG_\sigma)^\dagger(QG_\sigma)\bigr)^{-1/2}
	=QF_\sigma,
	\qquad Q\in\mathrm U(N).
	\]
	The left-unitary invariance of $G_\sigma$ therefore implies that
	$F_\sigma$ is Haar distributed among $r$-frames in $\mathcal H_\sigma$.
	
	Write
	\[
	G=G_+\oplus G_-,
	\qquad F=F_+\oplus F_-,
	\qquad
	\delta=\max_{\sigma=\pm}\delta_\sigma=\norm{F-G}.
	\]
	Across repeated queries, the replacement error will be accompanied by factors of the form $e^{Cq\delta}$.
	The next lemma controls both the tail of $\delta$ and the weighted moment needed to average these errors.
	
	\begin{lemma}[Frame-coupling tails and a weighted moment]
		\label{sim:frame-tail}
		There exist universal constants $c,C,C_0>0$ such that
		\begin{align}
			\Pr(\delta>u)
			&\le C\exp(Cr-cNu^2),
			&&u\ge0,
			\label{sim:unshifted-tail}\\
			\Pr\bigl(\delta>C_0\sqrt{r/N}+t\bigr)
			&\le Ce^{-cNt^2},
			&&t\ge0.
			\label{sim:shifted-tail}
		\end{align}
		Moreover, for every $h\ge0$,
		\begin{equation}\label{sim:weighted-tail}
			\E[\delta^2e^{h\delta}]
			\le C\frac{r+1}{N}
			\exp\left(
			Ch\sqrt{\frac{r+1}{N}}+C\frac{h^2}{N}
			\right).
		\end{equation}
	\end{lemma}
	
	\begin{proof}
		First, consider one sector and omit the sector index.
		For a fixed unit vector $v\in\mathbb C^r$, $N\norm{Gv}^2$ is a sum of $N$ independent exponential random variables of mean one.
		For one such variable $X$,
		\[
		\E e^{\lambda(X-1)}
		=\frac{e^{-\lambda}}{1-\lambda},
		\qquad \lambda<1,
		\]
		and the logarithm of this expression is at most $C\lambda^2$
		for $|\lambda|\le1/2$.
		Chernoff's bound therefore gives
		\[
		\Pr\bigl(|\norm{Gv}^2-1|>x\bigr)
		\le2e^{-cN\min\{x^2,x\}},
		\qquad x\ge0.
		\]
		
		Choose a $1/4$-net of the complex unit sphere in $\mathbb C^r$
		with cardinality at most $9^{2r}$.
		For any Hermitian matrix $A$, approximation by this net gives
		\[
		\norm A\le2\max_{v\text{ in the net}}|v^\dagger Av|.
		\]
		Applying this to $G^\dagger G-I$ and taking a union bound yields
		\[
		\Pr\bigl(\norm{G^\dagger G-I}>x\bigr)
		\le2\exp\bigl(Cr-cN\min\{x^2,x\}\bigr).
		\]
		The coupling error is the largest deviation of a singular value
		of $G$ from one.
		If $0<u\le1$ and $|s-1|>u$ for a singular value $s$, then
		$|s^2-1|>u$.
		If $u\ge1$, only an upper deviation is possible, and
		$s^2-1>u^2$.
		These observations give~\eqref{sim:unshifted-tail} for one sector.
		A union bound over the two sectors gives the same bound for $\delta$,
		with adjusted constants.
		Substituting $u=C_0\sqrt{r/N}+t$ and choosing $C_0$ sufficiently
		large proves~\eqref{sim:shifted-tail}.
		
		For the weighted moment, set
		\[
		a=C_0\sqrt{r/N},
		\qquad T=(\delta-a)_+.
		\]
		The shifted tail bound and tail integration imply
		\[
		\E T^4\le\frac{C}{N^2},
		\qquad
		\E e^{hT}
		=1+h\int_0^\infty e^{ht}\Pr(T>t)\,dt
		\le C e^{Ch^2/N}.
		\]
		The last inequality follows by completing the square in
		$ht-cNt^2$.
		By Cauchy--Schwarz,
		\[
		\E[T^2e^{hT}]
		\le(\E T^4)^{1/2}(\E e^{2hT})^{1/2}
		\le\frac{C}{N}e^{Ch^2/N}.
		\]
		Since $\delta\le a+T$,
		\[
		\begin{aligned}
			\E[\delta^2e^{h\delta}]
			&\le2e^{ha}
			\left(a^2\E e^{hT}+\E[T^2e^{hT}]\right)\\
			&\le C\left(a^2+\frac1N\right)
			e^{ha+Ch^2/N}.
		\end{aligned}
		\]
		Substituting the value of $a$ proves~\eqref{sim:weighted-tail}.
	\end{proof}
	
	We now bound the change in a single record query.
	Define $V_G,W_{G,\vartheta}$ by the same column formulas as
	$V_F,W_{F,\vartheta}$, replacing each $F_\sigma$ by $G_\sigma$
	and leaving the root columns unchanged.
	The fixed label-space embeddings are contractions, so
	\begin{equation}\label{sim:column-perturbation}
		\norm{V_G-V_F}
		+\norm{W_{G,\vartheta}-W_{F,\vartheta}}
		\le2\delta.
	\end{equation}
	
	Define the Gaussian-frame record query by the same ordered expression:
	\begin{equation}\label{sim:raw-frame-record}
		\widetilde{\mathcal U}_{G,\vartheta}
		=W_{G,\vartheta}V_G^\dagger
		+(I-W_{G,\vartheta}W_{G,\vartheta}^\dagger)
		\mathsf C(I-V_GV_G^\dagger).
	\end{equation}
	The tilde indicates that the Gaussian column maps have not been
	normalised to isometries.
	In particular, the factors surrounding $\mathsf C$ need not be
	orthogonal projections.
	
	Using~\eqref{sim:column-perturbation}, we have
	\[
	\norm{V_G},\norm{W_{G,\vartheta}}\le1+\delta.
	\]
	Subtracting the two query expressions one factor at a time and
	using~\eqref{sim:c-norm} gives
	\begin{equation}\label{sim:frame-query-difference}
		\norm{
			(\widetilde{\mathcal U}_{G,\vartheta}
			-\mathcal U_{F,\vartheta})\Pi_{n_c\le q}}
		\le Cc_q\delta(1+\delta)^3.
	\end{equation}
	For example, suppressing $\vartheta$,
	\[
	W_GW_G^\dagger-W_FW_F^\dagger
	=(W_G-W_F)W_G^\dagger
	+W_F(W_G-W_F)^\dagger.
	\]
	The completion term is treated by subtracting its left and right
	factors separately.
	Each difference contributes a factor of order $\delta$, while
	the remaining factors contribute at most $(1+\delta)^3$,
	because a query contains at most four frame factors.
	All these system matrices preserve completion occupation, so the
	restricted norm bound for $\mathsf C$ applies throughout.
	
	In particular, when $\delta\le1$,
	\[
	\norm{
		(\widetilde{\mathcal U}_{G,\vartheta}
		-\mathcal U_{F,\vartheta})\Pi_{n_c\le q}}
	\le Cc_q\delta,
	\qquad
	\norm{\widetilde{\mathcal U}_{G,\vartheta}\Pi_{n_c\le q}}
	\le c_q(1+C\delta).
	\]

	\subsection{Accumulating and averaging the frame error}
	\label{sim:accumulation-subsection}
	
	We now accumulate the single-query error over the adaptive protocol
	and average over the coupling $(F,G)$.
	Known instruments do not amplify vector errors: a complete instrument
	defines an isometric column map,
	\[
	\ket v\longmapsto\bigoplus_z(K_z\otimes I)\ket v,
	\qquad
	\sum_z\norm{(K_z\otimes I)v}^2=\norm v^2.
	\]
	Choosing these instruments according to orthogonal history flags
	also gives an isometry.
	
	Each query adds at most one $c$-particle, while known operations
	preserve completion occupation.
	Thus, every intermediate vector in a comparison involving at most
	$q$ queries lies in the domain $n_c\le q$.
	
	The global bound~\eqref{sim:frame-query-difference} and
	\eqref{sim:haar-frame-norm} imply, for every $\delta\ge0$,
	\[
	\begin{aligned}
		\norm{
			(\widetilde{\mathcal U}_{G,\vartheta}
			-\mathcal U_{F,\vartheta})\Pi_{n_c\le q}}
		&\le Cc_q\delta e^{C\delta},\\
		\norm{\widetilde{\mathcal U}_{G,\vartheta}\Pi_{n_c\le q}}
		&\le c_q e^{C\delta}.
	\end{aligned}
	\]
	Here we used
	$(1+\delta)^3\le e^{3\delta}$ and
	$1+C\delta(1+\delta)^3\le e^{C'\delta}$
	for a suitable universal constant $C'$.
	
	For two sequences of maps, the telescoping identity
	\[
	B_q\cdots B_1-A_q\cdots A_1
	=\sum_{t=1}^q
	B_q\cdots B_{t+1}(B_t-A_t)A_{t-1}\cdots A_1
	\]
	expresses the total difference as a sum of single-query differences.
	Apply it to the two record evolutions, including their common
	isometric instruments.
	For each coupled pair $(F,G)$, this gives
	\[
	\norm{\widetilde\Phi_G-\Phi_F}
	\le Cq\,c_q^q\,\delta e^{Cq\delta},
	\]
	where $\Phi_F$ and $\widetilde\Phi_G$ are the final joint vectors,
	including the completion records and history flags.
	Neither vector is assumed to be normalised.
	Since $D=d-p\ge d/2$,
	\[
	c_q^q
	=\left(1+\frac{p+q}{D}\right)^{q/2}
	\le\exp\left(C\frac{q(p+q)}d\right).
	\]
	Consequently,
	\begin{equation}\label{sim:conditional-vector-error}
		\norm{\widetilde\Phi_G-\Phi_F}
		\le Cq\delta
		\exp\left(Cq\delta+C\frac{q(p+q)}d\right).
	\end{equation}
	
	We next average this bound using Lemma~\ref{sim:frame-tail}.
	The assumptions $1\le r\le N=d/2-1$ give $N\ge d/4$.
	Together with $p\le2(r+1)$, they imply
	\begin{equation}\label{layer:eq:smallness}
		\frac{q(p+q)}d\le C\Delta,
		\qquad
		q\sqrt{\frac{r+1}{N}}\le C\Delta,
		\qquad
		\frac{q^2}{N}\le C\Delta.
	\end{equation}
	For the first inequality, use
	$(r+1)/d\le\sqrt{(r+1)/d}$.
	Squaring~\eqref{sim:conditional-vector-error} and applying
	\eqref{sim:weighted-tail} with $h=Cq$ yields
	\[
	\begin{aligned}
		\E\norm{\widetilde\Phi_G-\Phi_F}^2
		&\le
		Cq^2\exp\left(C\frac{q(p+q)}d\right)
		\E[\delta^2e^{Cq\delta}]\\
		&\le
		C\frac{q^2(r+1)}N
		\exp\left(
		C\frac{q(p+q)}d
		+Cq\sqrt{\frac{r+1}{N}}
		+C\frac{q^2}{N}
		\right).
	\end{aligned}
	\]
	By~\eqref{layer:eq:smallness} and $\Delta\le c_0$, the exponential
	factor is bounded by a universal constant. Hence
	\begin{equation}\label{sim:full-frame-ms}
		\E\norm{\widetilde\Phi_G-\Phi_F}^2
		\le C\frac{q^2(r+1)}d.
	\end{equation}
	
	To convert this vector estimate into an output estimate, use
	\[
	\ket x\bra x-\ket y\bra y
	=\ket{x-y}\bra x+\ket y\bra{x-y}.
	\]
	The trace norm of a rank-one operator then gives
	\begin{equation}\label{sim:vector-to-state}
		\norm{\ket x\bra x-\ket y\bra y}_1
		\le(\norm x+\norm y)\norm{x-y}.
	\end{equation}
	The Haar-frame record vector satisfies
	$\norm{\Phi_F}\le c_q^q=O(1)$.
	Moreover,
	\[
	\E\norm{\widetilde\Phi_G}^2
	\le2\E\norm{\Phi_F}^2
	+2\E\norm{\widetilde\Phi_G-\Phi_F}^2
	=O(1),
	\]
	because $q^2(r+1)/d\le\Delta^2$.
	Applying~\eqref{sim:vector-to-state} and Cauchy--Schwarz therefore gives
	\[
	\E\norm{
		\ket{\widetilde\Phi_G}\bra{\widetilde\Phi_G}
		-\ket{\Phi_F}\bra{\Phi_F}}_1
	\le Cq\sqrt{\frac{r+1}{d}}.
	\]
	Thus, replacing the Haar frames with Gaussian matrices changes the averaged output by at most
	\begin{equation}\label{sim:frame-output-error}
		Cq\sqrt{\frac{r+1}{d}}
	\end{equation}
	in trace norm. 
	Combining this estimate with the Haar-completion comparison, the
	trace-norm distance between the true output and the averaged output
	of the all-Gaussian polynomial model is at most
	\begin{equation}\label{sim:all-gaussian-error}
		C\left(\frac{q^2}{d}+q\sqrt{\frac{r+1}{d}}\right).
	\end{equation}

	\subsection{Complex Gaussian frame fields}
	\label{sim:mixed-field-subsection}
	
	The completion variables enter ket amplitudes only holomorphically.
	The frame variables also enter through their conjugates, because \eqref{sim:raw-frame-record} contains both $G$ and $G^\dagger$.
	We therefore need a record representation that preserves inner products of polynomials in both types of variables.
	
	For each coordinate $g_\mu\sim\mathcal N_{\mathbb C}(0,1/N)$,
	use the corresponding modes $a_\mu,b_\mu$ in the two frame record
	families and define
	\begin{equation}\label{sim:complex-field}
		M_\mu=\frac{a_\mu^\dagger+b_\mu}{\sqrt N},
		\qquad
		M_\mu^\dagger=\frac{a_\mu+b_\mu^\dagger}{\sqrt N}.
	\end{equation}
	These fields and their adjoints commute on finite-particle vectors:
	\[
	[M_\mu,M_\nu^\dagger]
	=\frac{-\delta_{\mu\nu}+\delta_{\mu\nu}}N=0,
	\qquad
	[M_\mu,M_\nu]=0.
	\]
	For a single coordinate,
	\[
	M^n\ket{\Omega}
	=N^{-n/2}(a^\dagger)^n\ket{\Omega},
	\]
	and hence
	\begin{equation}\label{sim:complex-field-moments}
		\bra{\Omega}(M^\dagger)^mM^n\ket{\Omega}
		=\delta_{mn}\frac{n!}{N^n}
		=\E[\overline g^{\,m}g^n].
	\end{equation}
	Commutativity allows every polynomial moment to be put in this form, and moments of distinct coordinates factor.
	Thus, the linear map
	\begin{equation}\label{sim:mixed-polynomial-map}
		T_{ab}f:=f(M,M^\dagger)\ket{\Omega}
		\qquad\text{satisfies}\qquad
		\ip{T_{ab}f}{T_{ab}h}
		=\E_G[\overline{f(G,\overline G)}\,h(G,\overline G)]
	\end{equation}
	for arbitrary polynomials $f,h$ in the frame coordinates and their
	conjugates.
	
	The creation terms record occurrences of a coordinate or its conjugate, while the annihilation terms reproduce their Gaussian
	contractions.
	For example,
	\[
	T_{ab}(g\overline g)
	=MM^\dagger\ket{\Omega}
	=\frac{\ket{\Omega}+a^\dagger b^\dagger\ket{\Omega}}N.
	\]
	The vacuum term, produced by the annihilation part of $M$,
	reproduces $\E[g\overline g]=1/N$.
	
	We now apply this substitution to the Gaussian-frame record query.
	Assemble the fields into an operator-valued matrix $M$ with the
	same sector and index structure as $G$.
	Define $V_M,W_{M,\vartheta}$ and
	$\widetilde{\mathcal U}_{M,\vartheta}$ by replacing $G$ with $M$
	and $G^\dagger$ with $M^\dagger$ in the ordered expressions
	defining $V_G,W_{G,\vartheta}$ and
	\eqref{sim:raw-frame-record}.
	Every occurrence of a given coordinate uses the same field, including occurrences in $I-W_GW_G^\dagger$ and $I-V_GV_G^\dagger$.
	
	Apply $T_{ab}$ to every amplitude of $\widetilde\Phi_G$, treating
	the system, history flags, and $c$ records as the coefficient space.
	If $\widetilde\Phi_M$ denotes the resulting joint vector, then
	\eqref{sim:mixed-polynomial-map} gives
	\[
	\tr_{ab}
	\ket{\widetilde\Phi_M}\bra{\widetilde\Phi_M}
	=\E_G
	\ket{\widetilde\Phi_G}\bra{\widetilde\Phi_G}.
	\]
	Thus, the frame-field representation reproduces the Gaussian average exactly and introduces no approximation error.
	All Gaussian randomness is now represented by the three record families $a,b,c$.
	
	\subsection{Replacing the Gaussian field by an isometric encoder}
	\label{sim:normalisation-subsection}
	
	The final approximation replaces the Gaussian frame fields with the isometric encoder $E$.
	We first bound their difference on states with bounded record occupation, then apply this replacement to the ordered query expression.
	
	In one physical sector, write
	\begin{equation}\label{sim:field-split}
		M=E_0+B,
		\qquad
		E_0=\frac1{\sqrt N}\sum_{i,\alpha}
		\ket i\bra\alpha\otimes a_{i\alpha}^\dagger,
		\qquad
		B=\frac1{\sqrt N}\sum_{i,\alpha}
		\ket i\bra\alpha\otimes b_{i\alpha}.
	\end{equation}
	These operators map a label register and records to a physical register and records.
	The creation part $E_0$ is the unnormalised encoder studied in Lemma~\ref{enc:compression}, and $B$ annihilates one $b$-particle.
	
	Let $\Pi_{\le R}^{\mathrm{rec}}$ denote the projector onto total record occupation at most $R$.
	We use the bounds
	\begin{equation}\label{sim:raw-encoder-estimates}
		\begin{aligned}
			\norm{(E_0^\dagger E_0-I)\Pi_{\le R}^{\mathrm{rec}}}
			&\le\frac RN,\\
			\max\left\{
			\norm{B\Pi_{\le R}^{\mathrm{rec}}},
			\norm{B^\dagger\Pi_{\le R}^{\mathrm{rec}}}
			\right\}
			&\le\sqrt{\frac{r+R}{N}}.
		\end{aligned}
	\end{equation}
	The first follows from~\eqref{enc:gram}.
	For the second, normal ordering on the $n_b$-particle sector gives
	\[
	BB^\dagger
	=\frac1N\left(
	rI+\sum_{\ell=1}^{n_b}
	\mathsf{Swap}_{\mathrm{physical},\ell}
	\right).
	\]
	Here $\mathsf{Swap}_{\mathrm{physical},\ell}$ exchanges the physical
	register with the physical coordinate of the $\ell$th $b$-particle,
	leaving its column label unchanged.
	The sum preserves the symmetric subspace and has norm at most $n_b$.
	Thus $B^\dagger$ on $n_b$ input particles has norm at most
	$\sqrt{(r+n_b)/N}$.
	Taking adjoints between adjacent particle sectors bounds $B$ on
	$n_b\ge1$ particles by $\sqrt{(r+n_b-1)/N}$, while $B$ vanishes
	on vacuum.
	This proves the second estimate and explains the contribution of $r$.
	
	Now use the global isometric encoder from Lemma~\ref{enc:compression},
	\[
	E=E_0(E_0^\dagger E_0)^{-1/2}.
	\]
	The inverse square root is defined separately on each fixed $a$-particle sector, where $E_0$ is injective for $r\le N$.
	The resulting polar factors together define a global isometry that increases $n_a$ by one and preserves the other occupations.
	
	For $R\le N/2$, Lemma~\ref{enc:compression} gives
	\[
	\norm{(E-E_0)\Pi_{\le R}^{\mathrm{rec}}}
	\le C\frac RN.
	\]
	The corresponding adjoint estimate follows by taking adjoints
	between adjacent $a$-particle sectors:
	on an input with $n_a\ge1$, it uses the encoder bound at $n_a-1$,
	and both adjoints vanish on $a$-vacuum inputs.
	Hence
	\[
	\norm{(E^\dagger-E_0^\dagger)\Pi_{\le R}^{\mathrm{rec}}}
	\le C\frac RN.
	\]
	Combining these estimates with~\eqref{sim:raw-encoder-estimates}
	yields
	\begin{equation}\label{sim:eta}
		\begin{aligned}
			\max\left\{
			\norm{(M-E)\Pi_{\le R}^{\mathrm{rec}}},
			\norm{(M^\dagger-E^\dagger)\Pi_{\le R}^{\mathrm{rec}}}
			\right\}
			&\le\eta_R,\\
			\eta_R
			&:=C\left(
			\sqrt{\frac{r+R}{N}}+\frac{R}{N}
			\right).
		\end{aligned}
	\end{equation}
	Taking the orthogonal direct sum over the two physical sectors
	gives the same bounds for the full frame field and encoder.

	We then replace $M$ by $E$ and $M^\dagger$ by $E^\dagger$ in the query expression.
	Using the column maps $V_E,W_{E,\vartheta}$ defined in~\eqref{layer:eq:encoders}, this produces
	\begin{equation}\label{sim:normalised-query}
		\mathcal U_{E,\vartheta}
		=W_{E,\vartheta}V_E^\dagger
		+(I-W_{E,\vartheta}W_{E,\vartheta}^\dagger)
		\mathsf C(I-V_EV_E^\dagger).
	\end{equation}
	This is precisely the record oracle constructed in
	Appendix~\ref{enc:section}.
	
	\subsection{Finite-particle domain and accumulated error}
	\label{sim:domain-subsection}
	
	The Gaussian fields are unbounded on the full Fock space.
	To apply the preceding norm estimates, we therefore track the record occupation of every intermediate vector.
	
	Each field $M$ or $M^\dagger$ changes total occupation by at most one.
	The planted term of $\widetilde{\mathcal U}_{M,\vartheta}$ contains
	at most two frame factors, while the completion term contains at most
	four frame factors and one $c$-creator.
	Thus a raw query increases total occupation by at most five.
	By Lemma~\ref{rec:norm-degree}, a query using the isometric encoder
	increases total occupation by at most one.
	Known system instruments preserve all record occupations.
	
	Starting from the vacuum, any sequence of at most $q$ queries, each chosen to be raw or normalised, therefore reaches a total occupation of at most $5q$.
	Expanding a single query and replacing its factors one at a time introduces at most five further occupation increases.
	Consequently, the common bound
	\begin{equation}\label{sim:envelope}
		R_*=5q+5
	\end{equation}
	covers all intermediate vectors used in the comparison, including
	inputs to adjoint factors.
	For sufficiently small $c_0$, the condition $\Delta\le c_0$ ensures
	$R_*\le N/2$, as required by~\eqref{sim:eta}.
	
	All cutoffs restrict the inputs of norm estimates only.
	We leave the operator products unchanged, retaining creation beyond
	the cutoff and the bosonic commutation relations used above.
	
	We can now compare the two queries by replacing their frame factors
	one at a time.
	The fixed label-space maps are contractions, and all frame factors
	have bounded norm on the intermediate domains, by
	\eqref{sim:raw-encoder-estimates} and $E^\dagger E=I$.
	Each replacement contributes at most $\eta_{R_*}$, while the completion recorder contributes at most $c_q$.
	Thus, for every reachable input $v$ under comparison,
	\begin{equation}\label{sim:field-query-difference}
		\begin{aligned}
			\norm{
				(\widetilde{\mathcal U}_{M,\vartheta}
				-\mathcal U_{E,\vartheta})v}
			&\le Cc_q\eta_{R_*}\norm v,\\
			\norm{\widetilde{\mathcal U}_{M,\vartheta}v}
			&\le c_q(1+C\eta_{R_*})\norm v.
		\end{aligned}
	\end{equation}
	The second inequality follows from the first and Lemma~\ref{rec:norm-degree}.
	
	Let $\widetilde\Phi_M$ and $\Phi_E$ be the final joint vectors for
	the raw-field and isometric-encoder evolutions, respectively,
	including all records and history flags.
	The telescoping argument from
	Appendix~\ref{sim:accumulation-subsection} gives
	\[
	\norm{\widetilde\Phi_M-\Phi_E}
	\le Cq\eta_{R_*}
	\exp\left(
	Cq\eta_{R_*}+C\frac{q(p+q)}d
	\right).
	\]
	Since $R_*=5q+5$ and $N=\Theta(d)$,
	\begin{equation}\label{sim:eta-to-delta}
		q\eta_{R_*}
		\le C\left(
		q\sqrt{\frac{r+q}{d}}
		+\frac{q(q+1)}d
		\right)
		\le C\Delta.
	\end{equation}
	The last inequality uses $q\ge1$.
	Together with~\eqref{layer:eq:smallness}, this bounds the exponential
	factor by a universal constant.
	The same product estimates bound both final vector norms by
	universal constants.
	Applying~\eqref{sim:vector-to-state} and then the common output channel gives a trace-norm error of at most \(C\Delta\).
	
	\begin{proof}[Proof of Proposition~\ref{sim:comparison}]
		Both Gaussian-to-record representations are exact.
		Combining the Haar-to-Gaussian bound~\eqref{sim:all-gaussian-error}
		with the field-to-encoder comparison just proved, the triangle
		inequality gives
		\[
		\norm{\rho_{\mathrm{true},\vartheta}
			-\rho_{E,\vartheta}}_1
		\le C\left(
		\frac{q^2}{d}
		+q\sqrt{\frac{r+1}{d}}
		+\Delta
		\right)
		\le C'\Delta.
		\]
	\end{proof}
	
	\section{From the record model to the ideal process}
	\label{ideal:sec:comparison}
	
	This appendix compares the weight of any final event in the record
	model with its probability in the ideal process.
	We first represent ideal measurement effects as operators on the
	system and records.
	We then show that one backward step through a record query and instrument approximately follows the ideal update rules.
	Finally, we accumulate this error through the adaptive protocol
	and evaluate the resulting effect on the initial state.
	Together with Proposition~\ref{sim:comparison}, this proves
	Theorem~\ref{layer:thm:simulation}.

	\subsection{Lifting ideal effects}
	
	An ideal effect specifies the probability of a future event as a
	linear functional of the ideal state.
	To compare this functional with the record evolution, we represent
	it on the joint system--record space using the root embedding $J$
	and hidden encoder $E$.
	
	\begin{definition}[Lifting an ideal effect]
		\label{ideal:def:lift}
		A pre-query ideal effect is a triple
		$A=(A_0,A_L,\lambda)$, where
		\[
		0\preceq A_0\preceq I_2,
		\qquad
		0\preceq A_L\preceq I_{\mathcal L},
		\qquad
		0\le\lambda\le1,
		\]
		and $A_L$ is block diagonal across the layers.
		Its lift is
		\begin{equation}\label{layer:eq:lift}
			\iota(A)
			=\lambda I
			+J(A_0-\lambda I_2)J^\dagger
			+E(A_L-\lambda I_{\mathcal L})E^\dagger.
		\end{equation}
		Tensor products with record identities are implicit.
		For a post-query effect $(A_L,\lambda)$, omit the $J$ term.
	\end{definition}
	
	Because $J$ and $E$ are isometries with orthogonal ranges,
	$\iota(A)$ acts as $A_0$ on the root, as $A_L$ on the encoded
	hidden space, and as $\lambda I$ on their orthogonal complement.
	Thus $0\preceq\iota(A)\preceq I$.
	The lift also preserves total record occupation.
	
	We propagate effects backward using the adjoints of the ideal maps.
	For the query,
	\[
	\tr[B\mathcal Q_\vartheta(\omega)]
	=\tr[\mathcal Q_\vartheta^*(B)\omega].
	\]
	For an instrument, the future effect may depend on its outcome, so
	\[
	\tr[\Phi_K^*(\{A_z\})\omega]
	=\sum_z\tr[A_z\Phi_{K,z}(\omega)].
	\]
	These pairings mean the sum of the layer traces and the residual
	scalar product.
	Since a query leaves no root block, the instrument pullback
	$\Phi_K^*(\{A_z\})$ is a pair $(H,g)$ on hidden and residual inputs.
	Pulling this pair through the preceding query gives a pre-query
	triple.

	\subsection{One backward step}
	
	Throughout this appendix, fix $\vartheta$ and work in the regime
	\begin{equation}\label{ideal:eq:regime}
		\Delta=\Delta(q,r,d)\le c,
		\qquad q\ge1,
		\qquad R=5q+5,
	\end{equation}
	where $c$ is a sufficiently small universal constant.
	As established in Appendix~\ref{sim:section}, this ensures
	$D\ge d/2$ and $R\le N/2$.
	
	The next lemma compares one record query followed by an instrument
	with the corresponding ideal operations.
	Its norm bound restricts only the input record occupation and controls the complete output of the error operator.
	This is the form needed to iterate the comparison.
	
	\begin{lemma}[One backward step reproduces the ideal rules]
		\label{ideal:lem:local}
		Let $\{K_z\}_z$ be a physical instrument satisfying
		$\sum_zK_z^\dagger K_z=I$, and assign each outcome a pre-query
		ideal effect $A_z=(A_{0,z},A_{L,z},\lambda_z)$.
		In the regime~\eqref{ideal:eq:regime}, there is a universal
		constant $C$ such that
		\begin{equation}\label{layer:eq:local}
			\left\|
			\left[
			\mathcal U_{E,\vartheta}^\dagger
			\left(\sum_zK_z^\dagger\iota(A_z)K_z\right)
			\mathcal U_{E,\vartheta}
			-\iota\!\left(
			\mathcal Q_\vartheta^*\Phi_K^*(\{A_z\})
			\right)
			\right]\Pi_{\le R-4}^{\mathrm{rec}}
			\right\|
			\le\zeta_R
			:=C\left(
			\frac R{\sqrt N}
			+\frac{R(R+1)}N
			+\frac{p+R}{D}
			\right).
		\end{equation}
	\end{lemma}
	
	\begin{proof}
		Suppress $\vartheta$ in the record embeddings and write
		\[
		V=V_E,\qquad W=W_{E,\vartheta},
		\qquad P_V=VV^\dagger,\qquad P_W=WW^\dagger.
		\]
		Set $B_z=A_{L,z}-\lambda_zI$, so that $\norm{B_z}\le1$.
		The instrument pullback splits as
		\begin{equation}\label{layer:eq:split}
			\begin{aligned}
				N_0
				&=\sum_zK_z^\dagger
				\bigl[\lambda_zI+
				J(A_{0,z}-\lambda_zI)J^\dagger\bigr]K_z,\\
				N_1
				&=\sum_zK_z^\dagger E B_zE^\dagger K_z,
				\qquad
				\mathsf N=N_0+N_1
				=\sum_zK_z^\dagger\iota(A_z)K_z.
			\end{aligned}
		\end{equation}
		Both $N_0$ and $\mathsf N$ are positive contractions.
		The operator $N_0$ acts only on the physical system, whereas $N_1$
		also acts on the records.
		
		Let
		\[
		\mathsf C_\perp=(I-P_W)\mathsf C(I-P_V).
		\]
		Expanding the two terms of the record query gives
		\[
		\begin{aligned}
			\mathcal U_{E,\vartheta}^\dagger
			\mathsf N\mathcal U_{E,\vartheta}
			={}&
			VW^\dagger\mathsf NWV^\dagger
			+\mathsf C_\perp^\dagger\mathsf N\mathsf C_\perp\\
			&+VW^\dagger\mathsf N\mathsf C_\perp
			+\mathsf C_\perp^\dagger\mathsf NWV^\dagger.
		\end{aligned}
		\]
		We estimate the planted term, the completion term, and the two
		cross terms separately.
		
		For the domain bookkeeping, $N_0,N_1,P_V,P_W$ preserve total
		record occupation.
		So do $E^\dagger K_zE$ and $E^\dagger K_z\mathsf C$:
		the latter exchanges one $a$-particle for one $c$-particle.
		The remaining encoders and record creators shift occupation by
		at most one.
		The margin between $R-4$ and $R$ covers the intermediate shifts
		in the products below.
		
		\emph{The planted term.}
		Define the two instrument columns and the outcome-dependent effect by
		\[
		X=\col_z(E^\dagger K_zE),
		\qquad
		Y=\col_zD_z,
		\qquad
		\mathcal B=\bigoplus_zB_z.
		\]
		By Lemma~\ref{enc:instrument},
		\[
		\norm X,\norm Y,\norm{\mathcal B}\le1,
		\qquad
		\norm{(X-Y)\Pi_{\le R}^{\mathrm{rec}}}
		\le C\frac R{\sqrt N}.
		\]
		Since these maps preserve record occupation,
		\[
		\norm{
			(X^\dagger\mathcal BX-Y^\dagger\mathcal BY)
			\Pi_{\le R}^{\mathrm{rec}}}
		\le2\norm{(X-Y)\Pi_{\le R}^{\mathrm{rec}}}.
		\]
		Combining this with the compression estimate for $N_0$ gives
		\begin{equation}\label{layer:eq:planted}
			\begin{aligned}
				E^\dagger\mathsf NE&=H+\mathcal E_H,\\
				H&=T(N_0)+\sum_zD_z^\dagger B_zD_z,\\
				\norm{\mathcal E_H\Pi_{\le R}^{\mathrm{rec}}}
				&\le C\frac R{\sqrt N}.
			\end{aligned}
		\end{equation}
		Using complete instrument columns is what avoids a separate error
		for each outcome.
		
		Substituting the definitions of $N_0$, $B_z$, and $w_{z\sigma}$,
		we obtain
		\begin{equation}\label{layer:eq:hidden-pullback}
			\begin{aligned}
				H={}&\sum_zD_z^\dagger A_{L,z}D_z\\
				&+\bigoplus_{\sigma=\pm}
				\left(
				\sum_z\bigl[
				\tr(A_{0,z}Q_{z\sigma})+\lambda_z w_{z\sigma}
				\bigr]
				\right)I_{\mathcal L_\sigma}.
			\end{aligned}
		\end{equation}
		The first term pairs the future effect with the same-layer output.
		The other two contributions pair it with the root and residual
		outputs.
		Thus, this is exactly the hidden-input adjoint rule of~\eqref{layer:eq:hidden-instrument}.
		In particular, $H$ is a layer-block-diagonal positive contraction.
		
		\emph{The completion term.}
		First consider $\mathsf C^\dagger\mathsf N\mathsf C$, before
		inserting the output projection $I-P_W$.
		For the system-only part, \eqref{comp:contraction} gives
		\[
		\norm{
			\left(\mathsf C^\dagger N_0\mathsf C
			-\frac{\tr N_0}{D}I\right)
			\Pi_{\le R}^{\mathrm{rec}}}
		\le\frac RD.
		\]
		For the encoded part, put
		\[
		Z=\col_z(E^\dagger K_z\mathsf C).
		\]
		Then $\mathsf C^\dagger N_1\mathsf C=Z^\dagger\mathcal BZ$.
		The column return estimate in Lemma~\ref{comp:return} therefore gives
		\begin{equation}\label{layer:eq:completion-pullback}
			\norm{
				\mathsf C^\dagger N_1\mathsf C
				\Pi_{\le R}^{\mathrm{rec}}}
			\le\norm{Z\Pi_{\le R}^{\mathrm{rec}}}^2
			\le C\frac{R(R+1)}N.
		\end{equation}
		
		Define
		\[
		g=\frac{\tr N_0}{d}.
		\]
		Since $0\preceq N_0\preceq I$ and $d-D=p$,
		\[
		\left|\frac{\tr N_0}{D}-g\right|
		\le\frac pD.
		\]
		Expanding $N_0$ gives
		\begin{equation}\label{layer:eq:residual-pullback}
			\begin{aligned}
				g=\sum_z\Bigg[
				&\tr\left(
				A_{0,z}\frac{J^\dagger K_zK_z^\dagger J}{d}
				\right)\\
				&+\lambda_z
				\frac{\tr(K_z^\dagger K_z)
					-\tr(J^\dagger K_zK_z^\dagger J)}d
				\Bigg].
			\end{aligned}
		\end{equation}
		These are precisely the root and residual contributions in
		\eqref{layer:eq:residual-instrument}.
		Consequently,
		\[
		(H,g)=\Phi_K^*(\{A_z\}).
		\]
		
		We next restore the output projection.
		Because $\operatorname{Ran}W\subseteq\operatorname{Ran}E$,
		Lemma~\ref{comp:return} gives
		\[
		\norm{P_W\mathsf C\Pi_{\le R}^{\mathrm{rec}}}
		\le\norm{E^\dagger\mathsf C\Pi_{\le R}^{\mathrm{rec}}}
		\le C\sqrt{\frac{R(R+1)}{ND}}.
		\]
		Also,
		\[
		\norm{\mathsf C\Pi_{\le R}^{\mathrm{rec}}}
		\le\sqrt{\frac{d+R}{D}}=O(1).
		\]
		Since $\mathsf N$ and $I-P_W$ are contractions, replacing
		$\mathsf C^\dagger\mathsf N\mathsf C$ by
		$\mathsf C^\dagger(I-P_W)\mathsf N(I-P_W)\mathsf C$
		costs at most
		\[
		2\norm{\mathsf C\Pi_{\le R}^{\mathrm{rec}}}
		\norm{P_W\mathsf C\Pi_{\le R}^{\mathrm{rec}}}.
		\]
		Finally, sandwiching by $I-P_V$ turns the leading scalar $gI$
		into $g(I-P_V)$.
		
		\emph{The cross terms.}
		For the system-only part, the single-operator return estimate gives
		\[
		\norm{E^\dagger N_0\mathsf C\Pi_{\le R}^{\mathrm{rec}}}
		\le C\sqrt{\frac{R(R+1)}{ND}}.
		\]
		For the encoded part,
		\[
		E^\dagger N_1\mathsf C=X^\dagger\mathcal BZ,
		\]
		so
		\[
		\norm{E^\dagger N_1\mathsf C\Pi_{\le R}^{\mathrm{rec}}}
		\le\norm{Z\Pi_{\le R}^{\mathrm{rec}}}
		\le C\sqrt{\frac{R(R+1)}N}.
		\]
		Inserting $I-P_W$ adds at most
		$\norm{P_W\mathsf C\Pi_{\le R}^{\mathrm{rec}}}$.
		Since $W=E(S_\vartheta\otimes I_{\mathcal F})$ and the remaining
		outer factors are contractions, these estimates control both
		cross terms in the query expansion.
		
		\emph{Identification with the ideal query.}
		Combining the preceding estimates, the leading operator is
		\begin{equation}\label{layer:eq:leading}
			VS_\vartheta^\dagger HS_\vartheta V^\dagger
			+g(I-P_V).
		\end{equation}
		To identify it, write $H=\bigoplus_{j=1}^{\ell}H_j$ and define
		\[
		\begin{aligned}
			\Theta_j
			&=S_{\vartheta,j}^\dagger H_{j+1}S_{\vartheta,j}
			&&(0\le j<\ell),\\
			\Theta_\ell&=0,
			&\Theta_L&=\bigoplus_{j=1}^{\ell}\Theta_j,
			&\Gamma_L&=\bigoplus_{j=1}^{\ell}\Gamma_j.
		\end{aligned}
		\]
		Each $\Theta_j$ is supported on $\operatorname{Ran}\Gamma_j$.
		Moreover, distinct input layers have distinct successor layers,
		and $H$ has no matrix elements between those layers.
		The column formulas~\eqref{layer:eq:encoders} therefore give
		\[
		P_V=JJ^\dagger+E\Gamma_LE^\dagger,
		\qquad
		VS_\vartheta^\dagger HS_\vartheta V^\dagger
		=J\Theta_0J^\dagger+E\Theta_LE^\dagger.
		\]
		The ideal query rule~\eqref{layer:eq:query} gives
		\[
		\mathcal Q_\vartheta^*(H,g)
		=\bigl(
		\Theta_0,\,
		\Theta_L+g(I_{\mathcal L}-\Gamma_L),\,
		g
		\bigr).
		\]
		Substituting this triple into~\eqref{layer:eq:lift} yields exactly
		\eqref{layer:eq:leading}.
		
		The total error is bounded by
		\[
		C\left(
		\frac R{\sqrt N}
		+\frac{R(R+1)}N
		+\frac{p+R}{D}
		\right)=\zeta_R.
		\]
		All estimates control the full output on inputs with occupation at most $R-4$, which  proves~\eqref{layer:eq:local}.
		The output-projection estimates used only
		$\operatorname{Ran}W_{E,\vartheta}\subseteq\operatorname{Ran}E$,
		so they allow the projection to depend on $\vartheta$.
	\end{proof}
	
	\subsection{Adaptive recursion and evaluation on the vacuum}
	
	We now apply the local comparison to a complete future event,
	propagate its error backward through the adaptive history tree,
	and evaluate the resulting effect on the initial state.
	
	\begin{proof}[Proof of Theorem~\ref{layer:thm:simulation}]
		Fix a final event.
		At each final classical history $h$, combine the final readout and
		any preceding known operations into one physical effect
		$0\preceq M_h\preceq I$.
		The corresponding ideal terminal effect is
		\[
		\left(T(M_h),\frac{\tr M_h}{d}\right),
		\]
		as specified in~\eqref{layer:eq:terminal}.
		To apply Lemma~\ref{ideal:lem:local} to the terminal measurement, represent the chosen event by the binary instrument
		\[
		K_1=M_h^{1/2},
		\qquad
		K_0=(I-M_h)^{1/2},
		\]
		where outcome $1$ means that the event occurs.
		Assign the constant ideal effects
		\[
		A_1=(I_2,I_{\mathcal L},1),
		\qquad
		A_0=(0,0,0).
		\]
		Their lifts are $\iota(A_1)=I$ and $\iota(A_0)=0$, so
		\[
		\sum_{z=0}^1K_z^\dagger\iota(A_z)K_z
		=M_h\otimes I_{\mathcal F}.
		\]
		The ideal instrument rules give
		\[
		\Phi_K^*(\{A_0,A_1\})
		=
		\left(
		\bigoplus_{\sigma=\pm}
		\frac{\tr(\Pi_\sigma M_h)}N I_{\mathcal L_\sigma},
		\frac{\tr M_h}{d}
		\right)
		=
		\left(T(M_h),\frac{\tr M_h}{d}\right).
		\]
		Lemma~\ref{ideal:lem:local} therefore applies directly and yields
		\[
		\left\|
		\left[
		\mathcal U_{E,\vartheta}^\dagger
		(M_h\otimes I)\mathcal U_{E,\vartheta}
		-\iota\!\left(
		\mathcal Q_\vartheta^*
		\left(T(M_h),\frac{\tr M_h}{d}\right)
		\right)
		\right]\Pi_{\le R-4}^{\mathrm{rec}}
		\right\|
		\le\zeta_R.
		\]
		This bound is uniform over the chosen event and final history.
		
		\emph{Future-event operators.}
		At a history $h$ with $t$ queries remaining, let $B_{t,h}$ be
		the record operator whose pairing with the current state gives
		the weight of the chosen event after the remaining evolution.
		For one remaining query,
		\[
		B_{1,h}
		=\mathcal U_{E,\vartheta}^\dagger
		(M_h\otimes I)\mathcal U_{E,\vartheta}.
		\]
		For $t\ge2$,
		\[
		B_{t,h}
		=\mathcal U_{E,\vartheta}^\dagger
		\left(
		\sum_zK_{z\mid h}^\dagger B_{t-1,hz}K_{z\mid h}
		\right)
		\mathcal U_{E,\vartheta}.
		\]
		These positive operators are defined on the finite-particle core.
		Because the record evolution need not preserve trace, they need
		not be bounded by the identity.
		
		Let $A_{t,h}$ be the corresponding ideal future effect:
		\[
		\begin{aligned}
			A_{1,h}
			&=\mathcal Q_\vartheta^*
			\left(T(M_h),\frac{\tr M_h}{d}\right),\\
			A_{t,h}
			&=\mathcal Q_\vartheta^*
			\Phi_{K_{\cdot\mid h}}^*
			\bigl(\{A_{t-1,hz}\}_z\bigr)
			&&(t\ge2).
		\end{aligned}
		\]
		The ideal maps preserve normalisation, so all $A_{t,h}$ are effects.
		Define the uniform error
		\[
		e_t
		=\sup_h
		\norm{
			(B_{t,h}-\iota(A_{t,h}))
			\Pi_{\le q-t}^{\mathrm{rec}}}.
		\]
		
		\emph{Propagation of the error.}
		By Lemma~\ref{rec:norm-degree}, a record query increases total
		occupation by at most one, and its adjoint never increases it.
		Thus a $t$-query future-event operator maps inputs of occupation
		at most $L$ to outputs of occupation at most $L+t$.
		Since the lifted ideal effect preserves occupation, every error
		operator appearing in the definition of $e_t$ has output occupation
		at most $q$.
		
		Suppose a node has $t+1$ queries remaining and put
		\[
		X_z=B_{t,hz}-\iota(A_{t,hz}).
		\]
		The propagated child error is
		\[
		\mathcal U_{E,\vartheta}^\dagger
		\left(\sum_zK_{z\mid h}^\dagger X_zK_{z\mid h}\right)
		\mathcal U_{E,\vartheta}.
		\]
		On input occupation at most $q-t-1$, the right query produces
		occupation at most $q-t$.
		The instrument preserves occupation, and its column
		$\col_zK_{z\mid h}$ is an isometry.
		The block-diagonal family $\bigoplus_zX_z$ therefore contributes
		at most $e_t$, with complete output supported on occupation at
		most $q$.
		
		The left adjoint query also has norm at most $c_q$ on this output
		domain.
		Indeed, its occupation monotonicity gives
		\[
		\mathcal U_{E,\vartheta}^\dagger
		\Pi_{\le q}^{\mathrm{rec}}
		=
		\Pi_{\le q}^{\mathrm{rec}}
		\mathcal U_{E,\vartheta}^\dagger
		\Pi_{\le q}^{\mathrm{rec}},
		\]
		and hence
		\[
		\norm{
			\mathcal U_{E,\vartheta}^\dagger
			\Pi_{\le q}^{\mathrm{rec}}}
		=
		\norm{
			\Pi_{\le q}^{\mathrm{rec}}
			\mathcal U_{E,\vartheta}
			\Pi_{\le q}^{\mathrm{rec}}}
		\le c_q.
		\]
		The right query has the same bound.
		Thus, the propagated child error is at most $c_q^2e_t$.
		Adding the local error from Lemma~\ref{ideal:lem:local} gives
		\begin{equation}\label{layer:eq:recursion}
			e_1\le\zeta_R,
			\qquad
			e_{t+1}\le c_q^2e_t+\zeta_R.
		\end{equation}
		The base case follows from the terminal estimate because
		$q-1\le R-4$.
		All subsequent input domains also lie within this cutoff.
		
		Since
		\[
		c_q^{2q}
		=\left(1+\frac{p+q}{D}\right)^q
		\le\exp\left(C\frac{q(p+q)}d\right)
		=O(1)
		\]
		by~\eqref{layer:eq:smallness}, summing the recursion yields
		\begin{equation}\label{layer:eq:accumulation}
			\begin{aligned}
				e_q
				&\le Cq\zeta_R\\
				&\le C\left(
				\frac{q^2}{\sqrt d}
				+\frac{q^3}{d}
				+\frac{q(p+q)}d
				\right)
				\le C\Delta.
			\end{aligned}
		\end{equation}
		Here we used $R=5q+5$, \eqref{layer:eq:smallness}, and
		$q/\sqrt d\le\Delta\le c$.
		
		\emph{Evaluation on the initial state.}
		The records start in vacuum, and $E^\dagger$ annihilates every
		input with vacuum records.
		Therefore, for any physical initial state $\rho$ and ideal effect
		$A=(A_0,A_L,\lambda)$,
		\begin{equation}\label{layer:eq:vacuum}
			\begin{aligned}
				\tr\!\left[
				\iota(A)(\rho\otimes\ket\Omega\bra\Omega)
				\right]
				&=
				\tr(A_0J^\dagger\rho J)
				+\lambda\bigl(1-\tr(R_0\rho)\bigr)\\
				&=\tr(A\omega_{\mathrm{in}}).
			\end{aligned}
		\end{equation}
		This agrees exactly with the ideal initialisation
		in~\eqref{layer:eq:initial}, including when $\rho$ has coherence
		between the root and its complement.
		Evaluating~\eqref{layer:eq:accumulation} on this initial state therefore bounds the difference between the record event weight
		and the ideal event probability by $C\Delta$.
		Initial classical randomness is handled by averaging this bound.
		
		Finally, Proposition~\ref{sim:comparison} bounds the difference
		between the physical event probability and its record weight by
		another $C\Delta$.
		The triangle inequality gives the same order of error between
		physical and ideal probabilities, uniformly over the chosen event.
		Both the physical and ideal output laws are normalised, so taking
		the supremum over events proves~\eqref{layer:eq:comparison}.
	\end{proof}

	\section{Ancilla-free lower bound for estimating multi-time correlators}
	\label{hard:section}

    We state and prove the formal version of the lower bound in Theorem~\ref{main:thm:advantage_single} for the case of consecutive time steps $t_j=j-1$ and a single known traceless dichotomic observable $O_j=O$, where $C(U)=\bra\psi (U^k)^\dagger(OU)^kO\ket\psi$.

\begin{theorem}[System-only lower bound for multi-time correlators, formal]
\label{hard:thm:formal}
Let $k\ge2$ and let $d\ge4k$ be even. For every known traceless dichotomic observable $O$, there exist a known unit vector $\psi$ and distributions $\mathcal D_s$ on $\mathrm U(d)$, $s\in\{+1,-1\}$, with $C(U)=s$ on the support of $\mathcal D_s$, such that any system-only protocol using at most $q$ forward queries satisfies
\begin{equation}\label{main:bound}
 \TV(\mathsf{Out}_+,\mathsf{Out}_-)
 \le\min\left\{1,\ q2^{1-k}
              +C\frac{q^2+q\sqrt k}{\sqrt d}\right\},
\end{equation}
where $\mathsf{Out}_s$ is its full classical output law averaged over one draw $U\sim\mathcal D_s$, and $C>0$ is a universal constant. Consequently, estimating $C(U)$ to additive error $1/3$ with success probability at least $2/3$ on every oracle in these supports requires
\begin{equation}\label{main:query-lower}
 q=\Omega\!\left(\min\{2^k,d^{1/4},\sqrt{d/k}\}\right).
\end{equation}
In particular, for $n\ge 4k$ qubits, any system-only protocol requires $q=\Omega(2^k)$ queries.
\end{theorem}

To establish this theorem, we construct two random oracle families with correlator values $+1$ and $-1$, and use the simulation theorem to compare their output distributions with those of the ideal process.
In the ideal process, the sign is detectable only through coherence between two paths.
We show that this coherence decays exponentially as it passes through the internal stages.
Combining this contraction with a hybrid argument and the simulation bound yields the claimed query lower bound.
	
	\subsection{The hard oracle family}
	\label{app:hard_oracle_multi_time}
	Fix $k\ge2$ and an even $d\ge4k$.
	Let $O$ be the known dichotomic observable from Theorem~\ref{hard:thm:formal}.
	Choose known unit vectors $e_\sigma$ satisfying
	$Oe_\sigma=\sigma e_\sigma$ for $\sigma\in\{+1,-1\}$, and set
	\[
	a_0=\psi=\frac{e_++e_-}{\sqrt2},
	\qquad
	b_0=O\psi=\frac{e_+-e_-}{\sqrt2},
	\qquad
	J=(a_0\ \ b_0),
	\qquad
	R_0=JJ^\dagger.
	\]
	The vectors $a_0,b_0$ form an orthonormal basis of the known root
	space.
	Let $\mathcal H_\sigma$ be the orthogonal complement of $e_\sigma$
	within the $\sigma$-eigenspace of $O$, with projector $\Pi_\sigma$.
	Then
	\[
	\mathcal H
	=\operatorname{Ran}J\oplus\mathcal H_+\oplus\mathcal H_-,
	\qquad
	\dim\mathcal H_\sigma=N=d/2-1.
	\]
	
	Set $r=2k-1\le N$ and draw independent Haar-distributed
	orthonormal $r$-frames $F_\sigma$ in the two sectors.
	Label the columns of each frame by
	\begin{equation}\label{hard:labels}
		A_1,B_1,A_2,B_2,\ldots,A_{k-1},B_{k-1},A_k.
	\end{equation}
	Denote the columns labeled $A_j$ and $B_j$ by $x_{j,\sigma}$
	and $y_{j,\sigma}$, respectively, and define
	\[
	a_j=\frac{x_{j,+}+x_{j,-}}{\sqrt2}
	\quad(1\le j\le k),
	\qquad
	b_j=\frac{y_{j,+}+y_{j,-}}{\sqrt2}
	\quad(1\le j<k).
	\]
	For each internal stage $1\le j<k$, the vectors
	$a_j,Oa_j,b_j,Ob_j$ are orthonormal, as is the endpoint pair
	$a_k,Oa_k$.
	Their spans and the root space are mutually orthogonal.
	
	For each sign $s\in\{+1,-1\}$, prescribe the transitions
	\begin{equation}\label{hard:arrows}
		\begin{aligned}
			Ua_{j-1}&=a_j
			&&(1\le j\le k),\\
			Ub_{j-1}&=Ob_j
			&&(1\le j<k),\\
			Ub_{k-1}&=sOa_k.
		\end{aligned}
	\end{equation}
	The $2k$ input vectors are orthonormal, as are their prescribed
	outputs.
	Let
	\[
	V_F=(a_0,b_0,a_1,b_1,\ldots,a_{k-1},b_{k-1}),
	\]
	and let $W_{F,s}$ contain the matching output columns
	from~\eqref{hard:arrows}.
	Thus
	\begin{equation}\label{hard:dimensions}
		p=2k,
		\qquad
		D=d-2k\ge d/2.
	\end{equation}
	Using the complement embeddings from~\eqref{layer:eq:oracle},
	define
	\begin{equation}\label{hard:unitary}
		U=W_{F,s}V_F^\dagger+B_{F,s}HC_F^\dagger,
		\qquad
		H\sim\operatorname{Haar}(\mathrm U(D)),
	\end{equation}
	with $H$ independent of the frames.
	This construction defines the distribution $\mathcal D_s$.
	
	The next lemma verifies that distinguishing the two oracle families
	is necessary for estimating the target correlator.
	
	\begin{lemma}[The encoded correlator]
		\label{hard:answer}
		Every $U$ in the support of $\mathcal D_s$ satisfies $C(U)=s$.
	\end{lemma}
	
	\begin{proof}
		The first sequence of transitions gives $U^k\psi=a_k$.
		For the second sequence, $O\psi=b_0$ and
		\[
		OUb_{j-1}=b_j\quad(1\le j<k),
		\qquad
		OUb_{k-1}=s a_k.
		\]
		Therefore $(OU)^kO\psi=s a_k$, and $C(U) =\bra\psi(U^k)^\dagger(OU)^kO\ket\psi =s$.
	\end{proof}
	
	\subsection{Reduction to the ideal process}
	
	We now express this family in the notation of
	Appendix~\ref{layer:section}.
	Take the label layers
	\begin{equation}\label{corr:eq:layers}
		\begin{aligned}
			\mathcal L_j
			&=\operatorname{span}\{
			\ket{A_j,+},\ket{A_j,-},
			\ket{B_j,+},\ket{B_j,-}\}
			&&(1\le j<k),\\
			\mathcal L_k
			&=\operatorname{span}\{\ket{A_k,+},\ket{A_k,-}\}.
		\end{aligned}
	\end{equation}
	Define
	\[
	\bar a_j
	=\frac{\ket{A_j,+}+\ket{A_j,-}}{\sqrt2}
	\quad(1\le j\le k),
	\qquad
	\bar b_j
	=\frac{\ket{B_j,+}+\ket{B_j,-}}{\sqrt2}
	\quad(1\le j<k).
	\]
	Let $\bar a_0,\bar b_0$ be the orthonormal root coordinates
	mapped by $J$ to $a_0,b_0$.
	For $0\le j<k$, choose $\mathcal I_j=\mathbb C^2$ and let
	$V_j$ embed its basis as $\bar a_j,\bar b_j$.
	The corresponding input projections are
	\[
	\Gamma_0=I,
	\qquad
	\Gamma_j
	=\ket{\bar a_j}\bra{\bar a_j}
	+\ket{\bar b_j}\bra{\bar b_j}
	\quad(1\le j<k),
	\qquad
	\Gamma_k=0.
	\]
	
	The prescribed label-space transitions are
	\begin{equation}\label{ideal:eq:arrows}
		\begin{aligned}
			S_{s,j}\bar a_j&=\bar a_{j+1}
			&&(0\le j<k),\\
			S_{s,j}\bar b_j&=Z_{\mathcal L}\bar b_{j+1}
			&&(0\le j<k-1),\\
			S_{s,k-1}\bar b_{k-1}&=sZ_{\mathcal L}\bar a_k.
		\end{aligned}
	\end{equation}
	Each map vanishes on the orthogonal complement of its prescribed input subspace, so $S_{s,j}^\dagger S_{s,j}=\Gamma_j$.
	Moreover, $Z_{\mathcal L}\mathcal L_j\subseteq\mathcal L_j$, thus every layer is invariant under $Z_{\mathcal L}$. 
	
	With $\ell=k$, $r=2k-1$, and $p=2k$,
	Theorem~\ref{layer:thm:simulation} gives, after adjusting the
	universal constants,
	\begin{equation}\label{corr:eq:comparison}
		\TV(\mathsf{Out}_s,\mathsf{Out}^{\mathrm{id}}_s)
		\le C\Delta_k,
		\qquad
		\Delta_k=\frac{q^2+q\sqrt k}{\sqrt d},
		\qquad
		\Delta_k\le c.
	\end{equation}
	The ideal process uses the initialisation, instruments, and terminal
	effects defined in Appendix~\ref{layer:section}.
	Only the last transition in~\eqref{ideal:eq:arrows} depends on $s$.
	
	\subsection{Contraction of coherence between the two paths}
	\label{sign:sec:proof}
	
	Changing the sign of the final transition leaves the populations of the two paths unchanged.
	It affects only their relative coherence.
	The following lemma bounds the extent to which this coherence survives each internal stage.
	
	\begin{lemma}[One internal stage contracts two-path coherence]
		\label{sign:lem:contraction}
		Fix a history and its chosen instrument, and write
		\[
		\alpha_z=\frac{\kappa_{z+}+\kappa_{z-}}2,
		\qquad
		\beta_z=\frac{\kappa_{z+}-\kappa_{z-}}2,
		\qquad
		D_z=\alpha_zI+\beta_zZ_{\mathcal L}.
		\]
		Then
		\begin{equation}\label{sign:eq:scalar}
			\sum_z(|\alpha_z|^2+|\beta_z|^2)\le1,
			\qquad
			\sum_z|\alpha_z\beta_z|\le\frac12.
		\end{equation}
		
		Immediately after a query, an internal layer $1\le j<k$
		is supported on
		\[
		\mathcal L_j^{\mathrm{live}}
		=\operatorname{span}\{\bar a_j,Z_{\mathcal L}\bar b_j\}.
		\]
		For an unnormalised state $\rho$ on this subspace, define
		\[
		c=\bra{\bar a_j}\rho\ket{Z_{\mathcal L}\bar b_j},
		\qquad
		\rho_{z,\mathrm{acc}}
		=\Gamma_jD_z\rho D_z^\dagger\Gamma_j.
		\]
		The coherence in the input accepted by the next query satisfies
		\begin{equation}\label{sign:eq:contraction}
			c'_z
			:=\bra{\bar a_j}\rho_{z,\mathrm{acc}}\ket{\bar b_j}
			=\alpha_z\overline{\beta_z}\,c,
			\qquad
			\sum_z|c'_z|\le\frac12|c|.
		\end{equation}
	\end{lemma}
	
	\begin{proof}
		Equation~\eqref{enc:kappa-completeness} gives
		$\sum_z|\kappa_{z\sigma}|^2\le1$ for each sector.
		Hence
		\[
		\sum_z(|\alpha_z|^2+|\beta_z|^2)
		=\frac12\sum_z
		\bigl(|\kappa_{z+}|^2+|\kappa_{z-}|^2\bigr)
		\le1.
		\]
		The inequality
		$2|\alpha_z\beta_z|\le|\alpha_z|^2+|\beta_z|^2$
		then proves~\eqref{sign:eq:scalar}.
		
		The four vectors
		$\bar a_j,Z_{\mathcal L}\bar a_j,
		\bar b_j,Z_{\mathcal L}\bar b_j$
		are orthonormal. Therefore
		\[
		\Gamma_jD_z\bar a_j=\alpha_z\bar a_j,
		\qquad
		\Gamma_jD_zZ_{\mathcal L}\bar b_j=\beta_z\bar b_j.
		\]
		Applying these identities to the two-dimensional input state
		proves the formula for $c'_z$.
		Summing its absolute value and using~\eqref{sign:eq:scalar}
		gives the contraction bound.
		
		
	\end{proof}

	We now bound the effect of changing the sign at a single query position.
	We then compare the two complete protocols by changing one position at a time.
	
	\begin{lemma}[A sign change at one query position has small effect]
		\label{sign:lem:one_position}
		Fix a query position $t$ in an ideal protocol.
		Compare two evolutions that use the same adaptive control policy
		and the same query signs at every position except $t$.
		Their output total variation distance is zero for $t<k$ and at
		most $2^{1-k}$ for $t\ge k$.
		The common signs at the other positions may be chosen arbitrarily.
	\end{lemma}
	
	\begin{proof}
		The two evolutions have the same state before query $t$.
		At history $h$, let $c_h$ be the coherence between the accepted
		last-layer inputs $\bar a_{k-1},\bar b_{k-1}$.
		Only this coherence can detect the sign change.
		On the endpoint basis
		$\bar a_k,Z_{\mathcal L}\bar a_k$, the difference between the
		two query outputs is
		\[
		\begin{pmatrix}
			0&2c_h\\
			2\overline{c_h}&0
		\end{pmatrix},
		\]
		whose trace norm is $4|c_h|$.
		Since histories occupy orthogonal classical blocks, the trace
		distance immediately after the query is
		\begin{equation}\label{sign:eq:local_distance}
			2\sum_h|c_h|.
		\end{equation}
		
		A hidden contribution arriving at the endpoint at query $t$
		must have left the root at query $t-k+1$.
		Each query advances it by one layer; instruments preserve its
		hidden layer or transfer its weight to the root or residual block.
		A residual input cannot enter a hidden layer.
		Since the ideal initialisation~\eqref{layer:eq:initial} has no
		hidden weight, no contribution can reach the endpoint when $t<k$.
		
		Now suppose $t\ge k$.
		For each history $g$ immediately before query $t-k+1$, let
		$\omega_g$ be its root block and put $p_g=\tr\omega_g$.
		Normalization gives $\sum_gp_g\le1$.
		Positivity of each root block implies
		\[
		\left|\bra{\bar a_0}\omega_g\ket{\bar b_0}\right|
		\le
		\sqrt{
			\bra{\bar a_0}\omega_g\ket{\bar a_0}
			\bra{\bar b_0}\omega_g\ket{\bar b_0}}
		\le\frac{p_g}{2}.
		\]
		The departure query transfers this coherence to the first layer
		without changing its magnitude.
		There are then $k-1$ internal instrument steps before the endpoint
		query.
		Applying Lemma~\ref{sign:lem:contraction} at each descendant
		history and summing over its outcomes gives
		\begin{equation}\label{sign:eq:coherence_mass}
			\sum_h|c_h|
			\le2^{-(k-1)}\sum_g\frac{p_g}{2}
			\le2^{-k}.
		\end{equation}
		Combining~\eqref{sign:eq:local_distance}
		and~\eqref{sign:eq:coherence_mass} bounds the immediate trace
		distance by $2^{1-k}$.
		The common adaptive suffix is a completely positive, trace-preserving map on the joint classical--quantum state.
		Contractivity of trace distance, followed by the final readout, proves the claim.
	\end{proof}
	
	Summing the single-position bounds gives a uniform bound for the entire ideal protocol.
	
	\begin{proposition}[Ideal sign indistinguishability]
		\label{sign:prop:ideal}
		Every $q$-query ideal protocol initialised by~\eqref{layer:eq:initial} satisfies
		\begin{equation}\label{sign:eq:ideal_tv}
			\TV(\mathsf{Out}^{\mathrm{id}}_+,
			\mathsf{Out}^{\mathrm{id}}_-)
			\le(q-k+1)_+\,2^{1-k}
			\le q\,2^{1-k},
		\end{equation}
		where $(x)_+=\max\{x,0\}$.
	\end{proposition}
	
	\begin{proof}
		Use $q+1$ hybrid evolutions, replacing the all-minus query-sign
		sequence by the all-plus sequence one position at a time.
		Neighboring hybrids have identical prefixes and a common adaptive
		suffix, and differ at exactly one query position.
		Lemma~\ref{sign:lem:one_position} bounds their output distance
		by zero at the first $k-1$ positions and by $2^{1-k}$ at each
		remaining position.
		The triangle inequality proves~\eqref{sign:eq:ideal_tv}.
	\end{proof}
	
	\subsection{Completion of the lower bound}
	
	\begin{proof}[Proof of Theorem~\ref{hard:thm:formal}]
		For $q\ge1$ and $\Delta_k\le c$, the triangle inequality,
		\eqref{corr:eq:comparison} for both signs, and
		Proposition~\ref{sign:prop:ideal} give
		\begin{equation}\label{sign:eq:true_tv}
			\TV(\mathsf{Out}_+,\mathsf{Out}_-)
			\le q\,2^{1-k}
			+C\frac{q^2+q\sqrt k}{\sqrt d}.
		\end{equation}
		For $q=0$, the two output distributions coincide.
		For $\Delta_k>c$, increasing the universal constant $C$ so that
		$Cc\ge1$ makes~\eqref{sign:eq:true_tv} follow from $\TV\le1$.
		Thus~\eqref{sign:eq:true_tv} holds for every $q$.
		Taking the minimum of its right-hand side and one
		proves~\eqref{main:bound}.
		
		To obtain the query lower bound, recall from
		Lemma~\ref{hard:answer} that the two oracle families have
		correlator values $+1$ and $-1$.
		An estimator with additive error at most $1/3$ and success
		probability at least $2/3$ for every oracle in either family
		therefore distinguishes the families with the same success
		probability by taking the sign of the real part of its estimate.
		Consequently,
		\[
		\TV(\mathsf{Out}_+,\mathsf{Out}_-)\ge\frac13.
		\]
		Combining this with~\eqref{sign:eq:true_tv} yields
		\[
		q=\Omega\!\left(
		\min\left\{2^k,d^{1/4},\sqrt{d/k}\right\}
		\right),
		\]
		which proves~\eqref{main:query-lower}.
	\end{proof}

	\section{Ancilla-free lower bound for simultaneously estimating nested correlators}
	\label{nest:sec:proof}

    This appendix gives a formal statement and proof of the system-only lower bound in Theorem~\ref{nest:thm:advantage_main}.
    
    \begin{theorem}[System-only lower bound for simultaneous nested-correlator estimation]
    \label{nest:thm:lower-formal}
    There are universal constants $c,C>0$ with the following property.
    For every $k\ge1$, every even dimension $d\ge4M$ with $M=2^k$,
    and every known traceless dichotomic observable $O$, there exist a known
    unit vector $\psi$ and distributions $\mathcal D_{\theta,s}$ over
    $\mathrm U(d)$, indexed by $\theta\in\{0,1\}^k$ and $s\in\{\pm1\}$,
    such that the correlators in~\eqref{nest:eq:targets}, specialized to
    $O_j=O$, $t_j=j-1$ and $\rho=\ket\psi\bra\psi$, obey
    \begin{equation}\label{nest:eq:onehot-target}
     F_\sigma(U)=\frac{3s}{8}\mathbf1_{\{\sigma=\theta\}}
    \end{equation}
    throughout the support of $\mathcal D_{\theta,s}$.
    Every system-only protocol satisfying
    \begin{equation}\label{nest:eq:accuracy}
     \Pr\!\left(\max_\sigma
     |\widehat F_\sigma-F_\sigma(U)|\le\frac13\right)\ge\frac23
    \end{equation}
    on every oracle in these supports requires
    \begin{equation}\label{nest:eq:lower-bound}
     q\ge c\min\{kM,d^{1/4},\sqrt{d/M}\}.
    \end{equation}
    In particular, $d\ge Ck^4M^4$ gives $q=\Omega(kM)$.
    \end{theorem}
    
    To prove Theorem~\ref{nest:thm:lower-formal}, we first express the nested
    correlators as sums of path overlaps and construct a random oracle family
    with exactly one nonzero correlator.
    We then apply the simulation theorem to obtain an ideal tree process.
    In this process, each traversal reaches the endpoint with a product state, and all information obtained during an endpoint visit can be represented by a single biased bit.
    We bound the information accumulated over successive visits and transfer the resulting estimation bound to the physical oracle.
	
	\subsection{Complementary-path expansion}
	
    We use the observables, times and input state of \eqref{nest:eq:targets},
    with $M=2^k$, and abbreviate $A_j=U^{-t_j}O_jU^{t_j}$.
    Let $\mathcal N_\sigma$ denote the unnormalised nested-bracket operator
    inside the trace in~\eqref{nest:eq:targets}.
    The lower-bound construction below uses the special case $O_j=O$,
    $t_j=j-1$ and $\rho=\ket\psi\bra\psi$.
    
    For $b\in\{0,1\}^k$, write $\bar b=1-b$ and
    $\chi_\sigma(b)=(-1)^{\sigma\cdot b}$.
    The paths and overlaps in~\eqref{nest:eq:main-paths} have the
    Heisenberg-picture expressions
    \begin{equation}\label{nest:eq:general-forward-path}
    	W_b
    	=U^T A_k^{b_k}\cdots A_1^{b_1},
    \end{equation}
    and
    \begin{equation}\label{nest:eq:general-overlap}
    	\begin{aligned}
    		Q_b
    		&=\tr\!\left(
    		A_1^{1-b_1}\cdots A_k^{1-b_k}A_{k+1}
    		A_k^{b_k}\cdots A_1^{b_1}\rho
    		\right).
    	\end{aligned}
    \end{equation}
    The following identity expresses every bracket pattern as a Fourier
    coefficient of these overlaps.

    \begin{lemma}[Complementary-path expansion]
	\label{nest:lem:expansion}
	For uniform $b\in\{0,1\}^k$,
	\begin{equation}\label{nest:eq:fourier}
		F_\sigma
		=\frac{i^{|\sigma|}}2
		\E_b[\chi_\sigma(b)Q_b].
	\end{equation}
	Moreover, $Q_{\bar b}=Q_b^*$, and $F_\sigma$ is real for every pattern $\sigma$.
\end{lemma}

    \begin{proof}
	In bracket $j$, let $b_j=1$ select right multiplication by $A_j$
	and $b_j=0$ select left multiplication.
	The corresponding coefficients are $1$ and $(-1)^{\sigma_j}$.
	Expanding the nested brackets gives
	\begin{equation}\label{nest:eq:word-expansion}
		\mathcal N_\sigma
		=\sum_b(-1)^{\sum_j\sigma_j(1-b_j)}
		A_1^{1-b_1}\cdots A_k^{1-b_k}A_{k+1}
		A_k^{b_k}\cdots A_1^{b_1}.
	\end{equation}
	Taking the trace against $\rho$, multiplying by the normalisation
	and phase in~\eqref{nest:eq:targets}, and using
	$(-1)^{\sum_j\sigma_j(1-b_j)}=(-1)^{|\sigma|}\chi_\sigma(b)$
	gives~\eqref{nest:eq:fourier}, since
	$(-i)^{|\sigma|}(-1)^{|\sigma|}=i^{|\sigma|}$.
	Taking the complex conjugate of~\eqref{nest:eq:general-overlap}
	gives $Q_{\bar b}=Q_b^*$.
	Together with
	$\chi_\sigma(\bar b)=(-1)^{|\sigma|}\chi_\sigma(b)$,
	this proves that $F_{\sigma}$ is real.
\end{proof}
	
	\subsection{A hidden tree with one nonzero correlator}
	\label{nest:sec:family}

	Let $d\ge4M$ be even, and let $O$ be the known traceless dichotomic
	observable.
	Choose known unit vectors $v_\pm$ with $Ov_\pm=\pm v_\pm$, and set
	\[
	e_\varnothing=\psi=\frac{v_++v_-}{\sqrt2},
	\qquad
	f_\varnothing=O\psi=\frac{v_+-v_-}{\sqrt2},
	\qquad
	J=(e_\varnothing\ \ f_\varnothing).
	\]
	These two vectors span the root space.
	As in Appendix~\ref{layer:section}, let $\mathcal H_\pm$ be the
	orthogonal complements of $v_\pm$ in the corresponding eigenspaces
	of $O$.
	Each has dimension $N=d/2-1$.
	Draw independent Haar-distributed frames with
	\[
	r=2M-2
	\]
	columns in each sector.
	Label $M-2$ columns by the binary words $w$ with
	$1\le |w|\le k-1$, and label the remaining $M$ columns by the
	words of length $k$.
	For an internal word $w$, denote its two sector columns by
	$x_{w,\pm}$ and define
	\[
	e_w=\frac{x_{w,+}+x_{w,-}}{\sqrt2},
	\qquad
	f_w=\frac{x_{w,+}-x_{w,-}}{\sqrt2}=Oe_w.
	\]
	The root pair and all internal pairs are mutually orthonormal.
	When $k=1$, there are no internal words.
	Let
	$F_{\mathrm{end},\pm}:\mathbb C^M\to\mathcal H_\pm$
	be the subframes formed by the endpoint columns.

	The hidden parameters are a coordinate
	$\theta\in\{0,1\}^k$ and a sign $s\in\{\pm1\}$.
	Fix $\lambda=3/4$, and define
	\[
	P_\theta
	=\bigotimes_{j=1}^k
	\begin{cases}
		X,&\theta_j=0,\\
		-Y,&\theta_j=1,
	\end{cases}
	\]
	where $X,Y$ are the standard Pauli matrices.
	Encode the parameters at the endpoint through
	\[
	R_{\theta,s}
	=F_{\mathrm{end},+}
	\sqrt{\frac{I+s\lambda P_\theta}{2}}
	+F_{\mathrm{end},-}
	\sqrt{\frac{I-s\lambda P_\theta}{2}}.
	\]
	Since $P_\theta$ is a Hermitian unitary and $\lambda<1$, both square roots are well defined.
	Orthogonality of the two sectors gives
	\begin{equation}\label{nest:eq:endpoint}
		R_{\theta,s}^\dagger R_{\theta,s}=I,
		\qquad
		R_{\theta,s}^\dagger O R_{\theta,s}=s\lambda P_\theta.
	\end{equation}
	Thus $R_{\theta,s}$ is an isometry from $\mathbb C^M$ into the
	$2M$-dimensional endpoint space.

	Prescribe the transitions
	\begin{equation}\label{nest:eq:arrows}
		\begin{aligned}
			Ue_w&=e_{w0},
			&
			Uf_w&=e_{w1}
			&& (0\le |w|\le k-2),\\
			Ue_w&=R_{\theta,s}\ket{w0},
			&
			Uf_w&=R_{\theta,s}\ket{w1}
			&& (|w|=k-1).
		\end{aligned}
	\end{equation}
	For $k=1$, only the second line applies, with
	$w=\varnothing$.
	Each query appends one bit to the path, and the $k$th query
	reaches the endpoint.
	There are
	\[
	p=2\sum_{j=0}^{k-1}2^j=2M-2
	\]
	prescribed input vectors.
	Their outputs consist of $M-2$ internal vectors and $M$ endpoint
	vectors, so both lists are orthonormal.
	Let $V_F$ and $W_{F,\theta,s}$ contain these input and output
	columns, respectively.
	Complete the prescribed transitions to a unitary by setting
	\begin{equation}\label{nest:eq:oracle}
		U
		=W_{F,\theta,s}V_F^\dagger
		+B_{F,\theta,s}H_D C_F^\dagger,
		\qquad
		H_D\sim\operatorname{Haar}(\mathrm U(D)),
		\qquad D=d-p.
	\end{equation}
	Here $C_F$ and $B_{F,\theta,s}$ are isometries onto the input
	and output complements, and $H_D$ is independent of the frames.
	The assumption $d\ge4M$ ensures $r\le N$ and $D\ge d/2$.
	The frames and completion are drawn once and reused at every query, and their joint draw defines $\mathcal D_{\theta,s}$.
	
	The next lemma verifies that this construction places the only nonzero
	correlator at coordinate $\theta$, with sign $s$.

	\begin{lemma}[Exact correlator vector]\label{nest:lem:onehot}
		Every oracle in the support of $\mathcal D_{\theta,s}$ satisfies
		\eqref{nest:eq:onehot-target}.
	\end{lemma}
	
	\begin{proof}
		For the target times, $W_b=(UO^{b_k})\cdots(UO^{b_1})$.
		Starting from $\psi=e_\varnothing$, the gate $O^{b_j}$ selects
		$e_w$ or $f_w$, and the next query appends $b_j$ to the path.
		\eqref{nest:eq:arrows} therefore gives
		\[
		W_b\psi=R_{\theta,s}\ket b.
		\]
		Since
		\[
		X\ket{b_j}=\ket{1-b_j},
		\qquad
		(-Y)\ket{b_j}=-i(-1)^{b_j}\ket{1-b_j},
		\]
		we have
		\[
		P_\theta\ket b
		=(-i)^{|\theta|}(-1)^{\theta\cdot b}\ket{\bar b}.
		\]
		Using~\eqref{nest:eq:endpoint}, the complementary-path overlap is
		\[
		Q_b
		=\bra{\bar b}R_{\theta,s}^\dagger
		O R_{\theta,s}\ket b
		=s\lambda(-i)^{|\theta|}(-1)^{\theta\cdot b}.
		\]
		Substituting into~\eqref{nest:eq:fourier} and using $\E_b(-1)^{(\sigma+\theta)\cdot b}
		=\mathbf1_{\{\sigma=\theta\}}$ gives
		\[
    	F_\sigma
    	=\frac{s\lambda}{2}\mathbf1_{\{\sigma=\theta\}}.
    	\]
    	The claim follows from $\lambda=3/4$.
	\end{proof}
	
	\subsection{Reduction to the ideal tree}
	\label{nest:sec:comparison}
	
	We now express the construction in the notation of
	Appendix~\ref{layer:section}.
	Take $\vartheta=(\theta,s)$, $\ell=k$, and $r=p=2M-2$.
	For $1\le j\le k$, define
	\[
	\mathcal L_j
	=\operatorname{span}
	\{\ket{w,\nu}:|w|=j,\ \nu=\pm1\},
	\qquad
	Z_{\mathcal L}\ket{w,\nu}=\nu\ket{w,\nu}.
	\]
	Thus $Z_{\mathcal L}$ acts within each layer.
	For $1\le |w|<k$, set
	\[
	\bar e_w=\frac{\ket{w,+}+\ket{w,-}}{\sqrt2},
	\qquad
	\bar f_w=\frac{\ket{w,+}-\ket{w,-}}{\sqrt2}
	=Z_{\mathcal L}\bar e_w.
	\]
	At the root, use the basis
	$\bar e_\varnothing,\bar f_\varnothing$ mapped by $J$ to
	$e_\varnothing,f_\varnothing$.
	
	For the endpoint sector inclusions
	$I_\pm:\mathbb C^M\to\mathcal L_k$, define
	\[
	\mathcal R_{\theta,s}
	=I_+\sqrt{\frac{I+s\lambda P_\theta}{2}}
	+I_-\sqrt{\frac{I-s\lambda P_\theta}{2}}.
	\]
	Then
	\begin{equation}\label{nest:eq:logical-endpoint}
		\mathcal R_{\theta,s}^\dagger\mathcal R_{\theta,s}=I,
		\qquad
		\mathcal R_{\theta,s}^\dagger
		Z_{\mathcal L}\mathcal R_{\theta,s}
		=s\lambda P_\theta.
	\end{equation}
	The ideal transitions are obtained from~\eqref{nest:eq:arrows}
	by replacing $e_w,f_w,R_{\theta,s}$ with
	$\bar e_w,\bar f_w,\mathcal R_{\theta,s}$.
	Take $\mathcal I_j=\mathcal L_j$ and $V_j=I$ for $0\le j<k$.
	Hence
	\[
	\Gamma_j=I_{\mathcal L_j}\quad(0\le j<k),
	\qquad
	\Gamma_k=0.
	\]
	Every root or internal input advances to the next layer, whereas
	an endpoint input is sent to the residual block.
	
	These choices satisfy the hypotheses of
	Theorem~\ref{layer:thm:simulation}.
	Consequently, for every allowed $q$-query protocol and each fixed
	$(\theta,s)$,
	\begin{equation}\label{nest:eq:comparison}
		\TV(\mathsf{Out}_{\theta,s},
		\mathsf{Out}^{\mathrm{id}}_{\theta,s})
		\le C\Delta,
		\qquad
		\Delta=\frac{q^2+q\sqrt{r+1}}{\sqrt d},
		\qquad
		q\ge1,\quad \Delta\le c_0.
	\end{equation}
	The physical output is averaged over the initial oracle draw.
	
	The ideal instrument rules are those of
	Appendix~\ref{layer:section}.
	In particular, the part of outcome $z$ that remains in a hidden
	layer applies
	\[
	D_z=\bigoplus_{\nu=\pm1}
	\kappa_{z\nu}I_{\mathcal L_\nu},
	\qquad
	\kappa_{z\nu}=\frac{\tr(\Pi_\nu K_z)}N.
	\]
	The other destinations are the root and residual blocks.
	Thus an instrument can retain the current layer or end the
	current traversal, but it cannot move directly to another hidden layer.
	Only the last transition, into the endpoint, depends on
	$(\theta,s)$.
	
	\subsection{Product states along a traversal}
	\label{nest:sec:ideal}
	
	To bound the information available in the ideal process, we may give
	the protocol additional classical information.
	Reveal its initial block and, after each instrument, whether the state
	remains in the current layer or moves to the root or residual block.
	These destinations are completely positive summands of the ideal
	instrument, so ignoring the additional flags recovers the original
	process.
	We allow later controls to depend on the flags, and an upper bound on
	success for this enlarged class also applies to the original protocol.
	The flags reveal the occupied layer, but no individual path label.
	
	A traversal starts when a query leaves the root.
	It ends upon a return to the root, entry into the residual block,
	or discard of the endpoint.
	Identify the root with one virtual qubit and each internal layer by
	\begin{equation}\label{nest:eq:virtual-basis}
		\bar e_w\longleftrightarrow\ket w\ket0,
		\qquad
		\bar f_w\longleftrightarrow\ket w\ket1,
		\qquad |w|=j.
	\end{equation}
	The first $j$ qubits encode the path, and the last is a flag.
	In these coordinates,
	\begin{equation}\label{nest:eq:flag-filter}
		Z_{\mathcal L}=I_{\mathrm{path}}\otimes X,
		\qquad
		D_z=I_{\mathrm{path}}\otimes L_z,
		\qquad
		L_z=a_zI+b_zX,
	\end{equation}
	where
	\[
	a_z=\frac{\kappa_{z,+}+\kappa_{z,-}}2,
	\qquad
	b_z=\frac{\kappa_{z,+}-\kappa_{z,-}}2.
	\]
	The next lemma shows that their reachable states have a product
	structure.
	
	\begin{lemma}[Product structure along a traversal]
		\label{nest:lem:product}
		Conditioned on a complete classical history $h$, a normalised state
		in layer $0\le j<k$ has the form
		\begin{equation}\label{nest:eq:product-state}
			\tau_h^{(j)}
			=\tau_{h,1}\otimes\cdots\otimes\tau_{h,j}\otimes\phi_h,
		\end{equation}
		where each factor is a one-qubit density matrix and $\phi_h$
		is the current flag state.
		In particular, the input to the endpoint transition is a product
		state on $k$ qubits:
		\begin{equation}\label{nest:eq:preendpoint-product}
			\tau_h
			=\tau_{h,1}\otimes\cdots\otimes\tau_{h,k-1}\otimes\phi_h,
			\qquad
			\tau_h\longmapsto
			\mathcal R_{\theta,s}\tau_h\mathcal R_{\theta,s}^\dagger.
		\end{equation}
		Forgetting classical refinements gives convex mixtures of product states.
	\end{lemma}
	
	\begin{proof}
		A root state is an arbitrary one-qubit density matrix $\phi_h$,
		which gives~\eqref{nest:eq:product-state} for $j=0$.
		
		Suppose the state has the stated form in an internal layer.
		A live instrument outcome acts only on its last factor.
		Writing $c_{h,z}=\tr(L_z\phi_hL_z^\dagger)$, its normalised output,
		when $c_{h,z}>0$, is
		\begin{equation}\label{nest:eq:product-filter}
			\frac{D_z\tau_h^{(j)}D_z^\dagger}
			{\tr(D_z\tau_h^{(j)}D_z^\dagger)}
			=
			\bigotimes_{a=1}^j\tau_{h,a}
			\otimes\frac{L_z\phi_hL_z^\dagger}{c_{h,z}}.
		\end{equation}
		For $0\le j<k-1$, the next query acts as
		\[
		\mathcal T_j:\ket w\ket b\longmapsto\ket{wb}\ket0.
		\]
		It therefore gives
		\begin{equation}\label{nest:eq:product-query}
			\mathcal T_j\tau_h^{(j)}\mathcal T_j^\dagger
			=
			\bigotimes_{a=1}^j\tau_{h,a}
			\otimes\phi_h\otimes\ket0\bra0.
		\end{equation}
		Thus a live instrument changes only the current flag, and a query
		stores that flag as the next path factor and appends a fresh flag.
		
		A return to the root discards all path factors.
		More explicitly, for a normalised hidden state $\omega$ with sector
		populations $\pi_\nu=\tr(P_{j\nu}\omega)$, the root branch of outcome
		$z$ has probability and normalised state
		\begin{equation}\label{nest:eq:root-reset}
			r_z(\omega)=\sum_\nu\pi_\nu\tr Q_{z\nu},
			\qquad
			\rho_{z,\mathrm{root}}(\omega)
			=\frac{\sum_\nu\pi_\nu Q_{z\nu}}{r_z(\omega)}
			\quad\text{if }r_z(\omega)>0.
		\end{equation}
		This is again a single root qubit.
		A residual branch retains only a scalar weight, and any later return from the residual block also prepares one root qubit.
		The induction, therefore, restarts after every reset.
		
		The final query applies $\mathcal R_{\theta,s}$ to the resulting
		$k$-qubit product state, proving~\eqref{nest:eq:preendpoint-product}.
		Averaging over forgotten classical refinements gives convex mixtures
		of product states.
	\end{proof}

	\subsection{Information from one endpoint visit}
	
	An endpoint visit begins upon arrival and ends when the state moves to the root or residual block, or when the final measurement is performed.
	Let $\Xi_\nu$ project onto sector $\nu\in\{\pm1\}$ of $\mathcal L_k$.
	For a normalised input $\tau$ to the endpoint transition,
	\eqref{nest:eq:logical-endpoint} gives the sector probabilities
	\begin{equation}\label{nest:eq:biasedbit}
		p_{\theta,s}(\nu\mid\tau)
		=\tr\!\left(
		\Xi_\nu\mathcal R_{\theta,s}\tau
		\mathcal R_{\theta,s}^\dagger
		\right)
		=\frac{1+\nu s\lambda\tr(P_\theta\tau)}2.
	\end{equation}
	The following lemma shows that these two probabilities suffice to
	simulate the entire classical record of a visit.
	
	\begin{lemma}[One-bit representation of an endpoint visit]
		\label{nest:lem:endpoint-bit}
		Fix a history $h$ preceding an endpoint arrival and let $\tau$
		be the input to the endpoint transition.
		The record $e$ of outcomes and destinations during the visit
		can be generated by first drawing a bit $S\in\{\pm1\}$
		with probabilities $p_{\theta,s}(\nu\mid\tau)$, then sampling $e$
		from a conditional distribution $K_h(e\mid S)$ that is independent
		of $\theta$, $s$, and $\tau$.
		Thus the probability $\mathbb P_{\theta,s}(e\mid h,\tau)$
		of the actual visit record satisfies
		\begin{equation}\label{nest:eq:endpoint-factorization}
			\mathbb P_{\theta,s}(e\mid h,\tau)
			=
			\sum_{\nu=\pm1}
			p_{\theta,s}(\nu\mid\tau)K_h(e\mid\nu).
		\end{equation}
	\end{lemma}
	
	\begin{proof}
		An instrument outcome that remains at the endpoint applies
		\[
		D_z=\kappa_{z,+}\Xi_++\kappa_{z,-}\Xi_-.
		\]
		This operator acts as a scalar within each sector.
		The root and residual maps and the final measurement probabilities
		depend only on the sector weights.
		Measuring the sector upon arrival, therefore, leaves the visit statistics unchanged.
		Its outcome $S=\nu$ has probability $p_{\theta,s}(\nu\mid\tau)$.
		Conditional on $S=\nu$, the instrument rules give
		\begin{equation}\label{nest:eq:endpoint-kernels}
			\begin{aligned}
				K(z,\mathrm{live}\mid\nu)&=|\kappa_{z\nu}|^2,\\
				K(z,\mathrm{root}\mid\nu)&=\tr Q_{z\nu},\\
				K(z,\perp\mid\nu)&=w_{z\nu}.
			\end{aligned}
		\end{equation}
		These probabilities sum to one:
		\[
		\sum_z\left(
		|\kappa_{z\nu}|^2+\tr Q_{z\nu}+w_{z\nu}
		\right)
		=\sum_zp_{z\nu}=1.
		\]
		A live outcome retains the same sector.
		A root outcome prepares $Q_{z\nu}/\tr Q_{z\nu}$ when its
		probability is nonzero, and a residual outcome ends the visit.
		A query also ends the visit by sending the state to the residual block.
		For a final POVM $\{M_y\}_y$, the conditional probabilities are
		\begin{equation}\label{nest:eq:endpoint-terminal-kernel}
			K(y\mid\nu)=\frac{\tr(\Pi_\nu M_y)}N.
		\end{equation}
		
		All these rules are parameter independent.
		Applying them according to the protocol's history-dependent policy
		defines the conditional distribution $K_h(e\mid\nu)$ of the entire
		visit record.
		Averaging over $S$ proves~\eqref{nest:eq:endpoint-factorization}.
	\end{proof}
	
	The one-bit representation isolates the parameter dependence of each
	visit in the bias of $S$.
	To bound the protocol's performance, we may reveal this bit and allow later controls to depend on it.
	This only gives the protocol more information, and ignoring the bit recovers the original ideal process.
	
	Given the enhanced history, the reset rules above determine the root state without knowing $(\theta,s)$.
	All transitions before the next endpoint's arrival are also parameter-independent.
	Thus, the next endpoint input is determined by the history alone and is a product state by Lemma~\ref{nest:lem:product}.
	The following bound applies to these states and their convex mixtures and controls the information gained from each visit.
	
	\begin{lemma}[Squared-bias bound]\label{nest:lem:bias}
		For every convex mixture $\tau$ of $k$-qubit product states,
		\begin{equation}\label{nest:eq:bias-budget}
			\sum_{\theta\in\{0,1\}^k}
			\bigl[\lambda\tr(P_\theta\tau)\bigr]^2
			\le\lambda^2.
		\end{equation}
	\end{lemma}
	
	\begin{proof}
		For a product state $\tau=\bigotimes_j\tau_j$, write
		$x_j=\tr(X\tau_j)$ and $y_j=\tr(Y\tau_j)$.
		Then
		\[
		\sum_\theta
		\bigl[\lambda\tr(P_\theta\tau)\bigr]^2
		=
		\lambda^2\prod_{j=1}^k(x_j^2+y_j^2)
		\le\lambda^2,
		\]
		because each qubit state satisfies $x_j^2+y_j^2\le1$.
		Since each bias is real and linear in $\tau$, the left-hand side
		of~\eqref{nest:eq:bias-budget} is convex in $\tau$.
		The bound therefore extends to convex mixtures of product states.
	\end{proof}
	
	Although there are $M=2^k$ possible hidden coordinates, the sum of their squared biases is bounded by the constant $\lambda^2$.
	Thus, a single endpoint input cannot produce large biases across many parameters at once.
	In the next subsection, we use this bound to control the information accumulated over adaptively chosen endpoint visits.
	
	\subsection{Accumulating information under adaptive control}
	\label{nest:sec:information}
	
	Each endpoint arrival requires a new root-to-endpoint traversal of $k$ queries.
	A protocol using at most $q$ queries therefore has at most 
	\begin{equation}\label{nest:eq:visit-count}
		L_{\mathrm{leaf}}:=\lfloor q/k\rfloor
	\end{equation}
	arrivals on any execution branch.
	For the analysis, append independent fair bits after termination if necessary, so that every branch has a list of exactly $L_{\mathrm{leaf}}$ bits.
	These additional bits do not affect the output.
	
	Let $\mathbb P_{\theta,s}$ denote the complete transcript law of the enhanced ideal process, including the destination flags, endpoint bits, instrument outcomes, classical randomness, and final estimates.
	Define a common reference law $\mathbb P_0$ by replacing every
	endpoint bit with a fair coin, while keeping the protocol's policy
	and all other conditional rules unchanged.
	
	At a fixed enhanced history $h$ preceding an endpoint arrival,
	the input $\tau_h$ is a product state determined by $h$ alone.
	Indeed, the recorded endpoint bits and outcomes determine the reset
	states, and the internal transitions are parameter independent.
	Under $\mathbb P_{\theta,s}$, the next endpoint bit has conditional
	mean
	\[
	b_{\theta,s}(h)
	=s\lambda\tr(P_\theta\tau_h),
	\qquad
	|b_{\theta,s}(h)|\le\lambda.
	\]
	For appended bits, set $b_{\theta,s}(h)=0$.
	Because the input at a fixed history is the same for all parameters,
	we can apply the squared-bias bound before averaging over histories.
	This gives the following bound on estimation success.
	
	\begin{proposition}[Success probability in the ideal tree]
		\label{nest:prop:ideal-success}
		For $\lambda=3/4$, every enhanced ideal protocol using at most
		$q$ queries satisfies
		\begin{equation}\label{nest:eq:idealsuccess}
			p_{\mathrm{id}}
			\le\frac12+\frac3{2\sqrt7}\sqrt{\frac{q}{kM}},
		\end{equation}
		where $p_{\mathrm{id}}$ is the probability that all correlator
		estimates have additive error at most $1/3$, averaged uniformly
		over $(\theta,s)$.
	\end{proposition}
	
	\begin{proof}
		We first compare the transcript laws with $\mathbb P_0$.
		Write $D(P\|Q)$ for relative entropy with natural logarithms.
		For a fair bit and a bit of bias $b$, with $|b|\le\lambda<1$,
		\[
		d(b)
		:=D\!\left(
		\operatorname{Bern}(1/2)
		\,\middle\|\,
		\operatorname{Bern}((1+b)/2)
		\right)
		=-\frac12\log(1-b^2)
		\le\frac{b^2}{2(1-\lambda^2)}.
		\]
		At a fixed history, only the endpoint-bit distributions differ
		between $\mathbb P_0$ and $\mathbb P_{\theta,s}$.
		The chain rule therefore gives
		\[
		D(\mathbb P_0\|\mathbb P_{\theta,s})
		=
		\sum_{\ell=1}^{L_{\mathrm{leaf}}}
		\E_{h_\ell\sim\mathbb P_0}
		d\bigl(b_{\theta,s}(h_\ell)\bigr),
		\]
		where $h_\ell$ is the history immediately before the $\ell$th bit.
		Using this direction of relative entropy ensures that all
		expectations are taken under the same reference law.
		
		At every actual endpoint arrival, Lemma~\ref{nest:lem:bias} gives
		\[
		\frac1{2M}\sum_{\theta,s}b_{\theta,s}(h)^2
		=
		\frac1M\sum_\theta
		\bigl[\lambda\tr(P_\theta\tau_h)\bigr]^2
		\le\frac{\lambda^2}{M}.
		\]
		The same bound holds for appended bits, whose biases are zero.
		Averaging the chain rule over $(\theta,s)$ yields
		\begin{equation}\label{nest:eq:KL}
			\frac1{2M}\sum_{\theta,s}
			D(\mathbb P_0\|\mathbb P_{\theta,s})
			\le
			\frac{L_{\mathrm{leaf}}\lambda^2}
			{2(1-\lambda^2)M}.
		\end{equation}
		Pinsker's inequality and Cauchy--Schwarz then imply
		\begin{equation}\label{nest:eq:TVideal}
			\begin{aligned}
				\frac1{2M}\sum_{\theta,s}
				\TV(\mathbb P_0,\mathbb P_{\theta,s})
				&\le
				\sqrt{
					\frac1{4M}\sum_{\theta,s}
					D(\mathbb P_0\|\mathbb P_{\theta,s})
				}\\
				&\le
				\frac{\lambda}{2}
				\sqrt{\frac{L_{\mathrm{leaf}}}{(1-\lambda^2)M}}.
			\end{aligned}
		\end{equation}
        To relate this distance to estimation success, we may take the
    	estimates $\widehat F_\sigma$ to be real, since taking real parts
    	cannot increase their errors.
    	By~\eqref{nest:eq:onehot-target}, simultaneous error at most
    	$1/3$ implies
    	\[
    	s\widehat F_\theta
    	\ge\frac38-\frac13=\frac1{24}.
    	\]
    	Thus success under parameter $(\theta,s)$ implies the event
    	\[
    	A_{\theta,s}
    	=\{s\widehat F_\theta\ge1/24\}.
    	\]
		For each $\theta$, the events $A_{\theta,+}$ and $A_{\theta,-}$
		are disjoint.
		Under the common reference law, this gives
		\[
		\frac1{2M}\sum_{\theta,s}
		\mathbb P_0(A_{\theta,s})\le\frac12.
		\]
		Consequently,
		\[
		\begin{aligned}
			p_{\mathrm{id}}
			&\le
			\frac1{2M}\sum_{\theta,s}
			\mathbb P_{\theta,s}(A_{\theta,s})\\
			&\le
			\frac12+
			\frac1{2M}\sum_{\theta,s}
			\TV(\mathbb P_0,\mathbb P_{\theta,s}).
		\end{aligned}
		\]
		Substituting $\lambda=3/4$ into~\eqref{nest:eq:TVideal} and using $L_{\mathrm{leaf}}\le q/k$
		proves~\eqref{nest:eq:idealsuccess}.
	\end{proof}
	
	\subsection{Completion of the lower bound}
	\label{nest:sec:finish}
	
	It remains to transfer the ideal success bound to the physical oracle families.
	
	\begin{proof}[Proof of Theorem~\ref{nest:thm:lower-formal}]
		Let $p_{\mathrm{true}}$ be the success probability averaged over uniform $(\theta,s)$ and the draw $U\sim\mathcal D_{\theta,s}$.
		By Lemma~\ref{nest:lem:onehot}, all oracles in a fixed family $\mathcal D_{\theta,s}$ have the same correlator vector.
		Success is therefore a fixed event in the classical output, to which~\eqref{nest:eq:comparison} applies.
		Combining this comparison with
		Proposition~\ref{nest:prop:ideal-success} and using $r=2M-2$ gives
		\begin{equation}\label{nest:eq:mainfinite}
			p_{\mathrm{true}}
			\le
			\frac12+
			\frac3{2\sqrt7}\sqrt{\frac{q}{kM}}
			+C\frac{q^2+q\sqrt M}{\sqrt d}
		\end{equation}
		for $q\ge1$ and $\Delta\le c_0$.
		For $q=0$, the output is parameter independent, and the
		disjoint-event argument above gives $p_{\mathrm{true}}\le1/2$.
		
		Suppose
		\[
		q<c_*\min\{kM,d^{1/4},\sqrt{d/M}\}.
		\]
		Then
		\[
		\sqrt{\frac{q}{kM}}<\sqrt{c_*},
		\qquad
		\frac{q^2}{\sqrt d}<c_*^2,
		\qquad
		\frac{q\sqrt M}{\sqrt d}<c_*.
		\]
		Choosing a sufficiently small universal $c_*>0$ ensures the
		simulation condition and gives
		\[
		p_{\mathrm{true}}
		\le
		\frac12+\frac3{2\sqrt7}\sqrt{c_*}
		+C(c_*^2+c_*)
		<\frac23.
		\]
		This contradicts a success probability of at least $2/3$ for
		every oracle in every support, proving~\eqref{nest:eq:lower-bound}.
	\end{proof}

\section{Lower bounds for coherent query number}
\label{depth:section}

In this appendix, we establish exponential query lower bounds for estimating either a single multi-time correlator $C(U)$ or a nested correlator $F_\theta(U)$ when the coherent query number is bounded by $L<k$. 
We consider a general adaptive model in which each experimental run may employ arbitrary ancillary registers and mid-circuit quantum instruments, but makes at most $L$ forward queries to the unknown unitary $U$. 
Between runs, all quantum memory is discarded or reset, while the full classical measurement record is retained to adaptively configure subsequent experiments.

\begin{theorem}[A threshold for the coherent query number, formal version of Theorem~\ref{thm:query_depth_informal}]
\label{depth:thm}
Let $k\ge2$, $M=2^k$, and let $d = 2^n \ge4M$. Fix a known
traceless dichotomic observable $O$. There exist a known unit vector
$\psi$ and oracle distributions $\mathcal D_{\theta,s}$, with
$\theta\in\{0,1\}^k$ and $s\in\{+1,-1\}$, such that for
$O_j=O$, $t_j=j-1$, and $\rho=\ket\psi\bra\psi$,
\begin{equation}
 C(U)=\frac{3s}{4}i^{|\theta|},\qquad
 F_\sigma(U)=\frac{3s}{8}\mathbf1_{\{\sigma=\theta\}}
 \quad\text{on }\operatorname{supp}\mathcal D_{\theta,s}.
 \label{depth:signals}
\end{equation}
For every fixed, publicly known $\theta$, any protocol in the above
model making at most $L<k$ forward queries per run and at most $q$ in total satisfies
\begin{equation}
 \TV(\mathsf{Out}_{\theta,+},\mathsf{Out}_{\theta,-})
 \le 32\sqrt{\frac{q^3\log d}{d}}
 +672q\sqrt{\frac{M}{d}},
 \label{depth:tv}
\end{equation}
where $\mathsf{Out}_{\theta,s}$ is the classical output distribution
averaged over a single oracle drawn from $\mathcal D_{\theta,s}$ and
reused throughout the protocol. Consequently, estimating either $C(U)$
or the prescribed $F_\theta(U)$ to within additive error $1/3$, with success
probability at least $2/3$ for every oracle in these supports, requires
\begin{equation}
 q\ge\frac1{4096}\min\left\{
 \left(\frac d{\log d}\right)^{1/3},\;
 \sqrt{\frac d{2^k}}\right\}.
 \label{depth:queries}
\end{equation}
In particular, $n\ge4k$ gives
$q=\Omega(2^{n/4})$.
\end{theorem}

Our proof builds on the hidden-tree ensemble of Appendix~\ref{nest:sec:family}, which encodes the unknown sign $s \in \{\pm 1\}$ solely at depth $k$. 
The core of the argument, detailed in the following subsections, compares any protocol making at most $L<k$ queries per run against an auxiliary process that truncates the tree to its first $L$ levels, rendering its evolution completely independent of $s$, and bounds the resulting total variation error across all $q$ queries.

\subsection{The hidden tree and its first levels}
\label{depth:prefixes}

We briefly recall the hidden-tree construction of Appendix~\ref{nest:sec:family} and isolate the prefix structure accessed by a shallow experiment. Throughout the proof, the target string $\theta \in \{0,1\}^k$ is fixed and publicly known, and the protocol only needs to distinguish the sign $s \in \{\pm 1\}$.

Choose known unit eigenvectors $v_\pm$ of $O$ satisfying $Ov_\pm=\pm v_\pm$,
and define the root basis
\[
 e_\varnothing=\psi=\frac{v_++v_-}{\sqrt2},\qquad
 f_\varnothing=\frac{v_+-v_-}{\sqrt2}.
\]
The orthogonal complement $\mathcal H_\pm$ of $v_\pm$ within each eigenspace has dimension
$N=d/2-1$. In each sector, we draw an independent Haar-distributed frame $F_\pm$ containing
one column $x_{w,\pm}$ for every binary word $w$ of length $1\le |w|\le k$, where $|w|$ denotes the word length and $w0,w1$ denote single-bit extensions. Each sector frame thus has
\[
 r=\sum_{j=1}^{k}2^j=(M-2)+M=2M-2
\]
columns, while the known root vectors are excluded from $F_\pm$. For each internal word $1\le|w|<k$, we set
\[
 e_w=\frac{x_{w,+}+x_{w,-}}{\sqrt2},\qquad
 f_w=\frac{x_{w,+}-x_{w,-}}{\sqrt2}.
\]
Writing $F=F_+\oplus F_-$, the oracle family of
Appendix~\ref{nest:sec:family} prescribes the tree transitions
\[
 \begin{aligned}
 Ue_w&=e_{w0}, & Uf_w&=e_{w1}
 &&(0\le|w|\le k-2),\\
 Ue_w&=R_{\theta,s}\ket{w0}, &
 Uf_w&=R_{\theta,s}\ket{w1}
 &&(|w|=k-1),
 \end{aligned}
\]
where $R_{\theta,s}$ is the endpoint isometry from
Appendix~\ref{nest:sec:family} (with $\lambda=3/4$) whose range lies in
the span of the length-$k$ frame columns. Crucially, only this final transition at $|w|=k-1$ depends on $(\theta,s)$.

Let $V_F$ collect the input vectors $(e_w,f_w)$ for $|w|<k$, and let
$W_{F,\theta,s}$ collect their corresponding outputs in the same order. Both form orthonormal frames of $p=2\sum_{j=0}^{k-1}2^j=2M-2$ columns.
Specifying $(F,\theta,s)$ fixes the action $UV_F=W_{F,\theta,s}$, while the action on $(\operatorname{Ran}V_F)^\perp$ is completed by an independent Haar-random unitary onto $(\operatorname{Ran}W_{F,\theta,s})^\perp$, as in Eq.~\eqref{nest:eq:oracle}. We refer to this as the \emph{Haar completion} of the prescribed transitions.

To model experiments of bounded coherent query number, for each $1\le j<k$ we define the \emph{depth-$j$ prefix} by retaining only the transitions on input words of length $|w|<j$. Its input and output frames are
\[
 V^{[j]}=(e_w,f_w)_{|w|<j},\qquad
 W^{[j]}=(e_{w0},e_{w1})_{|w|<j},
\]
ordered identically. Because the outputs of the depth-$j$ prefix reach word length at most $j<k$, this prefix never accesses the endpoint isometry $R_{\theta,s}$ and is completely independent of $(\theta,s)$.

Finally, for $0\le t<k$, let $Z_t$ collect the frame columns $x_{w,\pm}$ with $1\le|w|\le t$ (with $Z_0$ empty), comprising $a_t=\sum_{j=1}^t2^j=2^{t+1}-2$ columns per sector. 
We denote the orthogonal projector onto the remaining, deeper frame columns by
\[
 P_t=\sum_{\sigma=\pm}\ \sum_{t<|w|\le k}
       \ket{x_{w,\sigma}}\bra{x_{w,\sigma}}.
\]
Note that $Z_t$ and $P_t$ are introduced for later analysis and are unknown to the protocol. The following lemma summarises the exact correlator values of the ensemble and shows that truncating the tree at depth $t+1$ only alters transitions supported on $\operatorname{Ran}P_t$.

\begin{lemma}[Signals and geometry of the tree]
\label{depth:lem:geometry}
The oracle ensemble $\mathcal D_{\theta,s}$ satisfies Eq.~\eqref{depth:signals}. For each $1\le j<k$, the depth-$j$ prefix has $2^{j+1}-2$ input and output columns and is determined entirely by $Z_j$, independently of $(\theta,s)$. Moreover, for $0\le t\le k-2$, the transitions omitted from the depth-$(t+1)$ prefix correspond to input words of length $t+1\le|w|\le k-1$, all of whose inputs and outputs lie in $\operatorname{Ran}P_t$:
\[
 e_w,\ f_w,\ Ue_w,\ Uf_w\in\operatorname{Ran}P_t
 \qquad(t+1\le|w|\le k-1).
\]
Moreover, under the initial Haar distribution of $F_\pm$, conditioned on fixing the first $a_t$ columns $Z_t$, the remaining columns in
each sector form a Haar-distributed frame of rank $r-a_t$ in an ambient subspace of dimension
\begin{equation}
 N-a_t\ge D_*:=d/2-M/2+1\ge d/4.
 \label{depth:ambient}
\end{equation}
\end{lemma}

\begin{proof}
By Lemma~\ref{nest:lem:onehot}, the complementary-path overlaps satisfy
$Q_b=(3s/4)(-i)^{|\theta|}(-1)^{\theta\cdot b}$.
Evaluating this at $b=\mathbf1$ yields $C(U) = Q_{\mathbf 1}$ in Eq.~\eqref{depth:signals}, while the expression for $F_\sigma(U)$ follows directly from Eq.~\eqref{nest:eq:onehot-target}.
Next, the depth-$j$ prefix contains two input columns for each word of length $0\le |w|\le j-1$, giving $2\sum_{a=0}^{j-1}2^a=2^{j+1}-2$ columns in total. The prescribed tree transitions immediately imply that $(V^{[j]}, W^{[j]})$ depends only on $Z_j$ and that all inputs and outputs for $t+1\le |w|\le k-1$ are supported on the columns with $|w|>t$, i.e., in $\operatorname{Ran}P_t$.
Finally, conditioning a Haar-distributed frame on $a_t$ columns leaves the remaining $r-a_t$ columns Haar-distributed within the $(N-a_t)$-dimensional orthogonal complement in each sector. Since $t\le k-2$ implies $a_t\le 2^{k-1}-2 = M/2-2$, substituting $N=d/2-1$ and $d\ge 4M$ yields Eq.~\eqref{depth:ambient}.
\end{proof}

\subsection{Comparison with a process independent of the sign}
\label{depth:histories}

We now couple the full and prefix unitaries on a common probability space and use the prefix unitaries to define a sign-independent auxiliary process. 
Comparing the physical protocol against this auxiliary process then reduces to bounding the weight of each query input state on the unvisited tree subspace $\operatorname{Ran}P_t$.

\subsubsection{Coupling the Haar completions}

Because the full tree and its prefixes specify only partial isometries, we couple their random completions using a single Haar unitary $G\sim\operatorname{Haar}(\mathrm U(d))$ drawn independently of $F$. This joint construction ensures that each oracle retains its exact marginal completion law while allowing us to compare their actions directly on the same input state.

\begin{lemma}[Coupled construction of full and prefix oracles]
\label{depth:lem:coupling}
Let $G\sim\operatorname{Haar}(\mathrm U(d))$ be independent of $F$.
There exists a joint construction
\[
 U_{\theta,s}=U_{\theta,s}(F,G),\qquad
 U^{[j]}=U^{[j]}(Z_j,G),\quad 1\le j<k,
\]
such that, conditional on $F$, the unitary $U_{\theta,s}$ satisfies $U_{\theta,s}V_F=W_{F,\theta,s}$ and acts as a Haar-random unitary from
$(\operatorname{Ran}V_F)^\perp$ onto
$(\operatorname{Ran}W_{F,\theta,s})^\perp$.
Likewise, conditional on $Z_j$, each prefix oracle $U^{[j]}$ satisfies
$U^{[j]}V^{[j]}=W^{[j]}$ and acts as a Haar-random unitary between the corresponding orthogonal complements.

Define the frame overlap
\[
 \delta_{\theta,s}:=\norm{V_F^\dagger G^\dagger W_{F,\theta,s}}.
\]
For every $0\le t\le k-2$ and every system--ancilla vector $\phi$,
\begin{equation}
 \norm{(U_{\theta,s}-U^{[t+1]})\phi}
 \le2\bigl(\norm{P_t\phi}+\norm{P_tG\phi}\bigr)
       +32\delta_{\theta,s}\norm\phi,
 \label{depth:local}
\end{equation}
where system operators act as the identity on the ancilla. Moreover,
\begin{equation}
 \E_G[\delta_{\theta,s}\mid F]\le10\sqrt{M/d}.
 \label{depth:overlap}
\end{equation}
\end{lemma}
Thus, the single-query replacement error is controlled by the projections of $\phi$ and $G\phi$ onto the deeper tree subspace $\operatorname{Ran}P_t$, up to a small overlap correction $\delta_{\theta,s}$. We defer the proof of Lemma~\ref{depth:lem:coupling} to Sec.~\ref{depth:coupling}.

\subsubsection{Defining the auxiliary process}
The auxiliary process is identical to the physical protocol except for the oracle calls: within each experimental run, the $j$th query to $U_{\theta,s}$ is replaced by $U^{[j]}$. The step counter $j$ resets to $1$ after each complete quantum reset, whereas the shared random pair $\Xi=(F,G)$ is drawn once from its prior $\mu_\Xi$ and reused across all runs. Because every run makes at most $L<k$ queries, the auxiliary process only ever invokes $U^{[1]},\ldots,U^{[L]}$ and is therefore completely independent of $s$.

We number the query slots globally by $i=1,\ldots,q$ across all experiments. Immediately before a fixed slot, let $\mathsf H$ be the classical history, $h$ a realisation, and $t(h)$ the number of queries already made in the current experiment. For active histories, $t(h)\le L-1\le k-2$. 

The next lemma bounds the classical information accumulated in $\mathsf H$ and shows that the quantum state depends only on the tree columns accessed since the last reset.

\begin{lemma}[Information and state dependence after resets]
\label{depth:lem:memory}
Immediately before each global query slot of the auxiliary process,
the marginal of $\Xi$ is $\mu_\Xi$, and
\begin{equation}
 I(\Xi;\mathsf H)\le2q\log d.
 \label{depth:information}
\end{equation}
The reduced state of the query register has a normalised version
\begin{equation}
\sigma(\Xi,h)=\sigma(h,G,Z_{t(h)}),
 \label{depth:state}
\end{equation}
defined for every parameter--history pair. 
In particular, the auxiliary output distribution is independent of $s$.
\end{lemma}

\begin{proof}
Summing over all histories leaves
the marginal distribution of $\Xi$ unchanged at $\mu_\Xi$. 
For the information bound,
let $A\cong\mathbb C^d$ be the query register and $R$ be all other quantum
memory. A query acts only on $A$ and leaves the marginal on $\Xi\mathsf H R$ unchanged.
By the chain rule and $0\le I(\Xi;A\mid\mathsf H R)\le2\log d$, each query increases $I(\Xi;\mathsf H RA)$ by at most $2\log d$, while known operations cannot increase this information. Since the initial state
is independent of $\Xi$, Eq.~\eqref{depth:information} follows after
at most $q$ queries. 

For Eq.~\eqref{depth:state}, fix the history $h_0$ at the start
of the current experiment. The reset makes its initial state
known given $h_0$. Its first $t=t(h)$ queries use
$U^{[1]},\ldots,U^{[t]}$, which depend only on $(G,Z_t)$,
and its known operations are determined by the recorded outcomes.
Thus, the unnormalised state generated within this experiment
depends only on $h,G,Z_t$. The likelihood of reaching $h_0$,
which may depend on all of $\Xi$, contributes only a scalar
factor and cancels upon normalisation. This gives the claimed dependence. 
\end{proof}

\subsubsection{Bounding overlap with the remaining frame columns}

We will compare hybrid protocols that use the auxiliary rule for their first $i$ queries and the physical oracle thereafter. Because two consecutive hybrids share the same auxiliary evolution before the slot where they differ, their common input has reduced state $\sigma$, so the single-step error in Eq.~\eqref{depth:local} is governed by $\tr(P_t\sigma)$ and $\tr(P_tG\sigma G^\dagger)$. 
Although $\sigma$ may have large support on the first $t$ tree levels $Z_t$, the projector $P_t$ is orthogonal to $Z_t$ and projects solely onto the deeper, unvisited columns. As we now show, the expected overlap of $\sigma$ with these remaining Haar-random columns is bounded by the mutual information accumulated in the classical history.

\begin{lemma}[Information bounds overlap with the remaining columns]
\label{depth:lem:overlap}
Suppose $(\Xi,\mathsf H)$ has marginal $\Xi\sim\mu_\Xi$ and satisfies
$I(\Xi;\mathsf H)\le\mathcal I$. Let $t=t(h)\in\{0,\ldots,k-2\}$,
and let $\tau(h,G,Z_t)$ be the density matrix on the query register. Then
\begin{equation}
 \E\tr(P_t\tau)\le\frac{2(\mathcal I+r)}{D_*}.
 \label{depth:entropy-overlap}
\end{equation}
In particular, the auxiliary states $\sigma$ of Lemma~\ref{depth:lem:memory} satisfy
\begin{equation}
 \E\tr(P_t\sigma),\quad \E\tr(P_tG\sigma G^\dagger)
 \le B_q:= \frac{4(q\log d+M)}{D_*}.
 \label{depth:tail}
\end{equation}
\end{lemma}

\begin{proof}
Let $P$ denote the true joint distribution of $(\Xi,\mathsf H)$, and introduce the decoupled reference distribution $Q=\mu_\Xi\otimes P_{\mathsf H}$, under which the history $\mathsf H$ is independent of the oracle parameters $\Xi$. Under $P$, conditioning on a history $h$ generally skews the distribution of the remaining frame columns. Under $Q$, however, conditioning on $(h,G,Z_t)$ fixes the state $\tau(h,G,Z_t)$ while leaving the remaining columns of $F_\pm$ distributed according to the conditional Haar measure of Lemma~\ref{depth:lem:geometry}.

For a rank-$r'$ Haar-random projector $P'$ in dimension $D$ and any fixed unit vector $v$, 
\[
 \E\bra v P'\ket v^j=\frac{(r')_j}{(D)_j},\qquad
 (a)_j=a(a+1)\cdots(a+j-1),\quad(a)_0=1.
\]
For $D\ge D_*$ and $r'\le r$, we have $(r')_j/(D)_j \le (r')_j / D_*^j$, so expanding the exponential series gives
\[
 \E e^{(D_*/2)\bra v P'\ket v}
 \le\sum_{j\ge0}\frac{(r')_j}{2^j j!}=(1-1/2)^{-r'}=2^{r'}\le2^r.
\]
By convexity of the exponential function, the same bound holds when $\ket v\bra v$ is replaced by any density matrix, and extends across the direct sum of the two sectors $\mathcal H_\pm$ and the root space (where $P_t$ vanishes). Conditioning on $(h,G,Z_t)$ under $Q$ and averaging therefore gives
\[
 \E_Q\bigl[e^{(D_*/2)\tr(P_t\tau)}\bigr]\le2^r.
\]
Next, we invoke the Donsker–Varadhan variational formula for relative entropy and obtain
\[
 \lambda\E_P[X]\le D(P\|Q)+\log\E_Q\bigl[e^{\lambda X}\bigr],
\]
setting $X=\tr(P_t\tau)$ and $\lambda=D_*/2$. Since $D(P\|Q)=I(\Xi;\mathsf H)\le\mathcal I$, rearranging immediately yields Eq.~\eqref{depth:entropy-overlap}. Finally, substituting $\mathcal I=2q\log d$ from Eq.~\eqref{depth:information} and $r=2M-2<2M$ into Eq.~\eqref{depth:entropy-overlap} for the two choices $\tau=\sigma(h,G,Z_t)$ and $\tau=G\sigma(h,G,Z_t)G^\dagger$ establishes Eq.~\eqref{depth:tail}.
\end{proof}

\subsubsection{Accumulating the query errors}
We now sum the single-query replacement errors across all $q$ global query slots to bound the total variation distance between the physical and auxiliary output distributions. 

\begin{lemma}[Comparison with the auxiliary process]
\label{depth:lem:hybrid}
Let $\mathsf{Out}_{\rm aux}$ denote the classical output distribution of the auxiliary process. For either sign $s\in\{\pm1\}$,
\begin{equation}
 \TV(\mathsf{Out}_{\theta,s},\mathsf{Out}_{\rm aux})
 \le4q\sqrt{B_q}+320q\sqrt{M/d}.
 \label{depth:one-sign}
\end{equation}
Consequently,
\begin{equation}
 \TV(\mathsf{Out}_{\theta,+},\mathsf{Out}_{\theta,-})
 \le8q\sqrt{B_q}+640q\sqrt{M/d}.
 \label{depth:hybrid-bound}
\end{equation}
\end{lemma}
\begin{proof}
For $i=0,\ldots,q$, define the $i$th hybrid process to follow the auxiliary rule during the first $i$ global query slots and to query the physical oracle $U_{\theta,s}(\Xi)$ in all subsequent slots. Thus, the $0$th hybrid is the physical protocol and the $q$th hybrid is the auxiliary process.

Consider the $(i-1)$th and $i$th hybrids. They evolve identically prior to slot $i$, where their common state is that of the auxiliary process, and they apply the exact same conditional operations after slot $i$. Retaining $\Xi$ and the classical history as classical registers, the post-slot-$i$ evolution is a parameter-independent quantum channel, which contracts trace distance.

To bound the trace distance introduced at slot $i$, fix an active history $h$ (with $t=t(h)\le L-1\le k-2$) and purify the joint system--ancilla input state $\ket{\phi(\Xi,h)}$ so that its marginal on the query register is $\sigma(h,G,Z_t)$. By Fuchs--van de Graaf, the trace distance between the post-query states of the two hybrids is bounded by the Euclidean norm $\norm{(U_{\theta,s}-U^{[t+1]})\phi}$. By Lemma~\ref{depth:lem:coupling}, we have
\[
 \norm{(U_{\theta,s}-U^{[t+1]})\phi}
 \le 2\sqrt{\tr(P_t\sigma)} + 2\sqrt{\tr(P_tG\sigma G^\dagger)} + 32\delta_{\theta,s}.
\]
Taking the expectation over $(\Xi,\mathsf H)$ and applying Jensen's inequality along with Eq.~\eqref{depth:tail}, the two projector terms together contribute at most $4\sqrt{B_q}$. For the overlap term, because the marginal of $\Xi$ is $\mu_\Xi$, Eq.~\eqref{depth:overlap} gives
\[
 32\,\E_{\mu_\Xi}[\delta_{\theta,s}]
 \le 320\sqrt{M/d}.
\]
Thus, replacing the $i$th query changes the final output distribution by at most $4\sqrt{B_q}+320\sqrt{M/d}$ in total variation distance. Summing over $i=1,\ldots,q$ via the triangle inequality establishes Eq.~\eqref{depth:one-sign}. Because $\mathsf{Out}_{\rm aux}$ is independent of $s$, the triangle inequality yields Eq.~\eqref{depth:hybrid-bound}.
\end{proof}

\begin{proof}[Proof of Theorem~\ref{depth:thm}]
The exact correlator values in Eq.~\eqref{depth:signals} are established in Lemma~\ref{depth:lem:geometry}. To bound the total variation distance between the output distributions for $s=+1$ and $s=-1$, we substitute the expression of $B_q$ from Eq.~\eqref{depth:tail} into Eq.~\eqref{depth:hybrid-bound}:
\begin{align}
\TV(\mathsf{Out}_{\theta,+},\mathsf{Out}_{\theta,-})
 &\le 8q\sqrt{B_q}+640q\sqrt{\frac{M}{d}}\notag\\
 &\le 16q\sqrt{\frac{q\log d+M}{D_*}}
       +640q\sqrt{\frac{M}{d}}\notag\\
 &\le 32\sqrt{\frac{q^3\log d}{d}}+672q\sqrt{\frac{M}{d}},
 \label{depth:hybrid}
\end{align}
where the final inequality uses $D_*\ge d/4$ from Eq.~\eqref{depth:ambient} and $\sqrt{a+b}\le\sqrt a+\sqrt b$. This proves Eq.~\eqref{depth:tv}.

To translate Eq.~\eqref{depth:tv} into a query lower bound, note from Eq.~\eqref{depth:signals} that $|C(U)|=3/4$ and $|F_\theta(U)|=3/8$. Thus, an estimate $\widehat C$ of $C(U)$ with additive error at most $1/3 < 3/4$ uniquely determines $s$ via $\operatorname{sign}\operatorname{Re}(i^{-|\theta|}\widehat C)$. Likewise, an estimate $\widehat F_\theta$ of $F_\theta(U)$ within additive error $1/3 < 3/8$ determines $s$ via $\operatorname{sign}\operatorname{Re}\widehat F_\theta$. Any protocol that succeeds with probability at least $2/3$ on every oracle in $\operatorname{supp}\mathcal D_{\theta,+}\cup\operatorname{supp}\mathcal D_{\theta,-}$ therefore distinguishes $\mathcal D_{\theta,+}$ from $\mathcal D_{\theta,-}$ with success probability at least $2/3$, which requires $\TV(\mathsf{Out}_{\theta,+},\mathsf{Out}_{\theta,-}) \ge \frac13$.
On the other hand, if $q$ were strictly smaller than the right-hand side of Eq.~\eqref{depth:queries}, then Eq.~\eqref{depth:tv} would give
$\TV(\mathsf{Out}_{\theta,+},\mathsf{Out}_{\theta,-})
 < 32(4096)^{-3/2}+\frac{672}{4096}
 = \frac{672.125}{4096} < \frac13$,
a contradiction. This establishes Eq.~\eqref{depth:queries}.
\end{proof}

\subsection{Constructing the coupled Haar completions}
\label{depth:coupling}

This subsection proves Lemma~\ref{depth:lem:coupling}. Our goal is to couple the full oracle $U_{\theta,s}$ and the prefix oracles $U^{[j]}$ on a common probability space so that $U_{\theta,s}$ and $U^{[t+1]}$ act almost identically on any state supported outside the deeper tree levels $\operatorname{Ran}P_t$. Because $U_{\theta,s}$ and $U^{[t+1]}$ prescribe different numbers of tree transitions, their random completions act on subspaces of different dimensions and cannot be coupled by simply sharing a block matrix. Instead, we generate both unitaries from a single Haar unitary $G\sim\operatorname{Haar}(\mathrm U(d))$: for any prescribed input and output frames $(V,W)$, we right-multiply $G$ by a corrective unitary $R$ that enforces $GRV=W$. In high dimensions ($d\gg M$), the subspaces $\operatorname{Ran}V$ and $\operatorname{Ran}(G^\dagger W)$ are nearly orthogonal, which makes $R-I$ approximately a sum of rank-one terms over the columns of $V$ and $W$. 
Consequently, when we take the difference $U_{\theta,s}-U^{[t+1]}$, the contributions from the shared first $t+1$ tree levels cancel out, leaving an error supported only on the deeper levels $\operatorname{Ran}P_t$ as claimed in Eq.~\eqref{depth:local}.

\begin{proof}[Proof of Lemma~\ref{depth:lem:coupling}]
\emph{Step 1: Exact unitary completion from a shared Haar unitary.}
Let $V,W:\mathbb C^m\to\mathbb C^d$ be isometries ($V^\dagger V=W^\dagger W=I_m$) with $2m\le d$. This condition holds both for the full tree frames ($2p=4M-4\le d$) and for all prefix subframes. Given a unitary $G\sim\operatorname{Haar}(\mathrm U(d))$, we seek a unitary of the form $\mathcal U=GR$ that satisfies $\mathcal UV=W$. Pulling $G$ to the right-hand side, this condition is equivalent to requiring that $R$ map the frame $V$ to the pulled-back frame $B:=G^\dagger W$:
\[
 RV=B.
\]
To construct such a unitary $R$ explicitly, define the difference frame $X:\mathbb C^m\to\mathbb C^d$ and the $m\times m$ matrix $A$ by
\[
 X=V-B,\qquad A=I_m-B^\dagger V=X^\dagger V.
\]
Because $G$ is Haar distributed and $2m\le d$, the subspaces $\operatorname{Ran}V$ and $\operatorname{Ran}B$ have no nonzero intersection almost surely, thus $\norm{V^\dagger B}<1$ and $A$ is invertible almost surely. On this measure-one event, define
\[
 R(V,B)=I_d-XA^{-1}X^\dagger.
\]
Let us verify that $R=R(V,B)$ maps $V$ to $B$ and is unitary. First, using $X^\dagger V=A$, we have
\[
 RV=V-XA^{-1}(X^\dagger V)=V-X=B.
\]
Second, since $V^\dagger V=B^\dagger B=I_m$, expanding $X^\dagger X=(V-B)^\dagger(V-B)$ gives
\[
 X^\dagger X=2I_m-V^\dagger B-B^\dagger V=A+A^\dagger.
\]
Using this identity, we compute
\[
 R^\dagger R
 =(I_d-X(A^\dagger)^{-1}X^\dagger)(I_d-XA^{-1}X^\dagger)
 =I_d+X(A^\dagger)^{-1}(X^\dagger X-A-A^\dagger)A^{-1}X^\dagger
 =I_d,
\]
confirming that $R$ is unitary. Setting $\mathcal U(V,W;G):=GR(V,G^\dagger W)$ therefore yields a unitary that implements the prescribed transitions $V\mapsto W$.

We now verify that, conditioned on $(V,W)$, the action of $\mathcal U(V,W;G)$ from $(\operatorname{Ran}V)^\perp$ onto $(\operatorname{Ran}W)^\perp$ is Haar distributed. Let $K\in\mathrm U(d)$ be any unitary that acts as the identity on $\operatorname{Ran}W$ (so $KW=W$) and as an arbitrary unitary on $(\operatorname{Ran}W)^\perp$. Under the left translation $G\mapsto KG$, the distribution of $G\sim\operatorname{Haar}(\mathrm U(d))$ is invariant. Moreover, because $G^\dagger K^\dagger W=G^\dagger W=B$, the corrective unitary $R(V,G^\dagger W)$ is unchanged, so $\mathcal U$ transforms as
\[
 \mathcal U(V,W;KG)=KG\,R(V,G^\dagger K^\dagger W)=K\,\mathcal U(V,W;G).
\]
Thus, the conditional distribution of $\mathcal U(V,W;G)$ is left-invariant under all unitaries $K$ acting on $(\operatorname{Ran}W)^\perp$, which means its restriction from $(\operatorname{Ran}V)^\perp$ to $(\operatorname{Ran}W)^\perp$ is Haar distributed. Applying this construction with the same random matrix $G$ to the full frames and to each prefix defines the coupled oracles:
\[
 U_{\theta,s}=\mathcal U(V_F,W_{F,\theta,s};G),\qquad
 U^{[j]}=\mathcal U(V^{[j]},W^{[j]};G),\quad 1\le j<k.
\]
Since the prefix frames $(V^{[j]},W^{[j]})$ depend only on the first $j$ tree levels $Z_j$, the prefix oracle $U^{[j]}$ is completely determined by $(Z_j,G)$.

\emph{Step 2: Linearizing the corrective rotation.}
In the formula $R(V,B)=I_d-XA^{-1}X^\dagger$, the $m\times m$ inverse matrix $A^{-1}=(I_m-B^\dagger V)^{-1}$ couples different columns of $X$. However, when $V$ and $B$ are nearly orthogonal, $A$ is close to $I_m$. Replacing $A^{-1}$ by $I_m$ yields the simplified operator
\[
 R_0(V,B):=I_d-XX^\dagger
 =I_d-VV^\dagger-BB^\dagger+VB^\dagger+BV^\dagger.
\]
We bound the error of this approximation in terms of the frame overlap $\delta:=\norm{V^\dagger B}=\norm{I_m-A}$:
\begin{equation}
 \norm{R(V,B)-R_0(V,B)}\le16\norm{V^\dagger B}.
 \label{depth:rotation}
\end{equation}
To prove Eq.~\eqref{depth:rotation}, observe first that $\norm X\le\norm V+\norm B=2$. If $\delta\le1/2$, then $A^{-1}=(I_m-B^\dagger V)^{-1} = \sum_{t\ge0} (B^{\dagger}V)^t$ gives $\norm{A^{-1}-I_m}\le\delta/(1-\delta)\le 2\delta$, and therefore
\[
 \norm{R(V,B)-R_0(V,B)}
 =\norm{X(A^{-1}-I_m)X^\dagger}
 \le\norm X^2\norm{A^{-1}-I_m}\le 8\delta.
\]
If instead $\delta>1/2$, the triangle inequality gives the trivial bound
\[
 \norm{R(V,B)-R_0(V,B)}
 \le\norm{R(V,B)}+\norm{I_d}+\norm X^2
 \le 1+1+4=6<16\delta.
\]
Thus Eq.~\eqref{depth:rotation} holds for all $\delta\ge0$.

\emph{Step 3: Canceling the shared prefix columns.}
We now compare $U_{\theta,s}$ and $U^{[t+1]}$ for a fixed $0\le t\le k-2$. Set $j=t+1$ and partition the full frames into the depth-$j$ prefix columns (corresponding to input words of length $|w|\le t$) and the remaining deeper columns (corresponding to input words of length $t+1\le|w|\le k-1$):
\[
 V_F=\bigl(V^{[j]}\ \ V_{\rm rem}\bigr),\qquad
 W_{F,\theta,s}=\bigl(W^{[j]}\ \ W_{\rm rem}\bigr).
\]
Correspondingly, the difference frame $X_F:=V_F-G^\dagger W_{F,\theta,s}$ splits into two blocks of columns:
\[
 X_F=\bigl(X^{[j]}\ \ X_{\rm rem}\bigr),\qquad
 X^{[j]}:=V^{[j]}-G^\dagger W^{[j]},\qquad
 X_{\rm rem}:=V_{\rm rem}-G^\dagger W_{\rm rem}.
\]
Because $X_FX_F^\dagger = X^{[j]}(X^{[j]})^\dagger + X_{\rm rem}X_{\rm rem}^\dagger$, the shared prefix columns cancel out when we take the difference between the two linearized rotations:
\begin{equation}
 R_0(V_F,G^\dagger W_{F,\theta,s})
 -R_0(V^{[j]},G^\dagger W^{[j]})
 =-X_{\rm rem}X_{\rm rem}^\dagger.
 \label{depth:cancel-eq}
\end{equation}
Let us bound the action of $X_{\rm rem}X_{\rm rem}^\dagger$ on an arbitrary vector $\phi$. Since $V_{\rm rem}$ and $W_{\rm rem}$ are isometries, we have $\norm{X_{\rm rem}}\le\norm{V_{\rm rem}}+\norm{G^\dagger W_{\rm rem}}=2$, which gives
\[
 \norm{X_{\rm rem}X_{\rm rem}^\dagger\phi}
 \le\norm{X_{\rm rem}}\norm{X_{\rm rem}^\dagger\phi}
 \le 2\bigl(\norm{V_{\rm rem}^\dagger\phi}
             +\norm{W_{\rm rem}^\dagger G\phi}\bigr).
\]
By Lemma~\ref{depth:lem:geometry}, the columns of both $V_{\rm rem}$ and $W_{\rm rem}$ are orthonormal vectors lying entirely inside $\operatorname{Ran}P_t$. Hence $V_{\rm rem}V_{\rm rem}^\dagger\preceq P_t$ and $W_{\rm rem}W_{\rm rem}^\dagger\preceq P_t$, which implies $\norm{V_{\rm rem}^\dagger\phi}=\norm{V_{\rm rem}V_{\rm rem}^\dagger\phi}\le\norm{P_t\phi}$ and similarly $\norm{W_{\rm rem}^\dagger G\phi}\le\norm{P_tG\phi}$. Combining these bounds with Eq.~\eqref{depth:cancel-eq} yields
\[
 \norm{\bigl(R_0(V_F,G^\dagger W_{F,\theta,s})
 -R_0(V^{[j]},G^\dagger W^{[j]})\bigr)\phi}
 \le 2\bigl(\norm{P_t\phi}+\norm{P_tG\phi}\bigr).
\]
To pass from the linearized rotations $R_0$ back to the exact unitaries $U_{\theta,s}=GR(V_F,G^\dagger W_{F,\theta,s})$ and $U^{[j]}=GR(V^{[j]},G^\dagger W^{[j]})$, note that $(V^{[j]})^\dagger G^\dagger W^{[j]}$ is a submatrix of $V_F^\dagger G^\dagger W_{F,\theta,s}$, so its operator norm is also bounded by $\delta_{\theta,s}=\norm{V_F^\dagger G^\dagger W_{F,\theta,s}}$. Applying Eq.~\eqref{depth:rotation} to both $R(V_F,G^\dagger W_{F,\theta,s})$ and $R(V^{[j]},G^\dagger W^{[j]})$ adds at most $2\times 16\delta_{\theta,s}\norm\phi=32\delta_{\theta,s}\norm\phi$ in error. Multiplying on the left by the unitary $G$ preserves norms and establishes Eq.~\eqref{depth:local}. 
When $\phi$ is a joint system--ancilla vector, tensoring all operators with the ancilla identity leaves the derivation unchanged.

\emph{Step 4: Bounding the expected frame overlap.}
It remains to prove Eq.~\eqref{depth:overlap} by bounding $\E_G[\delta_{\theta,s}\mid F]$. Fix the frame $F$, so that $V_F,W_{F,\theta,s}:\mathbb C^p\to\mathbb C^d$ are fixed isometries. Choose a $(1/4)$-net $\mathcal N$ of the unit sphere in $\mathbb C^p$ of cardinality $|\mathcal N|\le 9^{2p}$. For any unit vectors $a,b\in\mathbb C^p$, choosing $a_0,b_0\in\mathcal N$ with $\norm{a-a_0}\le 1/4$ and $\norm{b-b_0}\le 1/4$ gives
\[
 |a^\dagger V_F^\dagger G^\dagger W_{F,\theta,s}b|
 \le |a_0^\dagger V_F^\dagger G^\dagger W_{F,\theta,s}b_0|
     +\bigl(\norm{a-a_0}+\norm{b-b_0}\bigr)\delta_{\theta,s}
 \le \max_{a_0,b_0\in\mathcal N}|a_0^\dagger V_F^\dagger G^\dagger W_{F,\theta,s}b_0|
     +\frac12\delta_{\theta,s}.
\]
Taking the supremum over unit vectors $a,b$ and rearranging yields
\[
 \delta_{\theta,s}
 \le 2\max_{a,b\in\mathcal N}
       |a^\dagger V_F^\dagger G^\dagger W_{F,\theta,s}b|.
\]
For each fixed pair $a,b\in\mathcal N$, both $V_Fa$ and $W_{F,\theta,s}b$ are fixed unit vectors in $\mathbb C^d$. Under $G\sim\operatorname{Haar}(\mathrm U(d))$, the random vector $G^\dagger(W_{F,\theta,s}b)$ is uniformly distributed on the unit sphere in $\mathbb C^d$, so the squared overlap $|(V_Fa)^\dagger G^\dagger(W_{F,\theta,s}b)|^2$ follows a $\operatorname{Beta}(1,d-1)$ distribution, whose tail probability is
\[
 \Pr_G\bigl(|a^\dagger V_F^\dagger G^\dagger W_{F,\theta,s}b|\ge u\bigm| F\bigr)
 =(1-u^2)^{d-1}\le e^{-(d-1)u^2},\qquad 0\le u\le 1.
\]
Taking a union bound over all $|\mathcal N|^2\le 9^{4p}$ pairs gives
\[
 \Pr_G(\delta_{\theta,s}\ge2u\mid F)
 \le\min\bigl\{1,\,e^{4p\log9-(d-1)u^2}\bigr\},\qquad 0\le u\le1.
\]
Finally, we integrate this tail bound to control the second moment of $\delta_{\theta,s}$:
\[
 \begin{aligned}
 \E_G[\delta_{\theta,s}^2\mid F]
 &=4 \int_0^4 \Pr_G(\delta_{\theta,s}\ge 2\sqrt v\mid F)\,dv\\
 &\le 4\int_0^\infty \min\{1,e^{4p\log9-(d-1)v}\}\,dv \\
 &\le 4u_0^2 + 4\int_{u_0^2}^{\infty} e^{4p\log 9-(d-1)s}\,ds
 \le\frac{4(4p\log9+1)}{d-1}\le\frac{100M}{d},
 \end{aligned}
\] 
where $u_0^2:=4p\log 9/(d-1)$ and the final inequality uses $p=2M-2$ and $d\ge 4M\ge 16$. By Cauchy--Schwarz, $\E_G[\delta_{\theta,s}\mid F]\le\bigl(\E_G[\delta_{\theta,s}^2\mid F]\bigr)^{1/2}\le 10\sqrt{M/d}$, completing the proof.
\end{proof}

\section{Single-ancilla simultaneous estimation of nested correlators}
\label{nest:sec:upper}

In this appendix, we state the formal performance guarantee for Algorithm~\ref{alg:nested} and provide the complete proof of the upper bound in Theorem~\ref{nest:thm:advantage_main}.

\begin{theorem}[Performance guarantee of Algorithm~\ref{alg:nested}]
\label{nest:thm:universal-upper}
Let $k\ge1$ and $M=2^k$.
For every unitary $U$, every sequence of known dichotomic observables $O_1,\ldots,O_{k+1}$, every sequence of integer observation times $0\le t_1\le\cdots\le t_{k+1}=T$, and every repeatedly preparable state $\rho$, Algorithm~\ref{alg:nested} produces estimates $\{\widehat F_\sigma\}_{\sigma\in\{0,1\}^k}$ of the nested correlators in Eq.~\eqref{nest:eq:targets} satisfying
\[
\Pr\!\left(
\max_{\sigma\in\{0,1\}^k}|\widehat F_\sigma-F_\sigma(U)|\le\varepsilon
\right)\ge1-\delta
\]
for any $\varepsilon>0$ and $0<\delta<1$, using one ancillary qubit and a single shared dataset of
\begin{equation}\label{nest:eq:upper-query-count}
N=\left\lceil\varepsilon^{-2}\log\frac{2M}{\delta}\right\rceil
=\left\lceil\varepsilon^{-2}\bigl(k+1+\log(1/\delta)\bigr)\right\rceil
\end{equation}
experimental repetitions, corresponding to $q=TN$ forward queries to $U$ in total.
\end{theorem}
    
\begin{proof}
In each repetition of Algorithm~\ref{alg:nested}, conditioned on a fixed path string $b\in\{0,1\}^k$, the $\ket0$ and $\ket1$ branches of the probe qubit undergo the effective evolutions $W_b$ and $W_{\bar b}$ defined in Eq.~\eqref{nest:eq:main-paths} (and Eq.~\eqref{nest:eq:general-forward-path}). Prior to the final measurements at time $T=t_{k+1}$, the joint probe--system state is
\begin{equation}\label{nest:eq:branch-state}
\Omega_b
=\frac12
\begin{pmatrix}
W_b\rho W_b^\dagger&W_b\rho W_{\bar b}^\dagger\\
W_{\bar b}\rho W_b^\dagger&W_{\bar b}\rho W_{\bar b}^\dagger
\end{pmatrix}.
\end{equation}
Measuring $(X+uY)/\sqrt2$ on the probe (with outcome $a\in\{\pm1\}$) and $O_{k+1}$ on the system (with outcome $x\in\{\pm1\}$) yields a product $ax\in\{\pm1\}$ whose conditional expectation value is
\begin{equation}\label{nest:eq:rotated-readout}
\E[ax\mid b,u]
=\tr\!\left[
\left(\frac{X+uY}{\sqrt2}\otimes O_{k+1}\right)\Omega_b
\right]
=\frac{\operatorname{Re} Q_b-u\operatorname{Im} Q_b}{\sqrt2},
\end{equation}
where $Q_b=\tr(O_{k+1}W_b\rho W_{\bar b}^\dagger)$. Averaging over the independent fair sign $u\in\{\pm1\}$ gives
\[
\E[ax\mid b]=\frac{\operatorname{Re} Q_b}{\sqrt2},
\qquad
\E[uax\mid b]=-\frac{\operatorname{Im} Q_b}{\sqrt2}.
\]
Consequently, the single-shot snapshot $\widehat Q=\sqrt2\,ax(1-iu)$ is an unbiased estimator of $Q_b$ conditioned on $b$, namely $\E[\widehat Q\mid b]=Q_b$.

For each target pattern $\sigma\in\{0,1\}^k$, let
\begin{equation}\label{nest:eq:shared-estimator}
T_\sigma
=\frac{(-1)^{\sigma\cdot b}}2
\operatorname{Re}\!\left(i^{|\sigma|}\widehat Q\right)
\end{equation}
denote the single-shot contribution to the estimator $\widehat F_\sigma = \frac{1}{N}\sum_{r=1}^N T_\sigma^{(r)}$ in Algorithm~\ref{alg:nested}. Taking the conditional expectation over $u,a,x$ for a fixed $b$ yields
\[
\E[T_\sigma\mid b]
=\frac{(-1)^{\sigma\cdot b}}2\operatorname{Re}\!\left(i^{|\sigma|}Q_b\right).
\]
Averaging over the uniformly drawn path $b\sim\{0,1\}^k$ and invoking the complementary-path duality of Lemma~\ref{nest:lem:expansion} then gives
\[
\E T_\sigma
=\frac12\operatorname{Re}\!\left(i^{|\sigma|}\E_b\bigl[(-1)^{\sigma\cdot b}Q_b\bigr]\right)
=F_\sigma(U),
\]
where we used the fact that $F_\sigma(U)$ is real.

Finally, to bound the sample complexity, observe that for every realisation of $(b,u,a,x)$ and every $\sigma\in\{0,1\}^k$, $\operatorname{Re}(i^{|\sigma|}\widehat Q) = \sqrt2\,ax$ when $|\sigma|$ is even and $\pm\sqrt2\,uax$ when $|\sigma|$ is odd. Thus each single-shot variable $T_\sigma$ takes values in $\{-1/\sqrt2,+1/\sqrt2\}$, a bounded interval of width $\sqrt2$. By Hoeffding's inequality, for each fixed $\sigma$,
\[
\Pr\!\left(
|\widehat F_\sigma-F_\sigma(U)|>\varepsilon
\right)
\le2\exp(-N\varepsilon^2).
\]
Applying the union bound over all $M=2^k$ bracket patterns $\sigma\in\{0,1\}^k$ yields
\begin{equation}\label{nest:eq:simultaneous-concentration}
\Pr\!\left(
\max_{\sigma\in\{0,1\}^k}|\widehat F_\sigma-F_\sigma(U)|>\varepsilon
\right)
\le2M e^{-N\varepsilon^2}.
\end{equation}
Setting the right-hand side to be at most $\delta$ is satisfied by choosing $N=\lceil\varepsilon^{-2}\log(2M/\delta)\rceil$. Since each run of Algorithm~\ref{alg:nested} invokes $T$ forward queries to $U$, the total query complexity is $q=TN$.
\end{proof}

\section{Ancilla-free estimation of a single nested correlator}
\label{nest:sec:single-free}

Here we present an ancilla-free implementation of the sequential-measurement scheme of Refs.~\cite{Dressel2018WeakMeasurementOTOC, DelRe2024RobustnPointCorrelation, Wang2025ComputingnTimeCorrelation}, which efficiently estimates any single nested correlator $F_\sigma(U)$ fixed prior to data collection. Using the known dichotomic observables $O_1,\ldots,O_{k+1}$, integer observation times $0\le t_1\le\cdots\le t_{k+1}=T$, and repeatedly preparable state $\rho$ specified in Eq.~\eqref{nest:eq:targets}, the procedure for a prescribed bracket pattern $\sigma\in\{0,1\}^k$ is detailed in Algorithm~\ref{alg:single-nested}. Each experimental run requires $T$ forward queries to $U$ and no ancillary quantum memory, producing a bounded single-shot random variable
\begin{equation}\label{nest:single:eq:sample}
    Z_\sigma=\frac{x}{2}\prod_{j=1}^k a_j\in\left\{-\frac12,+\frac12\right\}.
\end{equation}

\begin{algorithm}[h]
    \caption{Estimating a single nested correlator without ancillary qubits}
    \label{alg:single-nested}
    \begin{algorithmic}[1]
        \Require Initial state $\rho$, known observables $O_1,\ldots,O_{k+1}$, observation times $0\le t_1\le\cdots\le t_{k+1}$, target bracket pattern $\sigma\in\{0,1\}^k$, and number of repetitions $N$.

        \For{$r=1,\ldots,N$}
            \State Prepare the system in $\rho$ and evolve for $t_1$ steps under $U$.
            \Comment{Initialise}

            \For{$j=1,\ldots,k$}
                \Comment{Apply pattern-specific interventions}
                \If{$\sigma_j=0$}
                    \State Projectively measure $O_j$ on the system, obtaining outcome $a_j\in\{\pm1\}$.
                \ElsIf{$\sigma_j=1$}
                    \State Draw a fair sign $a_j\in\{\pm1\}$ and apply the unitary rotation $e^{-ia_j\pi O_j/4}$.
                \EndIf
                \State Evolve the system for $t_{j+1}-t_j$ steps under $U$.
            \EndFor

            \State Measure $O_{k+1}$ on the system, obtaining outcome $x\in\{\pm1\}$.
            \State Store $Z_\sigma^{(r)}=\frac{x}{2}\prod_{j=1}^k a_j$.
        \EndFor
        \State \Return $\displaystyle\widehat F_\sigma=\frac1N\sum_{r=1}^N Z_\sigma^{(r)}$.
    \end{algorithmic}
\end{algorithm}

\begin{theorem}[Ancilla-free estimation of a single nested correlator]
\label{nest:single:thm:upper}
For every fixed $\sigma\in\{0,1\}^k$, every $\varepsilon>0$, and every $0<\delta<1$, Algorithm~\ref{alg:single-nested} achieves
\[
    \Pr\!\left(
    |\widehat F_\sigma-F_\sigma(U)|\le\varepsilon
    \right)\ge1-\delta
\]
for every unitary $U$, every sequence of known dichotomic observables and integer observation times specified above, and every repeatedly preparable input state $\rho$, using no ancillary quantum memory and
\begin{equation}\label{nest:single:eq:query-bound}
    q=T\left\lceil
    \frac{\log(2/\delta)}{2\varepsilon^2}
    \right\rceil
\end{equation}
forward queries to $U$.
\end{theorem}

\begin{proof}
For a dichotomic observable $O$, the intermediate intervention conditioned on $\sigma_j=b\in\{0,1\}$ and outcome/sign $a\in\{\pm1\}$ is described by the Kraus operator
\begin{equation}\label{nest:single:eq:kraus}
    K_a^{(0)}(O)=\frac{I+aO}{2},
    \qquad
    K_a^{(1)}(O)=\frac{I-iaO}{2}
    =\frac1{\sqrt2}e^{-ia\pi O/4}.
\end{equation}
For any observable $B$, a direct expansion shows that weighting the outcomes by $a$ generates the corresponding bracket in the Heisenberg picture:
\begin{equation}\label{nest:single:eq:duality}
    \mathcal B_O^{(b)}(B)
    :=\sum_{a=\pm1}aK_a^{(b)}(O)^\dagger B K_a^{(b)}(O)
    =\frac{[B,O]_b}{2i^b}.
\end{equation}
Writing $A_j=(U^{t_j})^\dagger O_jU^{t_j}$, the intervention at time $t_j$ applies the Kraus operator $K_{a_j}^{(\sigma_j)}(A_j)$ in the Heisenberg picture. Propagating the signed expectation backward from the final measurement of $A_{k+1}$ yields
\begin{equation}\label{nest:single:eq:weighted-average}
    \E Z_\sigma
    =\frac12\tr\!\left[
    \bigl(
    \mathcal B_{A_1}^{(\sigma_1)}
    \circ\cdots\circ
    \mathcal B_{A_k}^{(\sigma_k)}
    \bigr)(A_{k+1})\,\rho
    \right].
\end{equation}
Each superoperator $\mathcal B_{A_j}^{(\sigma_j)}$ contributes the bracket $[\,\cdot\,,A_j]_{\sigma_j}$ together with a factor $1/(2i^{\sigma_j})$. Combined with the overall prefactor $1/2$, this reproduces the exact nested bracket and normalisation $2^{-(k+1)}(-i)^{|\sigma|}$ in Eq.~\eqref{nest:eq:targets}, confirming that
\begin{equation}\label{nest:single:eq:mean}
    \E Z_\sigma=F_\sigma(U).
\end{equation}

Since each sample $Z_\sigma^{(r)}$ takes values in $\{-1/2,+1/2\}$, Hoeffding's inequality gives
\begin{equation}\label{nest:single:eq:hoeffding}
    \Pr\!\left(
    |\widehat F_\sigma-F_\sigma(U)|>\varepsilon
    \right)
    \le2\exp(-2N\varepsilon^2).
\end{equation}
Setting $N=\lceil\log(2/\delta)/(2\varepsilon^2)\rceil$ guarantees a failure probability of at most $\delta$, and multiplying by $T$ forward queries per run yields Eq.~\eqref{nest:single:eq:query-bound}.
\end{proof}

\section{Single-ancilla estimation of multi-time correlators}
\label{corr:sec:upper}

For completeness, we reformulate the single-ancilla interferometric protocol of Pedernales et al.~\cite{Pedernales2014EfficientCorrelationFunction} in our notation and establish its forward-query complexity.
Let $U$ be an unknown unitary, $\rho$ a repeatedly preparable initial state, and $O_1,\ldots,O_{k+1}$ known dichotomic observables satisfying $O_j=O_j^\dagger$ and $O_j^2=I$, with $k\ge1$. For integer observation times $0\le t_1\le\cdots\le t_{k+1}=T$, write $O_j(t_j)=(U^{t_j})^\dagger O_jU^{t_j}$. The target is
\begin{equation}
\label{corr:upper:target}
    C(U)
    =\tr\!\left[O_{k+1}(t_{k+1})\cdots O_1(t_1)\rho\right]
    =Q_{\mathbf 1},
\end{equation}
where $\mathbf 1=(1,\ldots,1)$. In the trajectory notation of Eq.~\eqref{nest:eq:main-paths},
\begin{equation}
\label{corr:upper:paths}
\begin{aligned}
    W_{\mathbf 0}&=U^T,\\
    W_{\mathbf 1}&=U^{T-t_k}O_k\cdots U^{t_2-t_1}O_1U^{t_1},
\end{aligned}
\end{equation}
so that $C(U)=\tr(O_{k+1}W_{\mathbf 1}\rho W_{\mathbf 0}^\dagger)$. Each run interferes these two trajectories and measures the final observable $O_{k+1}$ jointly with a probe spin component. Algorithm~\ref{alg:multitime} implements this procedure using $T$ forward queries to $U$ per repetition.

\begin{algorithm}[h]
\caption{Estimating a general multi-time correlator with one extra qubit}
    \label{alg:multitime}
    \begin{algorithmic}[1]
        \Require Initial state $\rho$, known observables $O_1,\ldots,O_{k+1}$, observation times $0\le t_1\le\cdots\le t_{k+1}$, and number of repetitions $N$.
        \For{$r=1,\ldots,N$}
            \State Prepare the system in $\rho$ and the probe in $\ket+=(\ket0+\ket1)/\sqrt2$.
            \State Draw an independent fair sign $u_r\in\{\pm1\}$.
            \State Evolve the system for $t_1$ steps under $U$.
            \For{$j=1,\ldots,k$}
                \Comment{Fix $b=\mathbf 1$}
                \State Apply $\ket0\bra0\otimes O_j+\ket1\bra1\otimes I$.
                \State Evolve the system for $t_{j+1}-t_j$ steps under $U$.
            \EndFor
            \State Measure $(X+u_rY)/\sqrt2$ on the probe, obtaining $a_r\in\{\pm1\}$.
            \State Measure $O_{k+1}$ on the system, obtaining $x_r\in\{\pm1\}$.
            \State Store $\widehat Q_r=\sqrt2\,a_rx_r(1-iu_r)$.
        \EndFor
        \State \Return $\displaystyle\widehat C=\frac1N\sum_{r=1}^N\widehat Q_r$.
    \end{algorithmic}
\end{algorithm}

\begin{theorem}[Single-ancilla estimation of a general multi-time correlator]
\label{corr:upper:thm}
For every unitary $U$, repeatedly preparable state $\rho$, sequence of known dichotomic observables $O_1,\ldots,O_{k+1}$, and integer observation times $0\le t_1\le\cdots\le t_{k+1}=T$, Algorithm~\ref{alg:multitime} produces an unbiased estimate $\widehat C$ of the complex correlator in Eq.~\eqref{corr:upper:target}. For any $\varepsilon>0$ and $0<\delta<1$, choosing
\begin{equation}
\label{corr:upper:repetitions}
    N=\left\lceil8\varepsilon^{-2}\log\frac4\delta\right\rceil
\end{equation}
ensures
\[
    \Pr\!\left(|\widehat C-C(U)|\le\varepsilon\right)\ge1-\delta.
\]
The protocol uses one ancillary qubit and a total of
\begin{equation}
\label{corr:upper:query-cost}
    q=TN
    =T\left\lceil8\varepsilon^{-2}\log\frac4\delta\right\rceil
\end{equation}
forward queries to $U$.
\end{theorem}

\begin{proof}
First, telescoping the Heisenberg evolutions gives
\[
    W_{\mathbf 0}^\dagger O_{k+1}W_{\mathbf 1}
    =O_{k+1}(t_{k+1})\cdots O_1(t_1),
\]
which verifies the identity $C(U)=Q_{\mathbf 1}$. 
The controlled insertions in Algorithm~\ref{alg:multitime} apply $W_{\mathbf 1}$ to the $\ket0$ branch and $W_{\mathbf 0}$ to the $\ket1$ branch. Thus, for an arbitrary input density operator $\rho$, the joint probe--system state before the final measurements is
\begin{equation}
\label{corr:upper:joint-state}
    \Omega_{\mathbf 1}
    =\frac12
    \begin{pmatrix}
        W_{\mathbf 1}\rho W_{\mathbf 1}^\dagger
        &W_{\mathbf 1}\rho W_{\mathbf 0}^\dagger\\
        W_{\mathbf 0}\rho W_{\mathbf 1}^\dagger
        &W_{\mathbf 0}\rho W_{\mathbf 0}^\dagger
    \end{pmatrix}.
\end{equation}
The product of the two measurement outcomes therefore satisfies
\begin{equation}
\label{corr:upper:readout}
\begin{aligned}
    \E[ax\mid u]
    &=\tr\!\left[
        \left(\frac{X+uY}{\sqrt2}\otimes O_{k+1}\right)
        \Omega_{\mathbf 1}
    \right]\\
    &=\frac{\operatorname{Re}C(U)-u\operatorname{Im}C(U)}{\sqrt2}.
\end{aligned}
\end{equation}
Averaging over the fair sign $u$ yields
\[
    \E[ax]=\frac{\operatorname{Re}C(U)}{\sqrt2},
    \qquad
    \E[uax]=-\frac{\operatorname{Im}C(U)}{\sqrt2}.
\]
Consequently, each snapshot $\widehat Q=\sqrt2\,ax(1-iu)$ satisfies $\E\widehat Q=C(U)$, proving the unbiasedness of $\widehat C$.

Both $\operatorname{Re}\widehat Q$ and $\operatorname{Im}\widehat Q$ take values in $\{-\sqrt2,+\sqrt2\}$. Applying Hoeffding's inequality to their sample means, each at error $\varepsilon/\sqrt2$, and taking a union bound gives
\begin{equation}
\label{corr:upper:concentration}
    \Pr\!\left(|\widehat C-C(U)|>\varepsilon\right)
    \le4\exp(-N\varepsilon^2/8).
\end{equation}
The choice in Eq.~\eqref{corr:upper:repetitions} makes this probability at most $\delta$. Each repetition uses
\[
    t_1+\sum_{j=1}^{k}(t_{j+1}-t_j)=T
\]
forward queries, establishing Eq.~\eqref{corr:upper:query-cost}.
\end{proof}

\section{Details of the numerical experiments}
\label{app:numerical}

This appendix provides further details on the numerical experiments presented in Fig.~\ref{fig:num_exp}. We first specify the model and derive the expansion of the probe signal in terms of nested correlators. Then, we detail the implementation of Algorithms~\ref{alg:nested} and~\ref{alg:single-nested}, as well as the data analysis procedure.

\subsection{Model}
\label{num:sec:model}

We consider a chain of $n=10$ qubits ($d=2^{10}$) with the Hamiltonian and initial state given by
\begin{equation}
\label{num:eq:model}
H=\sum_{i=1}^{9}Z_iZ_{i+1}
  +1.05\sum_{i=1}^{10}X_i
  +0.5\sum_{i=1}^{10}Z_i,
\qquad
\rho=\frac{I+X_1}{d}.
\end{equation}
Thus, qubit $1$ is initially polarized along $x$, while the remaining qubits are maximally mixed. The unitary evolution is $U=e^{-iH\Delta}$ with $\Delta=0.4$. The observation times are 
\begin{equation}
\label{num:eq:times}
(t_1,\ldots,t_9)=(0,1,3,4,5,7,9,10,12).
\end{equation}
The evolution between consecutive probe--system contacts is therefore $U_{t_j,t_{j+1}} \coloneqq e^{-iH\Delta (t_{j+1}-t_j)}$.

The nested correlators $\{F_{\sigma}\}_{\sigma}$ in Eq.~\eqref{nest:eq:targets} are defined using the boundary observable $O_j=Z_1$. These choices yield $k=8$ intervals and $M=2^k=256$ correlators. Every run of either learning algorithm uses
\begin{equation}
\label{num:eq:budget}
t_1+\sum_{j=1}^{8}(t_{j+1}-t_j)=12
\end{equation}
forward queries to this same $U$. Accordingly, $N$ experimental runs correspond to a total query budget of $q=12N$ for either algorithm.

\subsection{Probe signal and correlator expansion}
\label{num:sec:signal}

The target experiment uses a probe initialised in $\ket{+}$.
At the first eight contacts, a boundary rotation and a
probe--system coupling are applied with the same random sign:
\begin{equation}
\label{num:eq:pulses}
V_{s_j,\alpha_j}
=e^{-is_j\alpha_j I_Q\otimes Z_1/2}
 e^{-is_j\beta_j Z_Q\otimes Z_1/2},
\qquad
\beta_j=\frac{\pi}{2}-\alpha_j.
\end{equation}
The final contact is
$V_{s_9,0}=e^{-is_9\pi Z_Q\otimes B/4}$.
All nine signs $\boldsymbol{s}=(s_1,\ldots,s_9) \in \{\pm 1\}^9$ are independent and uniform in $\{\pm1\}$.

Let $\Omega$ denote the joint probe--system state, and define its rescaled off-diagonal block
\begin{equation}
\label{num:eq:coherence}
\chi=2\,{}_Q\!\bra0\Omega\ket1_Q,
\qquad
L=\langle X_Q\rangle-i\langle Y_Q\rangle=\tr\chi.
\end{equation}
Initially, $\chi=\rho$. The prediction target is
\begin{equation}
\label{num:eq:sign-weighting}
S(\boldsymbol{\alpha})
=2^{-9}\sum_{\boldsymbol{s}\in\{\pm1\}^9}
\left(\prod_{j=1}^{9}s_j\right)
L(\boldsymbol{s};\boldsymbol{\alpha}).
\end{equation}
Operationally, its real part is the average of $(\prod_j s_j)x_Q$ over $X_Q$ readouts, and its imaginary part is the negative average of $(\prod_j s_j)y_Q$ over separate $Y_Q$ readouts, where $x_Q,y_Q\in\{\pm1\}$. Expanding $L$ as a polynomial in the binary signs shows that Eq.~\eqref{num:eq:sign-weighting} selects precisely the coefficient of $s_1\cdots s_9$. Contributions independent of any one sign cancel. This sign-weighting operation thus isolates the highest-order component of the signal.

To derive the correlator representation, we work in the interaction picture and write
\begin{equation}
\label{num:eq:Bj}
B_j=(U^{t_j})^\dagger Z_1 U^{t_j},
\qquad B_j^2=I.
\end{equation}
Conditioned on the sign $s$, contact $j$ transforms the off-diagonal block as
\begin{equation}
\label{num:eq:contact}
\Phi_{j,s}(X)
=e^{-is(\alpha_j+\beta_j)B_j/2}
 X e^{is(\alpha_j-\beta_j)B_j/2}.
\end{equation}
Introducing the linear maps
\begin{equation}
\label{num:eq:bracket-maps}
\mathcal A_j(X)=\frac{\{B_j,X\}}{2},
\qquad
\mathcal C_j(X)=\frac{[B_j,X]}{2i},
\end{equation}
and using the identity $e^{-i\gamma B_j}=\cos\gamma\,I-i\sin\gamma\,B_j$, we compute the signed average:
\begin{align}
\mathcal M_j(X)
&\coloneqq\frac12\sum_{s=\pm1}s\,\Phi_{j,s}(X)
\nonumber\\
&=-\frac{i}{2}\left[
(\sin\alpha_j+\sin\beta_j)B_jX
+(\sin\beta_j-\sin\alpha_j)XB_j\right]
\nonumber\\
&=-i\sin\beta_j\,\mathcal A_j(X)
+\sin\alpha_j\,\mathcal C_j(X).
\label{num:eq:signed-contact}
\end{align}
Substituting $\beta_j = \pi/2 - \alpha_j$, we obtain
\begin{equation}
\label{num:eq:weighted-maps}
\mathcal M_j=-i\cos\alpha_j\,\mathcal A_j
+\sin\alpha_j\,\mathcal C_j\quad (j\le8),
\qquad
\mathcal M_9=-i\mathcal A_9.
\end{equation}
The full sign-weighted signal is then generated by the composition of these maps:
\begin{equation}
S(\boldsymbol{\alpha})
=\tr\!\left[
\mathcal M_9\circ\cdots\circ\mathcal M_1(\rho)
\right]
=-i\,\tr\!\left[
B_9\,\mathcal M_8\circ\cdots\circ\mathcal M_1(\rho)
\right],
\label{num:eq:composed-signal}
\end{equation}
where the second equality follows from $\tr[\mathcal A_9(X)]=\tr(B_9X)$.

To connect this to the nested correlators, let $\mathcal D_j^{(0)}=\mathcal A_j$ and $\mathcal D_j^{(1)}=\mathcal C_j$. The cyclicity of the trace yields the duality relation
\begin{equation}
\label{num:eq:trace-duality}
\tr\!\left[Y\mathcal D_j^{(b)}(X)\right]
=\frac{1}{2i^b}\tr\!\left([Y,B_j]_b X\right),
\qquad b\in\{0,1\}.
\end{equation}
Applying Eq.~\eqref{num:eq:trace-duality} successively to Eq.~\eqref{num:eq:composed-signal} yields
\begin{equation}
\tr\!\left[
B_9\,\mathcal D_8^{(\sigma_8)}\circ\cdots\circ
\mathcal D_1^{(\sigma_1)}(\rho)\right]
=
\frac{(-i)^{|\sigma|}}{2^8}
\tr\!\left[
\rho\,\bigl[\cdots[B_9,B_8]_{\sigma_8}
\cdots,B_1\bigr]_{\sigma_1}\right]
=2F_\sigma(U).
\label{num:eq:maps-to-F}
\end{equation}
Expanding the first eight maps $\mathcal{M}_j$ in Eq.~\eqref{num:eq:composed-signal} therefore gives the exact representation:
\begin{equation}
\label{num:eq:signal-expansion}
S(\boldsymbol{\alpha})
=2\sum_{\sigma\in\{0,1\}^{8}}
w_\sigma(\boldsymbol{\alpha})F_\sigma(U),
\qquad
w_\sigma(\boldsymbol{\alpha})
=(-i)^{9-|\sigma|}
\prod_{j=1}^{8}
(\cos\alpha_j)^{1-\sigma_j}
(\sin\alpha_j)^{\sigma_j}.
\end{equation}
Crucially, the weights $w_\sigma(\boldsymbol{\alpha})$ are explicit functions of the control angles and do not depend on the system dynamics $U$. Furthermore, because $F_\sigma$ is real, correlators with an odd Hamming weight $|\sigma|$ contribute exclusively to $\operatorname{Re}S$, while those with an even $|\sigma|$ contribute to $\operatorname{Im}S$.

The learning data used to construct the $256$ correlators $F_{\sigma}$ are acquired using either Algorithm~\ref{alg:nested} or Algorithm~\ref{alg:single-nested}, as detailed below. Once estimated, these same correlators can be reused to predict the probe signal $S(\boldsymbol{\alpha})$ for any arbitrary choice of control angles.

\subsection{Implementation of learning algorithms}
\label{num:sec:sampling}

We first compute the exact correlator values $\{F_{\sigma}\}$ shown in Fig.~\ref{fig:num_exp}b by evaluating Eq.~\eqref{num:eq:maps-to-F} via direct matrix multiplication. These exact values serve as our reference baseline. They also determine the Born probabilities used to simulate the measurement records for both learning algorithms, as detailed below.

\subsubsection{Algorithm~\ref{alg:nested}: shared record}

For a path string $b\in\{0,1\}^8$, let
$\chi_\sigma(b)=(-1)^{\sigma\cdot b}$ and let $Q_b$ be the
complementary-path overlap defined in Eq.~\eqref{nest:eq:general-overlap}.
Inverting Eq.~\eqref{nest:eq:fourier} yields
\begin{equation}
\label{num:eq:overlap-inversion}
Q_b=2\sum_\sigma
\chi_\sigma(b)(-i)^{|\sigma|}F_\sigma.
\end{equation}
The algorithm draws a path $b$ and a fair sign $u$ uniformly, then
measures $(X_Q+uY_Q)/\sqrt2$ and $Z_1$ to obtain outcomes
$a,x\in\{\pm1\}$. Only their product $z=ax$ enters the estimator.
Following Eq.~\eqref{nest:eq:rotated-readout}, the joint distribution of these variables is
\begin{equation}
\label{num:eq:algorithm1-law}
p(b,u,z)=\frac{1}{4M}
\left[1+\frac{z}{\sqrt2}
\left(\operatorname{Re}Q_b-u\operatorname{Im}Q_b\right)\right].
\end{equation}
Recall from Eq.~\eqref{nest:eq:shared-estimator} that the single-record contribution to each target is
\begin{equation}
\label{num:eq:algorithm1-snapshot}
T_\sigma(b,u,z)
=\frac{\chi_\sigma(b)}{2}
\operatorname{Re}\!\left[
i^{|\sigma|}\sqrt2\,z(1-iu)\right]
\in\{-1/\sqrt2,+1/\sqrt2\},
\end{equation}
which satisfies $\mathbb E T_\sigma=F_\sigma$, as proven in Appendix~\ref{nest:sec:upper}.

To generate each dataset, we draw one count vector over the $4M=1024$
possible records and compute the shared estimator:
\begin{equation}
\label{num:eq:algorithm1-counts}
\{C_{buz}\}\sim\operatorname{Multinomial}
\!\left(N,\{p(b,u,z)\}\right),
\qquad
\widehat F^{(1)}_\sigma
=\frac1N\sum_{b,u,z}C_{buz}T_\sigma(b,u,z).
\end{equation}
The same count vector is used to estimate every $\sigma$.

\subsubsection{Algorithm~\ref{alg:single-nested}: separate estimation}
We allocate $m=N/M$ runs to every target $F_{\sigma}$.
At contact $j$, a zero bit $\sigma_j=0$ dictates a projective
measurement of $B$, while a unit bit $\sigma_j=1$ dictates the randomised
rotation $e^{-ia_j\pi B/4}$, as detailed in
Appendix~\ref{nest:sec:single-free}.
The retained output is
\begin{equation}
\label{num:eq:algorithm2-output}
Z_\sigma=\frac{x}{2}\prod_{j=1}^{8}a_j
\in\{-1/2,+1/2\},
\qquad \mathbb E Z_\sigma=F_\sigma.
\end{equation}
Because $Z_\sigma$ is binary, its distribution is given by
$\Pr(Z_\sigma=+1/2)=1/2+F_\sigma$. Independent counts for
the separate targets therefore follow
\begin{equation}
\label{num:eq:algorithm2-counts}
C_\sigma\sim\operatorname{Binomial}(m,1/2+F_\sigma),
\qquad
\widehat F^{(2)}_\sigma=\frac{C_\sigma}{m}-\frac12.
\end{equation}

\subsubsection{Error statistics and construction of Fig.~\ref{fig:num_exp}}
\label{num:sec:statistics}

For each algorithm $\ell\in\{1,2\}$ and each budget
$N\in\{2^{14},2^{16},\ldots,2^{26}\}$, we generate
$D=1024$ independent datasets. Define the root-mean-square (RMS) error $e_{\ell,r}(N)$ for a single dataset $r$, alongside the overall ensemble-averaged RMSE plotted in Fig.~\ref{fig:num_exp}c, as
\begin{equation}
\label{num:eq:rmse}
\begin{split}
e_{\ell,r}(N)
&=\left[\frac1M\sum_\sigma
\left(\widehat F^{(\ell,r)}_\sigma(N)-F_\sigma\right)^2
\right]^{1/2},\\
\operatorname{RMSE}_\ell(N)
&=\left[\frac1D\sum_{r=1}^{D}e_{\ell,r}(N)^2\right]^{1/2}
=\left[\frac1{DM}\sum_{r=1}^{D}\sum_\sigma
\left(\widehat F^{(\ell,r)}_\sigma(N)-F_\sigma\right)^2
\right]^{1/2}.
\end{split}
\end{equation}
The shaded bands in Fig.~\ref{fig:num_exp}c represent the empirical $2.5\%$ and $97.5\%$ quantiles of $e_{\ell,r}(N)$, describing statistical fluctuations across the independent datasets.

To analytically compare the performance of the two algorithms, we define $\mu_2=M^{-1}\sum_\sigma F_\sigma^2$. The expected mean-square errors then satisfy
\begin{equation}
\label{num:eq:population-errors}
\mathbb E[e_{1,r}(N)^2]=\frac{1/2-\mu_2}{N},
\qquad
\mathbb E[e_{2,r}(N)^2]=\frac{M(1/4-\mu_2)}{N}.
\end{equation}
For this model, $\mu_2=1.07718\times10^{-6}$. To achieve equal expected RMSEs, these expressions yield a cost ratio of
\begin{equation}
\label{num:eq:cost-ratio}
\frac{q_2}{q_1}=\frac{N_2}{N_1}
=M\frac{1/4-\mu_2}{1/2-\mu_2}
\simeq128.
\end{equation} 
Conversely, at an equal budget $N$, the corresponding RMSE ratio is approximately $\sqrt{128} \approx 11.31$. The finite-dataset RMSEs at $N=2^{24}$ are $1.72\times10^{-4}$ and $1.96\times10^{-3}$ for Algorithms~\ref{alg:nested} and~\ref{alg:single-nested}, respectively. This confirms a reduction of roughly two orders of magnitude in overall experimental cost.

Panels b and d of Fig.~\ref{fig:num_exp} show results from the first generated dataset ($r=0$) for each algorithm at $N=2^{24}$. In panel d, we set $\alpha_{1,3,5,7}=\theta$ and $\alpha_{2,4,6,8}=\phi$ on a uniform $121\times121$ grid $\mathcal{G}$ over $[-\pi/2,\pi/2]^2$. The exact signal map $S(\theta,\phi)$ is evaluated via Eq.~\eqref{num:eq:signal-expansion}, while the predicted maps are reconstructed from the estimated correlators as
\begin{equation}
\label{num:eq:signal-estimator}
\widehat S_\ell(\theta,\phi)
=2\sum_\sigma w_\sigma(\theta,\phi)
\widehat F^{(\ell)}_\sigma.
\end{equation}
To quantify the accuracy of these reconstructions, we define the relative RMS error over $\mathcal{G}$ as
\begin{equation}
\label{num:eq:relative-signal-error}
\epsilon_{S,\ell}
=\left[
\frac{\sum_{(\theta,\phi)\in\mathcal G}
|\widehat S_\ell(\theta,\phi)-S(\theta,\phi)|^2}
{\sum_{(\theta,\phi)\in\mathcal G}|S(\theta,\phi)|^2}
\right]^{1/2}.
\end{equation}
For the displayed maps, this yields $\epsilon_{S,1}=0.136$ for Algorithm~\ref{alg:nested} and $\epsilon_{S,2}=1.24$ for Algorithm~\ref{alg:single-nested}.

\end{document}